\documentclass[twocolumn]{aastex63}
\usepackage{natbib}
\usepackage{multirow}
\usepackage{tablefootnote}
\usepackage{textcomp}
\usepackage{graphicx}

\usepackage{amssymb}

\newcommand{\hg}{{H$\gamma$}}

\newcommand{\msun}{\ensuremath{\mathrm{M}_{\odot}}}
\newcommand{\lsun}{\ensuremath{\mathrm{L}_{\odot}}}
\newcommand{\msunyr}{{M$_{\sun}$ yr$^{-1}$}}

\newcommand{\kms}{{km s$^{-1}$}}

\def\ha{\mbox {H$\alpha$}}
\def\hb{\mbox {H$\beta$}}
\def\cm3{~cm$^{-3}$}

\newcommand{\mum}{\ifmmode{\rm \mu m}\else{$\mu$m}\fi}
\newcommand{\wba}{W$_{80}$}
\newcommand{\vwu}{{v$_{50}$}}
\newcommand{\vjiu}{{v$_{90}$}}
\newcommand{\vyi}{{v$_{10}$}}

\newcommand{\flxarc}{erg cm$^{-2}$ s$^{-1}$ arcsec$^{-2}$}
\newcommand{\oiia}{[O~{\sc ii}] $\lambda$3726}
\newcommand{\oiib}{[O~{\sc ii}] $\lambda$3729}
\newcommand{\oiiab}{[O~{\sc ii}] $\lambda$$\lambda$3726,3729}
\newcommand{\oiii}{[O~{\sc iii}] $\lambda$5007}
\newcommand{\oiiia}{[O~{\sc iii}] $\lambda$4959}
\newcommand{\oiiib}{[O~{\sc iii}] $\lambda$$\lambda$4959,5007}

\newcommand{\niiab}{[N~{\sc ii}] $\lambda$$\lambda$6548,6583}
\newcommand{\siia}{[S~{\sc ii}] $\lambda$6716}
\newcommand{\siib}{[S~{\sc ii}] $\lambda$6731}
\newcommand{\siiab}{[S~{\sc ii}] $\lambda$$\lambda$6716,6731}
\newcommand{\oi}{[O~{\sc i}] $\lambda$6300}
\newcommand{\oib}{[O~{\sc i}] $\lambda$6364}
\newcommand{\fex}{[Fe~{\sc x}] $\lambda$6375}
\newcommand{\nev}{[Ne~{\sc v}] $\lambda$3426}
\newcommand{\neiii}{[Ne~{\sc iii}] $\lambda$3869}
\newcommand{\oivv}{[O~{\sc i}] $\lambda$5577}

\newcommand{\loiii}{L$_{[O~{\sc III}]}$}
\newcommand{\loii}{L$_{[O~{\sc II}]}$}
\newcommand{\lagn}{L$_{AGN}$}

\newcommand{\hahb}{H$\alpha$/H$\beta$}

\newcommand{\oiiihb}{[O~{\sc iii}]/H$\beta$}
\newcommand{\niiha}{[N~{\sc ii}]/H$\alpha$}
\newcommand{\siiha}{[S~{\sc ii}]/H$\alpha$}
\newcommand{\oiha}{[O~{\sc i}]/H$\alpha$}
\newcommand{\heii}{He~{\sc ii} $\lambda$4686}
\newcommand{\oiioiiihb}{[O~{\sc ii}]/[O~{\sc iii}] vs [O~{\sc iii}]/H$\beta$}
\newcommand{\oiioiii}{[O~{\sc ii}]/[O~{\sc iii}]}

\newcommand{\mgb}{Mg I$b$}

\urldef{\kcwiurl}\url{https://github.com/Keck-DataReductionPipelines/KcwiDRP/blob/master/AAAREADME}
\urldef{\dimmurl}\url{http://mkwc.ifa.hawaii.edu/current/seeing/index.cgi}
\urldef{\irsaurl}\url{https://irsa.ipac.caltech.edu/cgi-bin/Gator/nph-scan?mission=irsa&submit=Select&projshort=2MASS}
\urldef{\nsaurl}\url{http://www.nsatlas.org/data}

\newcommand{\nuchi}{$\chi_{\nu}$}

\newcommand{\ta}{J0906$+$56}
\newcommand{\tb}{J0842$+$03}
\newcommand{\tc}{J0954$+$47}
\newcommand{\td}{J1005$+$12}
\newcommand{\te}{J0100$-$01}
\newcommand{\tf}{J1009$+$26}
\newcommand{\tg}{J0811$+$23}
\newcommand{\tx}{J0840$+$18}

\begin{document}

\title{Integral-Field Spectroscopy of Fast Outflows in Dwarf Galaxies with AGN}

\correspondingauthor{Weizhe Liu}\email{oscarlwz@gmail.com}

\author[0000-0003-3762-7344]{Weizhe Liu}\affiliation{Department of Astronomy, University of Maryland, College Park, MD 20742, USA}

\author[0000-0002-3158-6820]{Sylvain Veilleux} 
\affiliation{Department of Astronomy, University of Maryland, College Park, MD 20742, USA}
\affiliation{Joint Space-Science
  Institute, University of Maryland, College Park, MD 20742, USA}

\author[0000-0003-4693-6157]{Gabriela Canalizo} 
\affiliation{Department of Physics and Astronomy University of California Riverside, 900 University Avenue, CA 92521, USA}

 \author[0000-0002-1608-7564]{David S. N. Rupke}
 \affiliation{Department of Physics, Rhodes College, Memphis, TN 38112, USA}
 
 \author[0000-0002-5253-9433]{Christina M. Manzano-King} 
 \affiliation{Department of Physics and Astronomy University of California Riverside, 900 University Avenue, CA 92521, USA}

\author[0000-0002-4375-254X]{Thomas Bohn} 
\affiliation{Department of Physics and Astronomy University of California Riverside, 900 University Avenue, CA 92521, USA}

\author[0000-0002-1912-0024]{Vivian U}
\affiliation{Department of Physics and Astronomy, 4129 Frederick Reines Hall, University of California, Irvine, CA 92697, USA}

\begin{abstract}
Feedback likely plays a vital role in the formation of dwarf galaxies. While stellar processes have long been considered the main source of feedback, recent studies have revealed tantalizing signs of AGN feedback in dwarf galaxies. In this paper, we report the results from an integral-field spectroscopic study of a sample of eight dwarf galaxies with known AGN and suspected outflows. Outflows are detected in seven of them. The outflows are fast, with 50-percentile (median) velocity of up to $\sim$240 \kms\ and 80-percentile line width reaching $\sim$1200 \kms, in clear contrast with the more quiescent kinematics of the host gas and stellar components. The outflows are generally spatially extended on a scale of several hundred pc to a few kpc, although our data do not clearly resolve the outflows in three targets. The outflows appear to be primarily photoionized by the AGN rather than shocks or young, massive stars. The kinematics and energetics of these outflows suggest that they are primarily driven by the AGN, although the star formation activity in these objects may also contribute to the energy input. A small but non-negligible portion of the outflowing material likely escapes the main body of the host galaxy and contributes to the enrichment of the circumgalactic medium. Overall, the impact of these outflows on their host galaxies is similar to those taking place in the more luminous AGN in the low-redshift universe. 

\end{abstract}

\keywords{galaxies: active $-$ galaxies: dwarf $-$ galaxies: evolution $-$ galaxies: kinematics and dynamics $-$ ISM: jets and outflows}

\section{Introduction} \label{1}

While it is believed that supermassive black holes (SMBH, with masses $M_{BH} \simeq$ 10$^6$ $-$ 10$^9$ \msun) are ubiquitous in the centers of massive galaxies at the present epoch, the rate of incidence of (S)MBH in dwarf galaxies with stellar masses M$_{\star} \lesssim$ 10$^{9.5}$ \msun\ (roughly that of the Large Magellanic Cloud) is not well determined. The direct detection of SMBH in dwarf galaxies based on the stellar and gas dynamics within the gravitational sphere of influence of the SMBH is extremely challenging, although there have been recent efforts producing promising results \citep[][]{Nguyen2018,Nguyen2019}. Nevertheless, recent studies have revealed active galactic nuclei (AGN) in dwarf galaxies through diagnostics in the optical \citep[e.g.][]{Greene2007,Dong2012, Reines2013,Moran2014,Diecky2019,He2_10_2020,Mezcua2020}, near and mid-infrared \citep[e.g.][]{Sartori2015,Hood2017,Riffel2020}, X-rays \citep[e.g.][]{Pardo2016,Mezcua2018}, as well as from optical variability \citep[e.g.][]{Baldassare2018}, opening a new window for systematic studies of (S)MBH in dwarfs \citep[see][for a recent review]{Greene2019}. 

There is a general consensus that feedback processes likely play a vital role in the evolution of dwarf galaxies, given their shallow potential well \citep[e.g.][]{Veilleux2005,Veilleux2020}. Stellar processes have long been considered the main source of feedback in dwarf galaxies \citep[e.g.][]{Larson1974,Veilleux2005,Heckman2017,Martin2018}. However, it is still debated whether such stellar feedback is effective enough to reproduce the properties of the dwarf galaxies we see today \citep[e.g.][]{Garrison-Kimmel2013}. Given the growing number of AGN detected in dwarf galaxies, it is also important to consider the possible impact of AGN feedback. Few studies have explored this issue systematically. Plausible evidence of star formation quenching induced by AGN feedback in dwarf galaxies has been reported by \citet{Penny2018}. \citet{Bradford2018} have also found that the global HI content may be lower in dwarf galaxies with AGN, perhaps due to AGN feedback. In addition, radio observations have revealed radio jets in dwarf galaxies that are as powerful as those observed in more massive systems \citep{Mezcua2019b}. From the theoretical perspective, analytic analyses from \citet{Silk2017} and \citet{Dashyan2018} have pointed out the possibly significant effects of AGN feedback in dwarfs. New simulations by \citet{Koudmani2019,Koudmani2020} suggest that AGN boost the energetics of outflows in dwarf galaxies.

Powerful, kpc-scale outflows triggered by luminous AGN has been regarded as strong observational evidence of on-going AGN feedback \citep[e.g.][]{RupkeVeilleux2011,RupkeVeilleux2013a,RupkeVeilleux2013b,RupkeVeilleux2015,Rupke2017, Liu2013a,Liu2013b,Harrison2014,Westmoquette2013,RamosAlmeida2019}, which may impact even the circumgalactic medium \citep[e.g.][]{Veilleux2014,Lau2018,Liu2019}. It is thus interesting to explore if similar outflows can be found in dwarf galaxies with AGN. Recently, \citet{ManzanoKing2019} have observed a sample of 29 dwarf galaxies with AGN using Keck LRIS long-slit spectroscopy. Spatially extended (up to $\sim$2 kpc in radius), rapid outflows (median velocity offsets $\lesssim$180 km s$^{-1}$, 80-percentile widths \wba\ $\lesssim$ 1600 km s$^{-1}$) have been discovered in a third of the sources from the sample, suggesting that AGN feedback may be significant in these dwarf galaxies. More recently, a parsec-scale radio jet was reported in one of the targets with a reported outflow, adding evidence for AGN feedback in these dwarf galaxies \citep{Yang2020}. However, while the results from the long-slit spectra are tantalizing, they do not capture the two-dimensional morphology of the outflows. Integral field spectroscopy (IFS) that provides full two-dimensional coverage with high spatial resolution is needed to map the outflows and fully quantify the true impact of these outflows on the dwarf hosts. 

In this paper, we analyze newly obtained IFS data of eight dwarf galaxies with AGN showing the fastest and brightest outflowing gas in the sample studied by \citet{ManzanoKing2019}. The eight targets were observed with Keck/KCWI, and two of the targets were also observed with Gemini/GMOS. This paper is organized as follows. In Section \ref{2}, the data sets, physical properties of the targets measured from the IFS and ancillary data, and reduction procedures are described. The analysis techniques adopted in this paper are described in Section \ref{3}. The main results are presented in Section \ref{5} and detailed in Appendix \ref{4}. The implications of these results are discussed in Section \ref{6}, and the conclusions are summarized in Section \ref{7}. Throughout the paper, we assume a $\Lambda$CDM cosmology with $H_0$ = 69.3 km s$^{-1}$ Mpc$^{-1}$, $\Omega_{\rm m}$ = 0.287, and $\Omega_{\rm \Lambda} = 0.713$ \citep{wmap2013}.

\section{Sample, Observations, \& Data Reduction} \label{2}

\subsection{Sample} \label{21}
We observed 8 out of the 29 dwarf galaxies with AGN studied in \citet{ManzanoKing2019}. The 29 sources were originally selected from samples of dwarf galaxies with AGN in recent literatures based on Baldwin, Phillips \& Telervich and Veilleux \& Osterbrock 1987 \citep[hereafter BPT and VO87, respectively;][]{bpt,Veilleux1987} line ratio diagrams \citep{Reines2013,Moran2014} and mid-infrared diagnosis \citep{Sartori2015}. The readers are referred to \citet{ManzanoKing2019} for more details.

All targets are confirmed to host AGN based on the AGN-like line ratiosAll targets show AGN-like line ratios as measured from the Keck/LRIS long-slit spectra extracted from the central 1\arcsec\ region. Many of the targets show further evidence of hosting AGN, including i) the detection of strong \heii\ and \nev\  emission in the Keck LRIS long-slit spectra and KCWI spectra; ii) the detection of coronal emission lines in the near-infrared spectra of these objects (Bohn et al.\ 2020, in prep.). In addition, the highly ionized \fex\ line (I.P.$=$233.6 eV) is detected within the central 0.6\arcsec\ of target \ta\ based on the GMOS Integral Field Unit (IFU) spectra reported here; Targets \ta\ and \tc\ also show hard X-ray emission originating from AGN activity \citep{Baldassare2017}. The basic physical properties of the 8 targets in our sample, including those from the NASA-Sloan Altas\footnote{\nsaurl} (NSA), are summarized in Table \ref{tab:targets}.

\begin{deluxetable*}{cccc ccccc}[!htb]
\tablecolumns{9}
\tablecaption{Properties of the Targets\label{tab:targets}}
\tablehead{ \colhead{Name} & \colhead{Short Name} & \colhead{Redshift} & \colhead{log(M$_{\rm stellar}$/\msun)} &
\colhead{R$_{50}$} &
\colhead{log(L$_{[OIII]}$)} & \colhead{C$_{bol}$} & \colhead{log(L$_{AGN}$)} & \colhead{SFR}  \\
\colhead{(1)} & \colhead{(2)} & \colhead{(3)} & \colhead{(4)} & \colhead{(5)} & \colhead{(6)} & \colhead{(7)} & \colhead{(8)} & \colhead{(9)} 
}
\startdata
SDSS J010005.94$-$011059.0 & J0100$-$01 & 0.0517 & 9.47 & 1.2 & 40.96$^{+0.03}_{-0.04}$ & 142 & 43.5 & $<$0.6 \\
SDSS J081145.29$+$232825.7 & J0811$+$23 & 0.0159 & 9.02 & 0.6 & 39.63$^{+0.05}_{-0.06}$ & 87 & 42.0 & $<$0.01 \\
SDSS J084025.54$+$181858.9 & J0840$+$18 & 0.0151 & 9.28 & 1.0 & 39.96$^{+0.02}_{-0.02}$ & 87 & 42.0 & $<$0.01 \\
SDSS J084234.51$+$031930.7 & J0842$+$03 & 0.0291 & 9.34 & 1.0 & 40.51$^{+0.03}_{-0.03}$ & 142 & 43.1 & $<$0.3 \\
SDSS J090613.75$+$561015.5 & J0906$+$56 & 0.0467 & 9.36 & 1.5 & 41.15$^{+0.01}_{-0.01}$ & 142 & 43.7 & $<$0.3 \\
SDSS J095418.16$+$471725.1& J0954$+$47 & 0.0327 & 9.12 & 2.0  & 41.36$^{+0.02}_{-0.02}$ & 142 & 43.9 & $<$0.3 \\
SDSS J100551.19$+$125740.6 & J1005$+$12 & 0.00938 & 9.97 & 1.0 & 40.20$^{+0.05}_{-0.06}$ & 142 & 43.2 & $<$0.1 \\
SDSS J100935.66$+$265648.9 & J1009$+$26 & 0.0145 & 8.77 & 0.7 & 40.48$^{+0.01}_{-0.01}$ & 142 & 43.0 & $<$0.1 \\
\enddata
\tablecomments{Column (1): SDSS name of the target; Column (2): Short name of the target used in this paper; Column (3): Redshift of the target measured from the stellar fit to the spectrum integrated over the KCWI data cube; Column (4): Stellar mass from the NSA; Column (5): Half-light radius from the NSA, in unit of kpc;
Column (6): Total \oiii\ luminosity based on the observed total \oiii\ fluxes within the field of view of the KCWI data without extinction correction, in units of erg s$^{-1}$; Column (7): [O III]-to-bolometric luminosity correction factor adopted from \citet{Lamastra2009};
Column (8): Bolometric AGN luminosity, based on the extinction-corrected [O~III] luminosity, in units of erg s$^{-1}$; Column (9): Upper limit on the star formation rate based on the extinction-corrected \oiiab\ flux from the KCWI data, in units of \msunyr. Here we assume that 1$/$3 of the \oiiab\ emission is from the star formation activity, following \citet{Ho2005}.}
\end{deluxetable*}

\subsection{Observations} \label{22}

\subsubsection{GMOS Observations} \label{221}
\ta\ and \tb\ were observed through Gemini fast-turnaround (FT) programs GN-2019A-FT-109 and GS-2019A-FT-105 (PI S.\ Veilleux). The GMOS IFU \citep{GMOS,GMOSS} data were taken on 2019-04-04 and 2019-04-05 at Gemini-N for \ta, and on 2019-04-28 and 2019-04-29 at Gemini-S for \tb. The GMOS IFU 1-slit, B600 mode was used for both targets, and the spectral resolution was $\sim$100 \kms\ FWHM at 4610 \AA. The field of view of this GMOS setup is 3.5\arcsec$\times$5\arcsec. The details of the observations are summarized in Table \ref{tab:obs}.

\begin{deluxetable*}{ccccc ccccc}
\tablecolumns{10}
\tabletypesize{\scriptsize}
\tablecaption{Summary of Observations\label{tab:obs}}
\tablehead{\colhead{Name} & Telescope/Instrument & \colhead{Dates} & \colhead{Grating(Slicer)} & \colhead{t$_{exp}$} &  \colhead{PSF} & \colhead{Range} & \colhead{PA} & \colhead{FOV} & 5-$\sigma$ detection \\
& & & & &  & & & & limit ($\times$10$^{-17})$
}
\colnumbers
\startdata
\te &    Keck/KCWI                           & 2020-01-30            & BL(Small)   & 1200$+$600      & 1.2\arcsec  & 3500--5500 \AA &  51.0   & 8\arcsec$\times$20\arcsec & 9 \\
\tg &    Keck/KCWI                           & 2020-01-30            & BL(Medium)   & 4$\times$1200      & 1.2\arcsec  & 3500--5500 \AA &  0.0   & 16\arcsec$\times$20\arcsec & 1 \\
\tx &    Keck/KCWI                           & 2020-01-30            & BL(Medium)   & 3$\times$1200      & 1.2\arcsec  & 3500--5500 \AA & 101.0    & 16\arcsec$\times$20\arcsec &  1 \\
\tb &    Gemini/GMOS                          & 2019-04-28,29 & B600 & 8$\times$1125     & 0.55\arcsec & 3750--7070 \AA & 122.0 & 3.5\arcsec$\times$5\arcsec & 1 \\ 
\tb &    Keck/KCWI                           & 2020-01-31            & BL(Small)   & 2$\times$1200      & 0.9\arcsec  & 3500--5500 \AA & 290.0    &  8\arcsec$\times$20\arcsec & 1 \\
 \ta & Gemini/GMOS & 2019-04-04,05 & B600 & 8$\times$1155     & 0.6\arcsec  & 3880--7200 \AA \tablenotemark{a} & 273.0 & 3.5\arcsec$\times$5\arcsec & 3 \\
\ta & Keck/KCWI   & 2020-01-31            & BL(Small)   & 2$\times$1200+280 & 0.9\arcsec  & 3500--5500 \AA &  0.0   & 8\arcsec$\times$20\arcsec & 2 \\
\tc &     Keck/KCWI                          & 2020-01-30            & BL(Small)   & 5$\times$1200     & 1.2\arcsec  & 3500--5500 \AA &  0.0   & 8\arcsec$\times$20\arcsec & 2  \\
\td &   Keck/KCWI                            & 2020-01-30            & BL(Small)   & 6$\times$600      & 1.2\arcsec  & 3500--5500 \AA & 60.0  & 8\arcsec$\times$20\arcsec & 3 \\
\tf &   Keck/KCWI                            & 2020-01-30            & BL(Small)   & 7$\times$600      & 1.2\arcsec  & 3500--5500 \AA & 45.5    & 8\arcsec$\times$20\arcsec & 5
\enddata
\tablecomments{Column (1): Short name of the target; Column (2): Telescope and instrument used for the observations; Column (3): Date of the observation; Column (4): Grating adopted in the observation, slicer configuration adopted for the corresponding KCWI observation is also shown in the bracket; Column (5): Exposure time of the observation in seconds; Column (6): FWHM of the PSF measured from the acquisition image (GMOS data) or IFU observation of the spectrophotometric standard star (KCWI data); Column (7): Spectral coverage of the data set; Column (8): Position angle of the IFU in degrees measured East of North; Column (9): Full field of view of the IFU. (10): 5-$\sigma$ detection limit for a \oiii\ emission line with FWHM of 1000 \kms, in units of  \flxarc. The typical uncertainty of the listed values is $\sim$30\%}
\tablenotetext{a}{The data with wavelength shorter than 5000 \AA\ were discarded in the analysis due to the low S/N.}
\end{deluxetable*}

We measured the point spread function (PSF) of the IFS data by fitting single 2-D Gaussian profiles to bright stars in the acquisition images of each target. The mean values of the measured FWHM (0.60\arcsec\ for \ta\ and 0.55\arcsec\ for \tb) were used as the empirical Gaussian PSF for the IFS data. Whether these PSF are a good approximation for our analysis can be checked by comparing the PSF of the acquisition images of the standard stars with those of the IFS frames on the stars themselves. We find that the former is more extended than the latter, i.e., the average FWHM of the PSF for the acquisition images is $\sim$90\% larger than that of the IFS frames in arcseconds, although the former is only $\sim$15\% larger than the latter in unit of image pixel size. This suggests that the FWHM of the PSF determined from the acquisition images overestimate those of the science observations. Thus, the use of PSF measurements derived from the acquisition images in our analysis conservatively overestimates the true size of the PSF in the IFS observations on our targets.  

\subsubsection{KCWI Data} \label{222}

All targets were observed with KCWI \citep{KCWI} through Keck program 2019-U217 (PI G.\ Canalizo) on 2020-01-31 and 2020-02-01. All targets were observed with BL grating. \tg\ and \tx\ were observed with the medium-slicer setup (spectral resolution $\sim$160 \kms\ FWHM at 4550 \AA), while the others were observed with the small-slicer setup (spectral resolution $\sim$80 \kms\ FWHM at 4550 \AA). The details of the observations are summarized in Table \ref{tab:obs}. 

We measured the PSF of these IFU observations from the observations of spectrophotometric standard stars taken before, in between, and after the on-target observations, where single 2-D Gaussian profiles were fit to the narrow-band images (5000--5100 \AA) of those standard stars reconstructed from the data cubes. For one of the targets, \tb, a nearby bright star fell in the field-of-view and was thus observed simultaneously with the target in one science exposure. The same 2-D Gaussian fit was applied to it and the results were compared with other PSF measurements. For each night, all individual measurements of the PSF described above broadly agree with each other, and the median FWHM of these best-fit Gaussian profiles were adopted as the FWHM of the PSF for further analysis. Notice that we do not have measurements for the PSF taken at the same time of the on-target science observations, therefore, the variations in the size of the actual PSF may be larger. This speculation is based on the variation of the DIMM seeing measured by the Mauna Kea Weather Center\footnote{\dimmurl}, which ranges from 0.4\arcsec\ to 0.8\arcsec\ throughout the two observation nights.  

\subsection{Data Reduction} \label{23}

\subsubsection{GMOS Data} \label{231}
Both GMOS data sets were reduced with the standard Gemini Pyraf package (v1.14), supplemented by scripts from IFSRED library \citep{ifsred}. We followed the standard processes listed in the GMOS data reduction manual, except that we did not apply scattered light removal for the science frames. This was based on the fact that i) there was no clear features indicative of scattered light in the raw data and ii) the attempt to apply scattered light removal led to significant and unphysical wiggles in the extracted spectra. 

The final data cubes were generated by combining individual exposures of each target using script IFSR\_MOSAIC from the IFSRED library. The wavelength solutions were further verified by checking the sky emission lines (mainly \oivv, and also weaker \oi\ and \oib). For J0906$+$56, the differences between the measured line centers of the sky emission and the reference values are between $-$10 \kms\ and 10 \kms. The differences are randomly distributed across the data cube and no pattern is seen. Therefore, no further correction was applied to the wavelength calibration. 

However, for target \tb, shifts of up to $\sim$5 \AA\ between the measured and reference line centers of the sky line \oivv\ were seen.  The arc exposure for this target was taken eleven days after the science observations, perhaps explaining these large shifts. Additional corrections were applied to modify the wavelength solutions: i) For each exposure, the zero-point shifts of the spectra were corrected using the sky emission \oivv; ii) for the final combined data cube, small ($\lesssim$ 0.8 \AA), wavelength dependent shifts in the wavelength solution were further corrected by adding shifts $\Delta(\lambda)$, where $\Delta(\lambda/{\mbox{\normalfont\AA}})= 0.0016(\lambda/{\mbox{\normalfont\AA}}) - 9.06$ is the best-fit linear fit to the shifts between the measured line centers and the expected ones calculated from the emission-line redshift determined from the Keck/LRIS spectrum (Manzano-King, private communication). The strong optical emission lines \oiii, \ha, \niiab, and \siiab\ were included in the fit. We further required that $\Delta(\lambda_{[O~I]  \lambda5577}/{\mbox{\normalfont\AA}})$ = 0, i.e., zero shifts at the wavelength of sky emission line \oivv. The residuals of the best-fit are $\lesssim$ 0.15 \AA\ in general.

\subsubsection{KCWI Data} \label{232}

The KCWI data sets were reduced with the KCWI data reduction pipeline and the IFSRED library. We followed the standard processes listed in the KCWI data reduction manual\footnote{\kcwiurl} for all targets. The data cubes generated from individual exposures were resampled to 0.15\arcsec\ $\times$ 0.15\arcsec\ (small-slicer setup) or 0.29\arcsec\ $\times$ 0.29\arcsec\ square spaxels (medium-slicer setup) using IFSR\_KCWIRESAMPLE. The resampled data cubes of the same target were then combined into a single data cube using IFSR\_MOSAIC.  

\section{Analysis} \label{3}

\subsection{Voronoi Binning} \label{31}

The data cubes were mildly, spatially binned using the Voronoi binning method \citep{voronoi}. As our aim is to characterize the broad, blueshifted components in the emission lines (especially \oiii) which trace the outflows, we binned the data cube according to the signal-to-noise ratio (S/N) of the blue wing of the \oiii\ emission line (calculated in the target-specific, 200 \kms-wide velocity window). The spaxels with S/N of the blue wing less than 1 were excluded from the binning, and each final spatial bin was required to reach a minimum S/N of 3.  

\begin{figure*}[!htb]
\begin{minipage}[t]{0.5\textwidth}
\includegraphics[width=\textwidth]{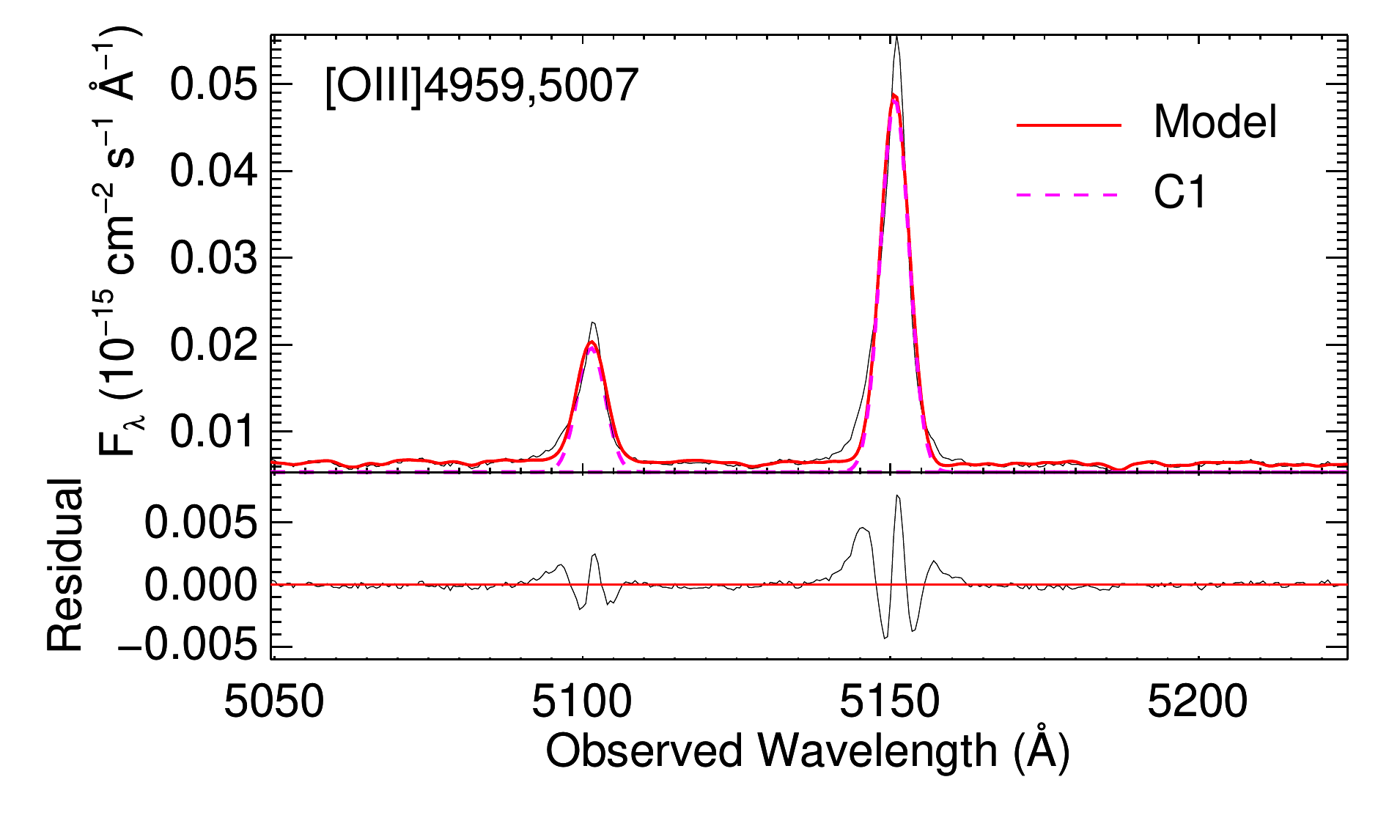}
\end{minipage}
 \begin{minipage}[t]{0.5\textwidth}
\includegraphics[width=\textwidth]{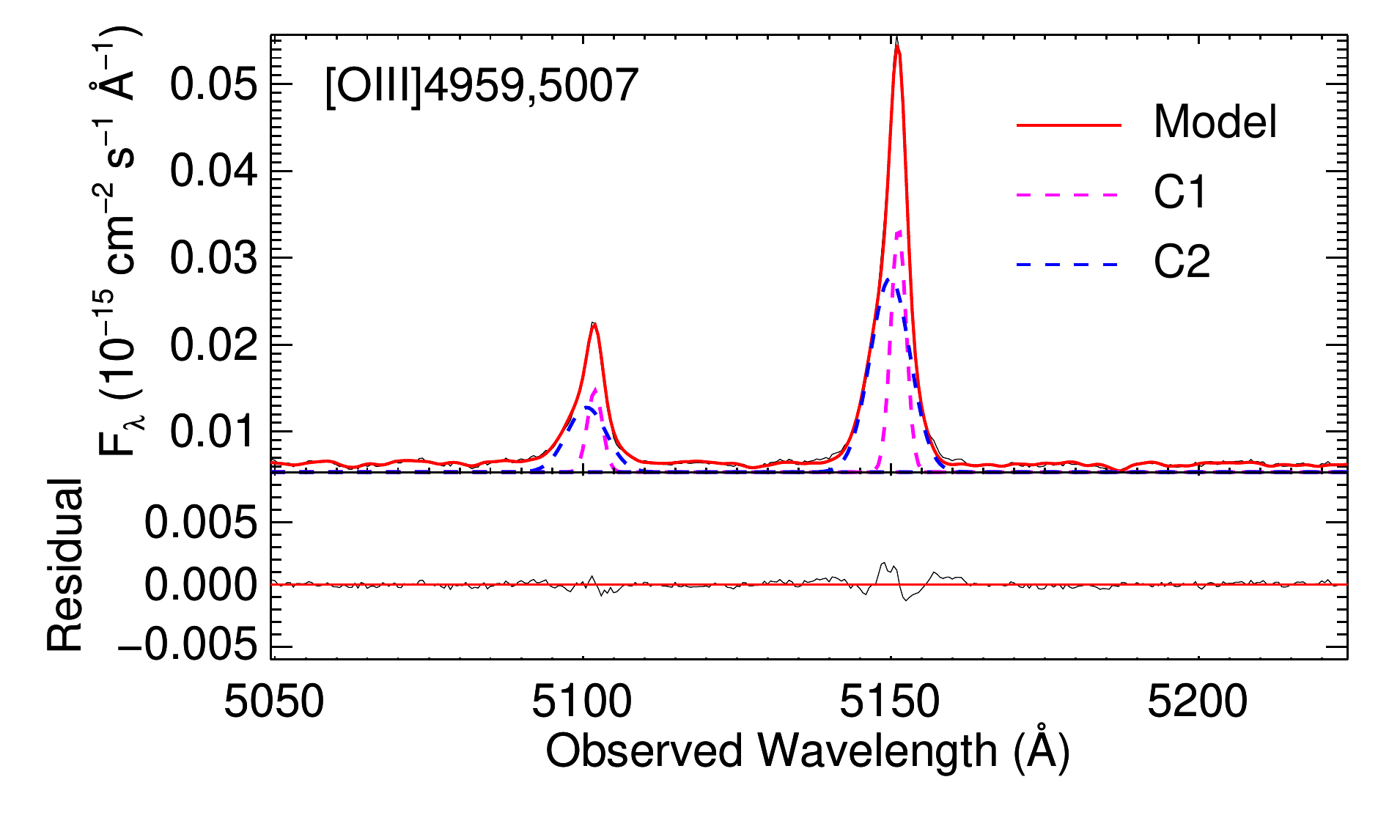}
\end{minipage}
\caption{Examples of fits to the \oiiib\ line profiles for \tb\ using (left panel) one Gaussian component and (right panel) two Gaussian components.  In each panel, the top spectrum in black is the observed data, while the solid red curve is the best fit model and the dashed curves represent the individual Gaussian components (C1, C2). The residuals after subtraction of the best-fit models from the data are shown in solid black curve at the bottom, and the y$=$0 line is shown in red.
\label{fig:J0842o3fit}}
\end{figure*}

\begin{figure*}[!htb]
\begin{minipage}[t]{0.5\textwidth}
\includegraphics[width=\textwidth]{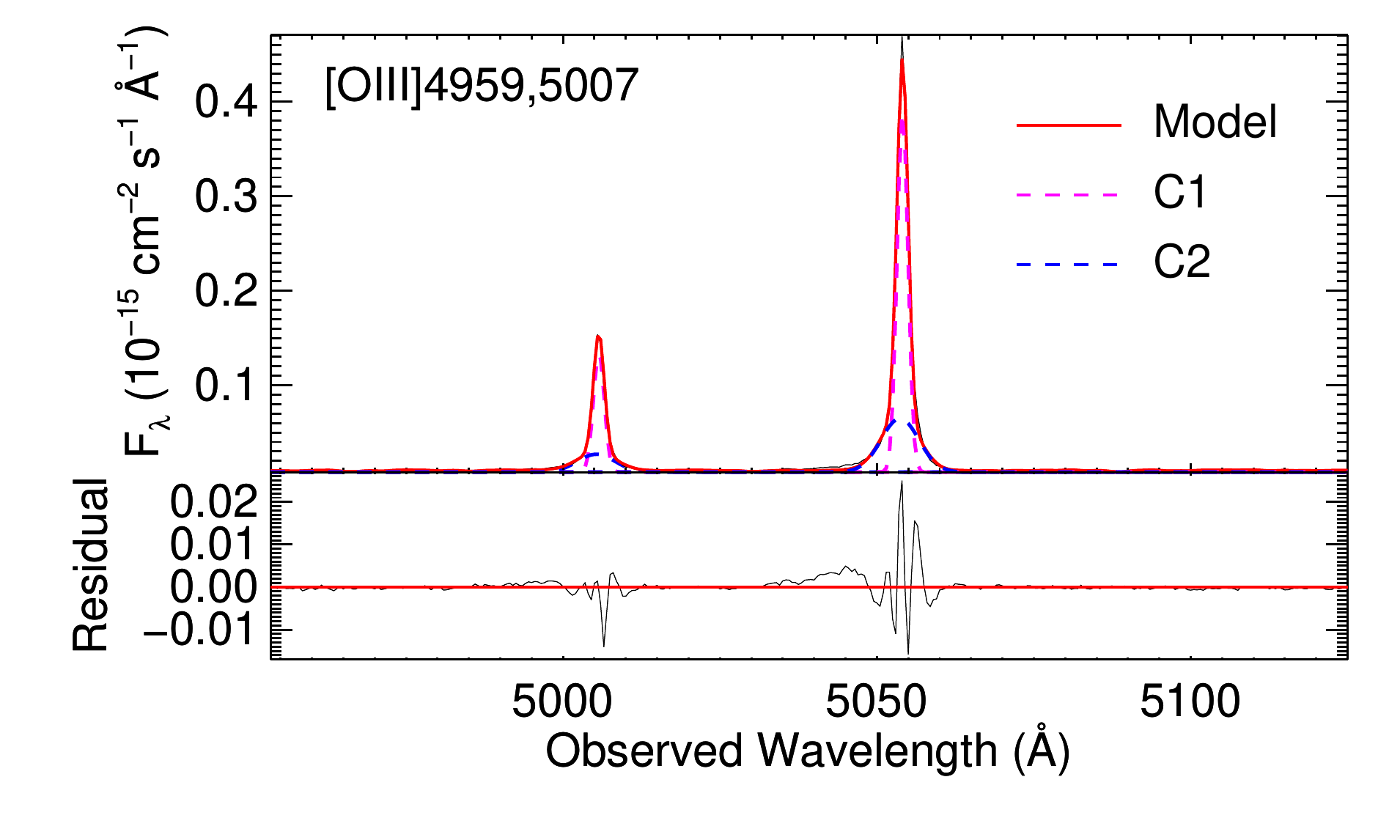}
\end{minipage}
\begin{minipage}[t]{0.5\textwidth}
\includegraphics[width=\textwidth]{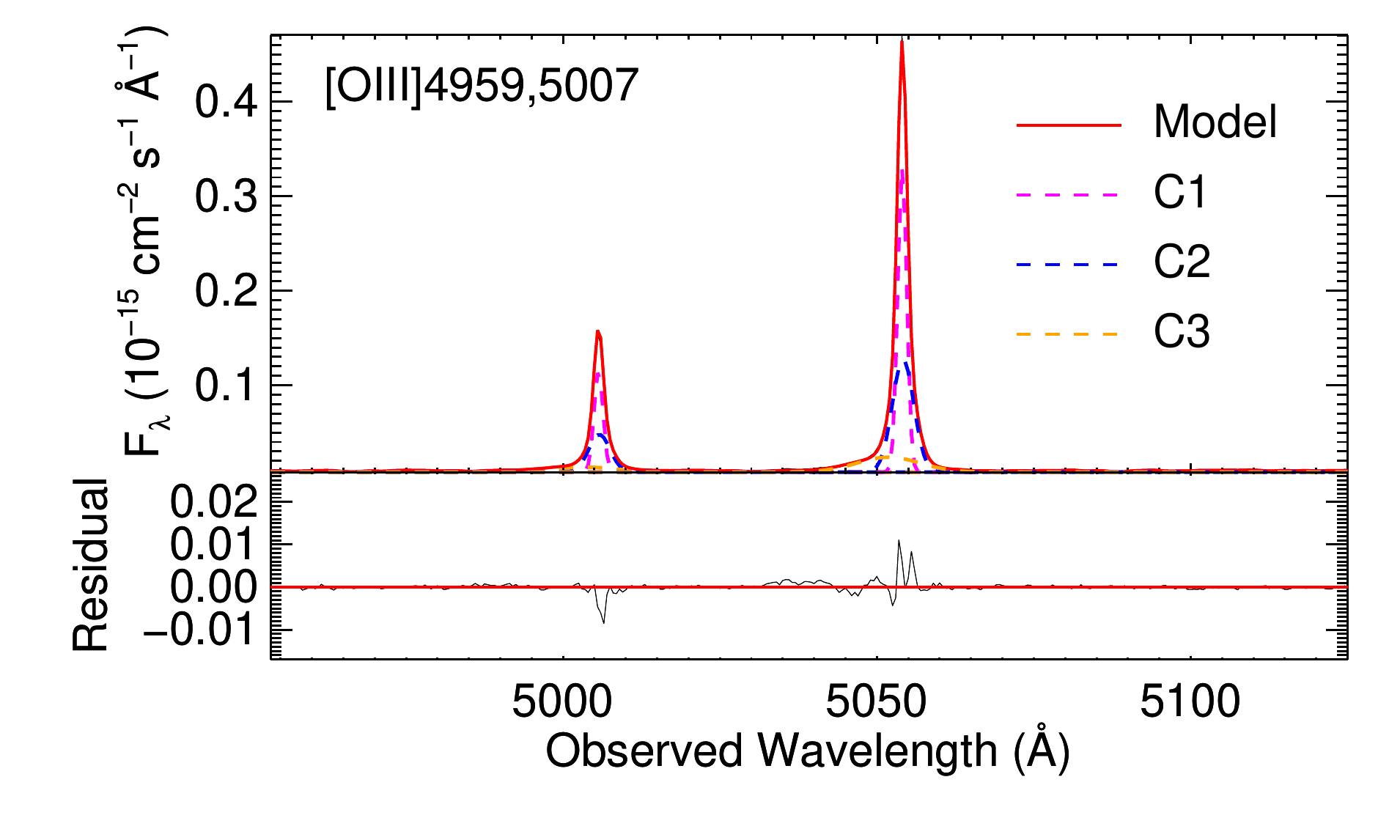}
\end{minipage}
\caption{Examples of fits to the \oiiib\ line profiles for \td\ using (left panel) two Gaussian components and (right panel) three Gaussian components.   The presentation of the data, fits, and residuals is the same as in Fig. \ref{fig:J0842o3fit}. 
\label{fig:J1005o3fit}}
\end{figure*}

\subsection{Spectral Fits} \label{32}
The spectral fits utilized IDL library IFSFIT \citep{ifsfit}, supplemented by customized python scripts. 

\subsubsection{Fits to the \oiiib\ Emission} \label{321}

The \oiiib\ line emission from our targets shows the strongest blueshifted wings among all of the emission line tracers of the ionized outflow. In addition, the absence of other strong emission and absorption features in the vicinity of \oiiib\ makes the faint [O~III] wing components easier to analyze. In order to capture the faintest signal from the outflows traced by those faint emission line wing, we started by solely fitting the \oiiib\ line emission. With the emission lines masked out, the stellar continuum was fit using the public software pPXF \citep{ppxf} with 0.5$\times$ solar metallicity stellar population synthesis (SPS) models from \citet{Delgado2005}. Polynomials of order up to 4 were added to account for any non-stellar continua. 

The continuum-subtracted \oiiib\ emission lines were then fitted with multiple Gaussian components using the IDL library MPFIT \citep{mpfit}. The line centers and line widths of the corresponding Gaussian components of both lines were tied together, and only the amplitudes were allowed to change freely. We did not fix the relative amplitude ratios of the doublet so that a fit was allowed when a Gaussian component was only detected in \oiii\ but not in \oiiia. We checked the flux ratios of the doublet from the best-fit results afterwards when applicable and found that they were very close to the theoretical expectation (within 2\%). We allowed a maximum of three Gaussian components in the fits, and the required number of components in each spaxel was determined by a combination of software automation and visual inspection: 
An additional component was added to the best-fit model when 1) it was broader than the spectral resolution; 2) it had a S/N $>$ 2; 3) it was not too broad to be robustly distinguished from the continuum (i.e., the peak S/N of individual spectral channel was required to be greater than 1.5 when the line width \wba\ was greater than 800 \kms). The best-fit parameters from the continuum and emission line fits were adopted as initial parameters for a second fit to check for convergence of the fit. 

In order to check how the uncertainties on the fit to the stellar continuum might affect the results on the \oiiib\ emission lines, we also tried fitting the continuum with a straight line through the continuum-only windows adjacent to the \oiiib\ emission lines. The differences of the best-fit parameters of the \oiiib\ emission lines between the two continuum fitting schemes were on average less than 2\%, indicating that the best-fit results were not sensitive to the choice of continuum fitting function in most cases.

Examples of the multi-Gaussian fits, using the KCWI spectra of targets \tb\ and \td,  are shown in Figs.\ \ref{fig:J0842o3fit} and \ref{fig:J1005o3fit}, respectively.

For \tb, a model with one Gaussian component cannot fit the spectra well (\nuchi\ $>>$ 1). Two Gaussian components, the narrower C1 component, and the broader C2 component, are enough to describe the \oiiib\ emission profiles. For \td, neither a model with one Gaussian component nor one with two Gaussian components can fit the \oiiib\ profiles well (\nuchi\ $>>$ 1 and \nuchi\ $=$ 3.36, respectively). Three Gaussian components are needed to properly fit the \oiiib\ line emission: the narrowest component (C1), the intermediate-width component (C2) and the broadest component (C3). For the rest of the paper, we name the individual velocity components with the same rule adopted here, i.e., the C1, C2, and C3 components are defined by their increasing line widths. 

The results from these fits are discussed in detail in Appendix \ref{4} and summarized in Section \ref{5}.

\subsubsection{Emission Line Fits to the Full Spectral Range} \label{322}

Emission line fits to the full spectral range were also carried out where all of the strong emission lines (\ha, \hb, \oiiib, \niiab, \siiab, and \oi\ in the GMOS data, \hb, \hg, \oiiab, \neiii, and \oiiib\ in the KCWI data) were fit simultaneously. The continuum-subtracted spectra obtained from Section \ref{321} were adopted for these fits. Following the routine adopted for the fit of the \oiiib\ emission lines alone, all of the emission lines were fitted with multiple Gaussian components, where the line centers and widths of the corresponding Gaussian components for each line were tied together. For each target, the maximum number of Gaussian components used in the fit was determined from the best fits of \oiiib\ emission described in Section \ref{321}. Based on the best-fit results obtained above, we did not detect additional, distinct broad hydrogen Balmer line emission that can be attributed to a genuine broad-line-region (BLR) in any of the eight targets.

\subsection{Non-Parametric Measurements of the Emission Line Profiles} \label{33}

\begin{figure}[!htb]   
\plotone{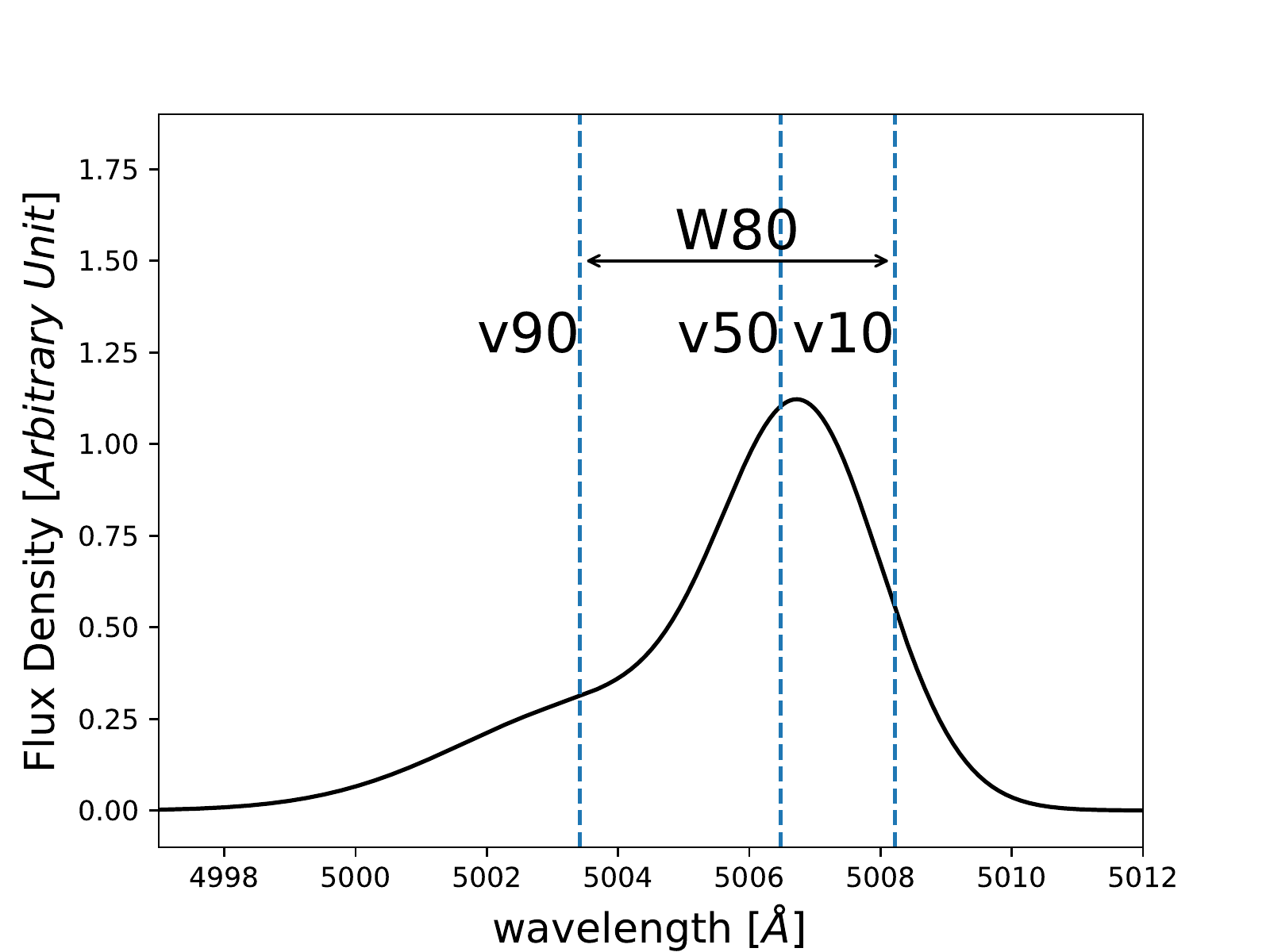}
\caption{Example of a line profile illustrating the various non-parametric kinematic parameters used in this paper. The vertical dashed lines mark the locations of \vyi, \vwu, and \vjiu\ for the mock emission line profile shown in the figure. \wba\ is the line width between \vjiu\ and \vyi.}
\label{fig:nonpar}
\end{figure}

Non-parametric line profile measurements were utilized to describe the gas kinematics for both the individual Gaussian components and the overall line profiles. The details are described below, and an example is shown in Fig. \ref{fig:nonpar}.

i. \vyi\ and \vjiu\ are the velocities at the 10th and 90th percentiles of the total flux, respectively, calculated starting from the red side of the line.

ii. \wba\ is the line width defined to encompass 80 percent of the total flux such that \wba=\vyi $-$ \vjiu. 

iii. \vwu\ is the median velocity, the velocity at the 50th percentile of the total flux.

\subsection{AGN Luminosities} \label{34}

The bolometric AGN luminosities (\lagn) of our targets were calculated from the extinction-corrected \oiii\ luminosities integrated over the entire IFS data cubes (\loiii) \footnote{Based on the [O II]/[O III] vs [O III]/\hb\ diagrams drawn from the KCWI data, at least $\sim$90\% of the spaxels show AGN-like line ratios in each target. Consistently, all of our targets show AGN-like line ratios in the BPT and VO87 diagrams based on the Keck/LRIS spectra extracted from the central 1\arcsec\ box regions. Moreover, for targets \tb\ and \ta\ where the BPT and VO87 diagrams can be derived from the GMOS IFU data, we find that the spaxels with AGN-like line ratios contribute at least $\sim$95\% of the [O III] flux. Overall, the [O III] luminosities integrated over the entire data cubes are thus at most slight overestimates of the [O III] luminosities originating from the AGN.}. The extinction correction was determined from the Balmer decrement based on the spatially-integrated spectrum, assuming an intrinsic \ha/\hb\ ratio of 2.87\footnote{While studies have shown that the intrinsic \hahb\ ratio of AGN is 3.1 \citep[][]{Osterbrock2006}, we adopt the value 2.87 since (1) the intrinsic Balmer line ratios of AGN in these dwarf galaxies are poorly constrained due to a lack of dedicated studies; (2) in Section \ref{63}, we will compare our results of outflow energetics with those from some previous studies \citep[e.g.][]{Harrison2014,Rupke2017} where they adopted the value 2.87. Nevertheless, if we adopt instead an intrinsic \hahb\ value of 3.1 in our calculations, the derived AGN luminoisity will only decrease by $\sim$0.1 dex for our targets.} for the GMOS data, or an intrinsic \hb/\hg\ ratio of 2.13 for the KCWI data \citep[Case B, T=10$^4$ K;][]{Osterbrock2006} and the \citet{ccm89} extinction curve with $R_V$ $=$ 3.1. For \te\ and \tg, where \hg\ is too weak to be measured robustly, the Balmer decrement was determined from the \ha/\hb\ ratio measured from the SDSS spectra. We adopted the empirical bolometric correction factors in \citet{Lamastra2009}: \lagn $=$ 142 \loiii\ and \lagn $=$ 87 \loiii\ for 40 $<$ log(\loiii) $<$ 42 and 38 $<$ log(\loiii) $<$ 40 in cgs units, respectively. Note that the AGN luminosities calculated here may be affected by relatively large systematic errors since the intrinsic Balmer line ratio, the shape of the extinction curve, and the \loiii\ to \lagn\ correction factor in systems like our targets are uncertain. The observed \loiii\ and derived \lagn\ are summarized in Table \ref{tab:targets}.

\subsection{Upper Limits on the Star Formation Rates} \label{35}

Robust star formation rate (SFR) measurements of our targets cannot be obtained due to the lack of sensitive far-infrared data. None of the targets is detected in IRAS and AKARI all sky survey. An order-of-magnitude estimate of SFR for our targets can be derived by dividing the stellar mass with the Hubble time, assuming a constant star formation rate. For a stellar mass of log($M_{\rm stellar}$/\msun) $=$ 9.5, this gives a SFR on the order of 0.2 \msunyr, an order of magnitude lower than the upper limits derived from the far-infrared data. 

Star formation rates may also be estimated from \oiiab\ luminosities (\loii) in AGN \citep[e.g.][]{Ho2005}. The derived SFR are in principle upper limits on the intrinsic SFR since the AGN contributes to the \oiiab\ fluxes. Adopting equation (10) in \citet{Kewley2004}, we follow the same recipe in \citet[][]{Ho2005}, where 1/3 of the [O~II] emission comes from the star formation activity. The \loii\ was measured from the spatially-integrated KCWI spectra, and was corrected for extinction in the same way as that for \loiii. The gas-phase metallicity of the targets adopted in the calculations above were assumed to be solar (This is based on our ionization diagnosis in Section \ref{53}. Given that the \oiiab\ flux is dominated by the nuclear region and contaminated by AGN emission, a metallicity higher than the prediction from the stellar mass--metallicity relation is not surprising). These results are summarized in Table \ref{tab:targets}. Instead, if we use 0.5 $\times$ solar (LMC-like) metallicity \citep[e.g.][]{LMCmetal1999} in the calculations, the upper limits on SFR will be $\sim$20\%\ lower. Therefore, the upper limits recorded in Table \ref{tab:targets} are conservatively high.

To assess the upper limits on SFR derived above, we have also compared them to the median SFR listed in the MPA-JHU DR7 catalog based on SDSS data \citep{Brinchmann2004}. One possible caveat of the SFR from MPA-JHU DR7 catalog is that they misclassify 6 out of the 8 targets studied here as starburst/star-forming galaxies. Therefore, for these 6 targets, there could be significant systematic errors in the SFR listed in the catalog. Moreover, even for the two targets classified as AGN (\tg\ and \td), the treatment of AGN contamination to the SFR measurements might still introduce certain systematic errors to the SFR. Nevertheless, from the comparison we find that (a) the median SFR measured within the SDSS fibers in the MPA-JHU catalog are all below our [O II]-based upper limits except for \tg\ (SFR $\simeq$ 0.02 \msunyr\ from fiber SFR in catalog vs SFR $<$ 0.01 \msunyr\ from our [O II] data); (b) Even if we consider the total SFR (corrected for fiber loss) listed in the MPA-JHU catalog, only three targets show clearly higher SFR in the catalog than our [O II]-based upper limits (the largest difference is seen for \ta: total SFR $\simeq$ 0.74 \msunyr\ in the catalog vs SFR $<$ 0.3 \msunyr\ from our data), while the SFR of \tb\ in the catalog is only 1/10 of the upper limit measured from our [O II] data. These differences are likely caused by the fact that the AGN emission in these targets are not modelled properly in the MPA-JHU catalog. In general, our [O II]-based upper limits are not systematically lower than the values from MPA-JHU catalog.

\section{Outflows Detected in the Sample} \label{5}

The main results from our analysis of the IFS data are summarized in this section. The target-specific maps of the \oiii\ flux and kinematics, globally and for each velocity component, the stellar kinematics, and the radial profiles of the fluxes from individual velocity components are discussed in Appendix \ref{4} (Fig. \ref{fig:o3map1}--\ref{fig:radial6}). In addition, line ratio maps and the spatially resolved BPT and VO87 diagrams are shown for \tb\ (Fig. \ref{fig:J0842bpt1} and \ref{fig:J0842bpt2}) and \ta\ (Fig. \ref{fig:J0906bpt2} and \ref{fig:J0906bpt3}). In all cases, the systematic velocities of our targets are determined from the stellar velocities measured from the spectra integrated over the whole KCWI data cubes.

\begin{deluxetable*}{ccccrrrrrrr}[!htb]
\tablecolumns{11}
\tablecaption{Kinematic Properties of the Targets\label{tab:kinematics}}
\tablehead{
\colhead{Name} & \colhead{N$_{comp}$} & \colhead{Component} & \colhead{Data Set} & \colhead{Median \vwu}  &   \colhead{Min \vwu} & \colhead{Max \vwu}        &  \colhead{Median \wba}   & \colhead{Max \wba} & \colhead{\vwu, int.} & \colhead{\wba, int.} \\
& & & & \colhead{[\kms]} & \colhead{[\kms]} & \colhead{[\kms]} & \colhead{[\kms]} & \colhead{[\kms]} & \colhead{[\kms]} & \colhead{[\kms]} \\
\colhead{(1)} & \colhead{(2)} & \colhead{(3)} & \colhead{(4)} & \colhead{(5)} & \colhead{(6)} & \colhead{(7)} & \colhead{(8)} & \colhead{(9)} & \colhead{(10)} & \colhead{(11)} 
}
\startdata
\hline
\te & 2  & C1    & KCWI & $-$20  & $-$60  & 0      & 120 & 210  &  ...      & ...    \\
    &    & C2    & KCWI & $-$40  & $-$240 & 50     & 310 & 650  & ...       & ...    \\
    &    & Total & KCWI & $-$20  & $-$130 & 0      & 220 & 440  & $-$20  & 150 \\
\\
\tg & 1  & C1    & KCWI & $-$40  & $-$60  & $-$20  & 140 & 220  & $-$40  & 150 \\
\\
\tx & 1  & C1    & KCWI & $-$10  & $-$30 & 20     & 50   & 130  & $-$10  & 50  \\
\\
\tb & 2 & C1    & GMOS & $-$80  & $-$110 & $-$20  & 130 & 250  &   ...     &  ...   \\
    &   & C2    & GMOS & $-$160 & $-$220 & $-$110 & 500 & 650  & ...       & ...     \\
    &   & Total & GMOS & $-$110 & $-$150 & $-$80  & 400 & 520  & $-$120 & 420 \\
    &   & C1    & KCWI & $-$30  & $-$60  & 10     & 150 & 220  & ...       &  ...   \\
    &   & C2    & KCWI & $-$110 & $-$160 & $-$40  & 500 & 750  &  ...      &  ...   \\
    &   & Total & KCWI & $-$70  & $-$110 & $-$20  & 400 & 700  & $-$60  & 320 \\ 
\\
\ta & 3 & C1    & GMOS & $-$10  & $-$30  & 30     &  30\tablenotemark{a} & 30\tablenotemark{a}   &   ...     &  ...   \\
    &   & C2    & GMOS & 30     & $-$10  & 60     & 350 & 410  &  ...      &  ...   \\
    &   & C3    & GMOS & $-$50  & $-$100 & 40     & 920 & 1200 &  ...      &  ...   \\
    &   & Total & GMOS & 0      & $-$20  & 20     & 550 & 650  & 10     & 570 \\
    &   & C1    & KCWI & $-$10  & $-$50  & 50     & 110 & 140  & ...       & ...  \\
    &   & C2    & KCWI & 60     & 30     & 90     & 430 & 680  & ...      &   ...  \\
    &   & C3    & KCWI & $-$70  & $-$150 & 10     & 980 & 1250 &  ...    & ...  \\
    &   & Total & KCWI & 10     & $-$50  & 50     & 520 & 670  & 20     & 420 \\
\\
\tc & 3 & C1    & KCWI & 10     & 0      & 20     & 70  & 100  &  ...      &  ...   \\
    &   & C2    & KCWI & 0      & $-$70  & 20     & 260 & 430  & ...       &  ...   \\
    &   & C3    & KCWI & $-$60  & $-$80  & 0      & 730 & 1100 &  ...      &  ...   \\
    &   & Total & KCWI & 0      & $-$10  & 10     & 240 & 530  & 0      & 220 \\ 
\\
\td   & 3  & C1    & KCWI & $-$20  & $-$40  & 10      & 80 & 120 & ...       & ...    \\
      &    & C2    & KCWI & $-$30  & $-$100 & 50     & 440 & 710  & ...       & ...    \\
      &    & C3    & KCWI & $-$140 & $-$200 & $-$60  & 730 & 1200 &  ...      &  ...   \\
      &    & Total & KCWI & $-$30  & $-$60  & 10     & 300 & 680  & $-$30  & 260 \\
\\
\tf   & 2 & C1    & KCWI & $-$10  & $-$30  & 0      & 80 & 100  &   ...     &   ...  \\
      &   & C2    & KCWI & $-$20  & $-$60  & 40     & 210 & 480  &   ...     &  ...   \\
      &   & Total & KCWI & $-$10  & $-$50  & 10     & 90  & 150  & $-$20  & 90 
\enddata
\tablecomments{Column (1): Short name of the target; Column (2): Number of velocity components required by the best-fit results from Section \ref{32}; (3): Individual velocity components (C1,C2,C3) and overall emission line profiles (Total) from the best fits; Column (4): Instrument used for the observations; Columns (5)-(7): Median, minimum and maximum values of \vwu\ measured across the whole data cube. The spaxels with the highest and lowest 5\%\ of \vwu\ are excluded in the calculations. The values listed are rounded to the nearest 10 \kms; Columns (8)--(9): Median and maximum values of \wba\ measured across the whole data cube. The spaxels with the highest and lowest 5\%\ of \wba\ are excluded in the calculations. The values listed are rounded to the nearest 10 \kms; Columns (10)--(11): \vwu\ and \wba\ of the overall emission line profiles from the spatially-integrated spectra of the whole data cubes. The values listed are rounded to the nearest 10 \kms.}
\tablenotetext{a}{Compared with the KCWI data, the GMOS data has a poorer spectral resolution (FWHM $\simeq$ 100 \kms\ vs 80 \kms) and a shallower depth (see Table \ref{tab:obs}). The significantly smaller line width of C1 component measured in the GMOS data is thus most likely due to that the decomposition of the emission line profile is less constrained in the GMOS data. Therefore, for this target, we adopt the KCWI-based line width measurements of the C1 components as the fiducial values in our analysis instead.}
\end{deluxetable*}

\begin{figure*}[!htb]   
\epsscale{1.25}
\plotone{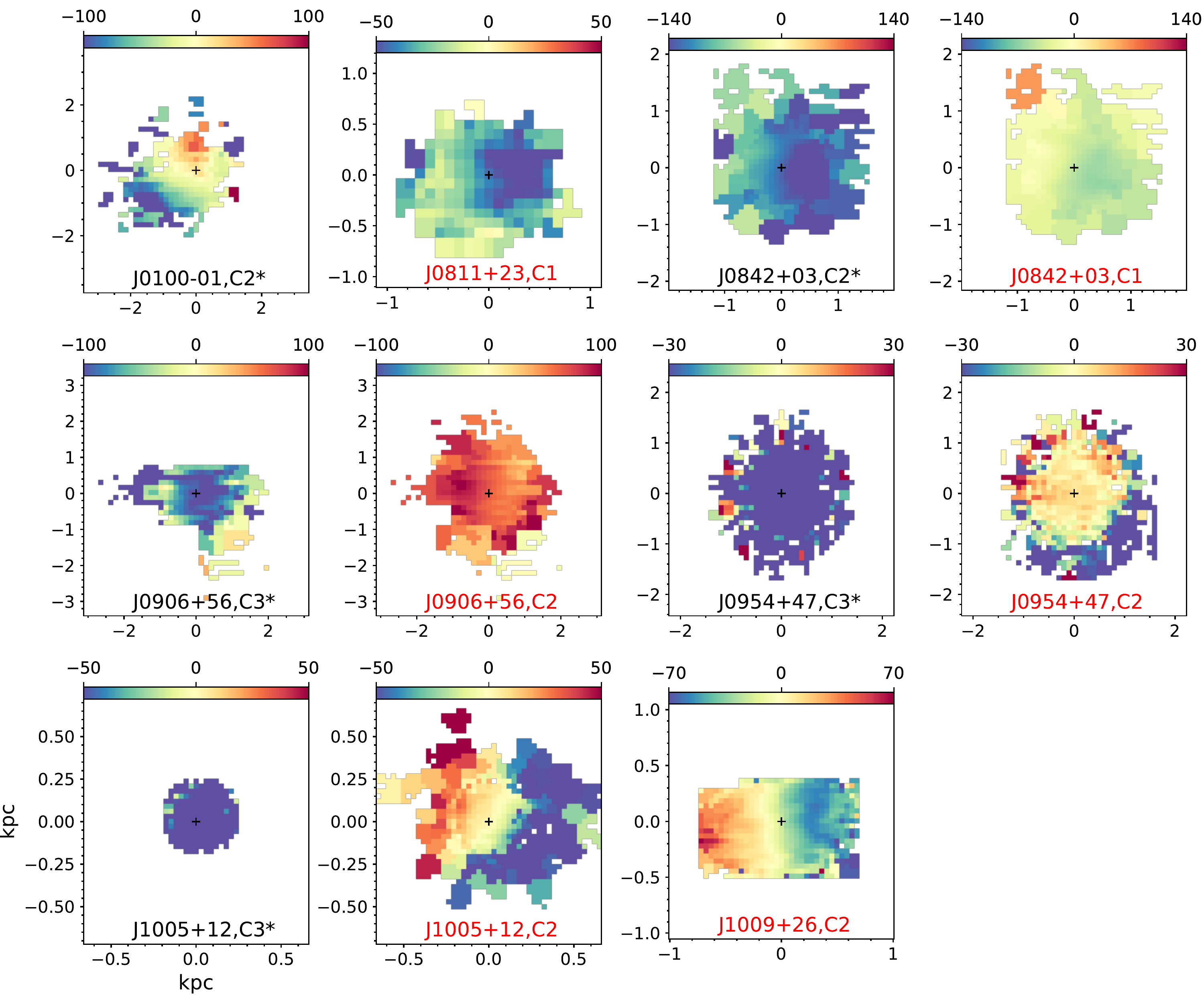}
\caption{Median velocity (\vwu) maps (in units of \kms) for the velocity components of \oiii\ emission showing evidence of outflows in the seven targets with detected outflows. An overview of these outflow components is presented in Section 4.1 and the detailed analyses of these components in individual targets are presented in Appendix \ref{4}. The name of the target and the corresponding velocity component is noted at the bottom of each panel: The components showing strong evidence for outflows are labelled in black and marked with asterisks, whereas those with relatively more uncertain origins are labelled in red, as stated in Section \ref{51} and discussed in detail in the Appendix \ref{4}. The color scale of each panel is set to be the same as that of the corresponding target-specific map in Appendix \ref{4}, except for those of the C2 and C1 components of \tb, where the color scales are centered on 0 \kms\ instead. The black cross in each panel denotes the spaxel where the peak of the total \oiii\ emission line flux falls.}
\label{fig:snapshot1}
\end{figure*}

\begin{figure*}[!htb]   
\epsscale{1.25}
\plotone{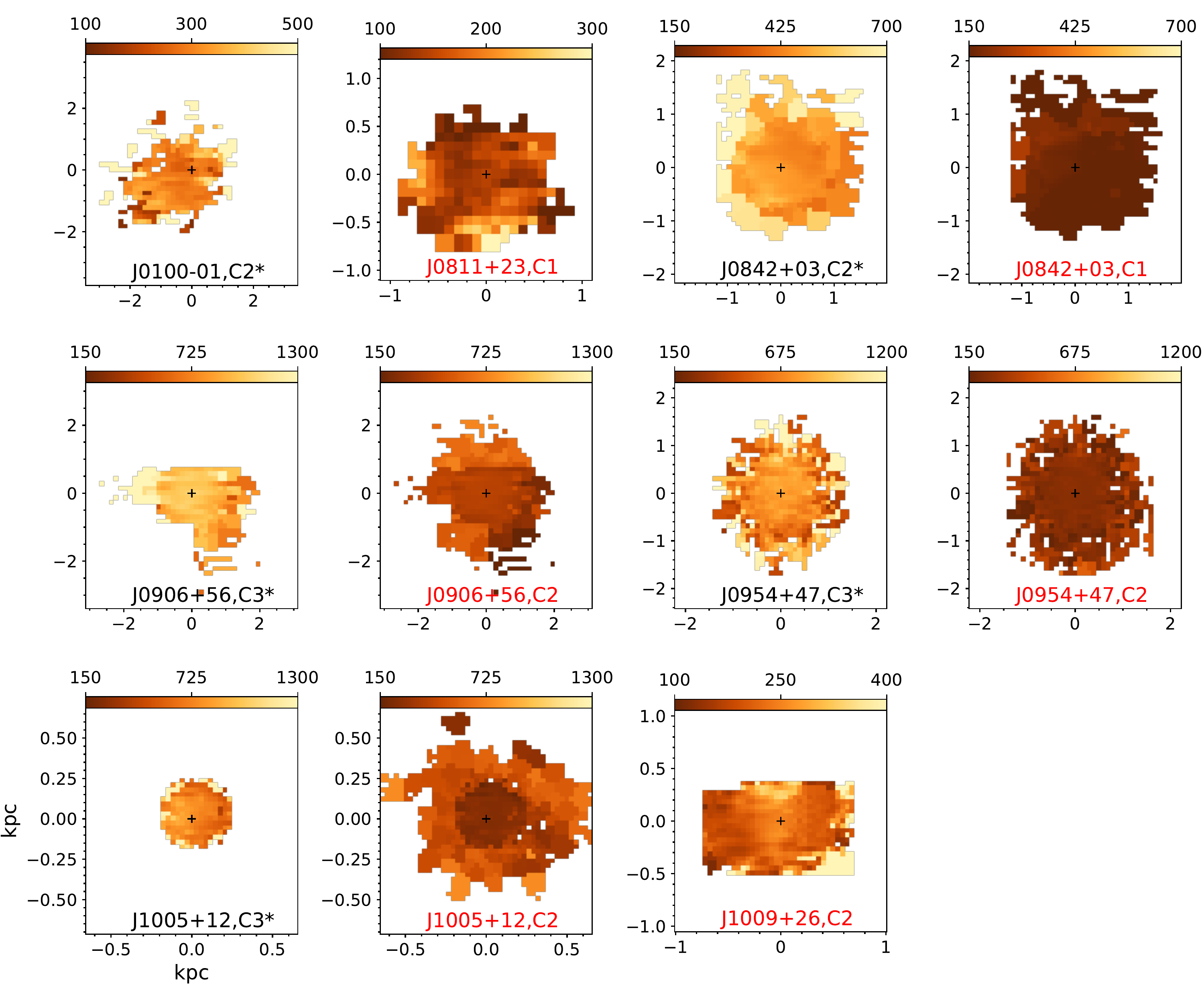}
\caption{Same as Fig. \ref{fig:snapshot1} but for line width \wba\ (in units of \kms).}
\label{fig:snapshot2}
\end{figure*}

\subsection{Gas Kinematics across Our Sample} \label{51}
The gas kinematic properties of the galaxies in our sample are summarized in Table \ref{tab:kinematics}. This includes basic statistics (min, max, median) on \vwu\ and \wba\ for individual velocity components and the entire \oiii\ line emission across the data cubes, as well as measurements of \vwu\ and \wba\ from the spatially-integrated spectra.

Overall, we find that the number of velocity components needed to adequately fit the emission line profiles in our targets ranges from three for \ta, \tc, and \td, two for \te, \tb, and \tf, and one for targets \tg\ and \tx.

The kinematic properties of the C3 components in \ta, \tc, and \td, and the C2 components in \te\ and \tb, show strong evidence for outflows since they are very broad and/or significantly blueshifted with respect to the stellar velocity field derived from the same data (Their names are shown in black and marked with asterisks in Fig.\ \ref{fig:snapshot1} and \ref{fig:snapshot2}. The kinematic properties of the C2 components in \ta, \tc, and \td, as well as the C1 component in \tb\ also suggest that they are at least part of, or affected by, the outflows in these systems. In addition, given the peculiar kinematics of the C2 component in \tf\ and C1 component in \tg\ relative to that of the stellar component, we argue in Appendix \ref{4} that they also likely represent outflowing gas in these objects (These last two groups of velocity components have relatively more ambiguous origins than the first group, so their names are shown in red in Fig.\ \ref{fig:snapshot1} and \ref{fig:snapshot2} to distinguish them from the first group. In the following discussion, we associate all of these velocity components with the outflows in these seven objects, and will refer to them as outflow components by default. In the end, only \tx\ does not show any sign of outflowing gas in our IFS data, so it is omitted from the following discussion of the outflows, except when mentioned explicitly. 

The kinematic properties of the outflows in these seven targets span a relatively large range in terms of line width and median velocities. Quantitatively, the maxima of \wba\ range from $\sim$220 \kms\ to $\sim$1200 \kms, and the minima of \vwu\ range from $\sim$30 \kms\ to $\sim$$-$240 \kms\ based on our IFU data. The apparent morphology of these outflow-tracing components are in general symmetric with respect to the galaxy center, except for \te\ and \tf, which show biconical morphology in projection. In addition, significant non-radial velocity gradients/structures are also seen for the outflow components of targets \tg\ and \tb, as well as the C2 components of targets \tc\ and \td. (see Fig. \ref{fig:snapshot1} and \ref{fig:snapshot2} for snapshots of the \vwu\ and \wba\ maps of these components; see Appendix \ref{4} for additional target-specific flux and kinematic maps).

\subsection{Spatial Extents of the Outflows} \label{52}

A key question is whether the outflows detected in our targets are extended on galactic scale. As shown in Fig. \ref{fig:o3map1}--\ref{fig:radial6} of Appendix \ref{4} and discussed in this Appendix, the analysis of the IFS data has revealed spatially resolved structures in the velocity fields of the outflow components, as well as excess flux relative to the PSF, in \te, \tg, \tb, and \tf, strongly suggesting that the outflows in these galaxies are spatially resolved. 
Similarly, the C2 components in \ta, \tc, and \td, and the C1 component in \tb\ are also probably spatially resolved. However, the results of our analysis are inconclusive for the C3 components of \ta, \tc, and \td. 

An independent constraint on the spatial extent of the outflow components in \ta, \tc, and \td\ may be derived from a more formal deconvolution of the data cubes. 
For this, we follow a procedure explained in detail below, which is a simplified version of the deconvolution scheme introduced in \citet{Rupke2017}. First, we assume that the flux in the spaxel with the peak emission line flux (a 0.2\arcsec$\times$0.2\arcsec\ box for the GMOS data and a 0.15\arcsec$\times$0.15\arcsec\ one for the KCWI data) is dominated by AGN emission. The spectrum from this spaxel is treated as an AGN emission template from a point source. Next, we fit each spaxel n with this AGN template $+$ smooth exponential continuum functions $+$ host emission lines, according to: 

\begin{equation}
I^n_{total} = C_{AGN}I^n_{AGN} + I^n_{exp, continuum}+I^n_{emission}
\end{equation}

The scaling factor C$_{AGN}$ for the AGN emission template and the exponential continuum functions in the equation are each the sum of four exponentials, so eq. (1) can be re-written as:
\begin{equation}
I^n_{total} = \sum_{i=1}^{4}I_i^n I^n_{AGN} + \sum_{j=1}^{4}I_j +I^n_{emission}
\end{equation}

and the four exponentials are :

\begin{eqnarray}
I_1^n = a_1^n e^{-b_1^n<\lambda>} 
\end{eqnarray}
\begin{eqnarray}
I_2^n = a_2^n e^{-b_{2}^n(1-<\lambda>)}
\end{eqnarray}
\begin{eqnarray}
I_3^n = a_3^n(1-e^{-b_{3}^n<\lambda>})
\end{eqnarray}
\begin{eqnarray}
I_4^n = a_4^n({1-e^{-b_{4}^n(1-<\lambda>)}}),
\end{eqnarray}
where ${a_i^n}\geq{0}$;  ${b_i^n}\geq{0}$; ${<}\lambda{>}{=}\frac{\lambda-\lambda_{min}}{\lambda_{max}-\lambda_{min}}$; and $[\lambda_{min},\lambda_{max}]$ is the fit range. These exponentials are adopted because they are monotonic and are positive-definite. The four exponentials allow for all combinations of concave/convex and monotonically increasing/decreasing. We have not used stellar templates in the fits above, since the stellar absorption features are not strong enough in individual spaxels to constrain the extra free parameters and the fits become divergent.

 The host emission lines are modeled with a maximum of two Gaussian components. The fits are iterative. In step 1), the cores of the emission lines are masked and the continuum is fit with the AGN template $+$ exponential continuum terms. In step 2), the best-fit model from step 1) is subtracted from the original spectrum, and the emission lines are fit. In 3), the best-fit emission line models are used to determine a better emission line mask window in the continuum fit, and then steps 1) through 3) are repeated until the best-fit results are stable.

The results of this analysis on \ta, \tc, and \td\ indicate 1) clear evidence for spatially extended narrow line emission originating on the scale of the host galaxy in all three targets; 2) blueshifted, broad line emission with a S/N of $\sim$3--8 tracing the outflows in the host galaxy in the spatially-stacked spectra for all three targets. The line widths of these components fall in between those of the C3 and C2 components in these targets; 3) but inconclusive (S/N $\lesssim$ 2 in general) evidence for spatially resolved line emission from the outflow components. The same analysis conducted on the other four targets confirms the presence of spatially resolved, blueshifted and/or redshifted velocity components from the host galaxy, corresponding to the outflow components detected in our more detailed kinematic analysis (Appendix \ref{4}).

Before concluding this section, it is important to repeat that the PSF deconvolution scheme described here relies on the assumption that the spectra used as AGN templates for these targets are indeed pure AGN emission (and thus from an unresolved point source). While the line ratios measured from these spectra fall in the AGN region in the BPT/VO87 diagrams (for the GMOS data of \ta) or the \oiioiiihb\ diagram (for the KCWI data), there are reasons to believe that emission from the host galaxies themselves still contributes significantly to the spectra. First and foremost, weak to moderate (S/N $\simeq$ 2--9) \mgb\ absorption features of stellar origin are detected in these spectra. In addition, we carried out a separate, power-law continuum $+$ stellar templates fit to the continuum emission of these spectra (in the ranges of $\sim$5000--7000 \AA\ for the GMOS data, and $\sim$3600--5500 \AA\ for the KCWI data). An AGN-like power-law continuum component is not formally needed in the best-fit results. Our PSF deconvolution procedure thus almost certainly overestimates (underestimates) the contribution from the unresolved AGN emission (resolved host emission), so the S/N of the spatially resolved outflow emission in \ta, \tc, and \td\ obtained above should therefore be considered conservative lower limits.

\begin{figure*}[!htb]
\begin{minipage}[t]{0.33\textwidth}
\includegraphics[width=\textwidth]{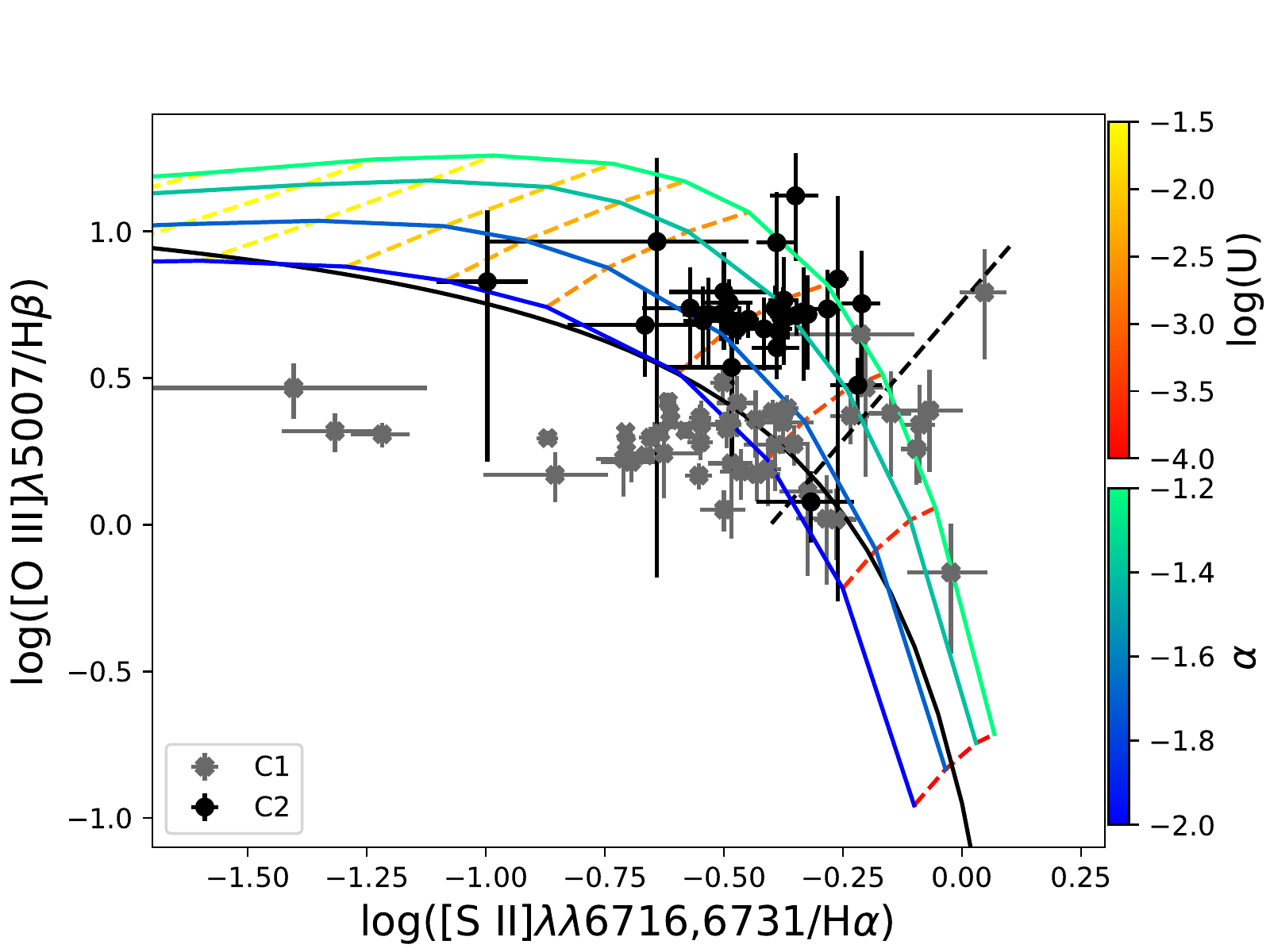}
\end{minipage}
\begin{minipage}[t]{0.33\textwidth}
\includegraphics[width=\textwidth]{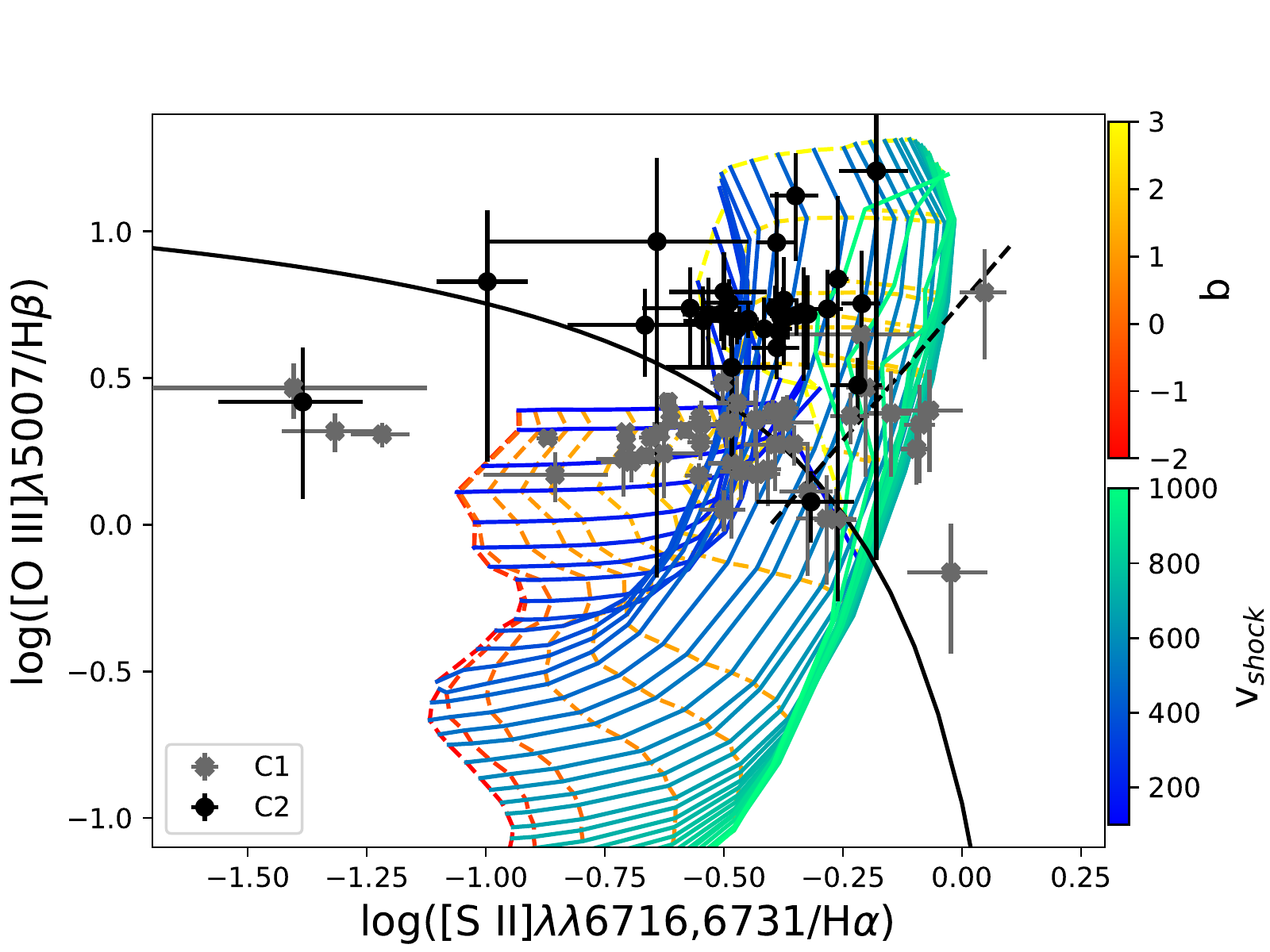}
\end{minipage}
\begin{minipage}[t]{0.33\textwidth}
\includegraphics[width=\textwidth]{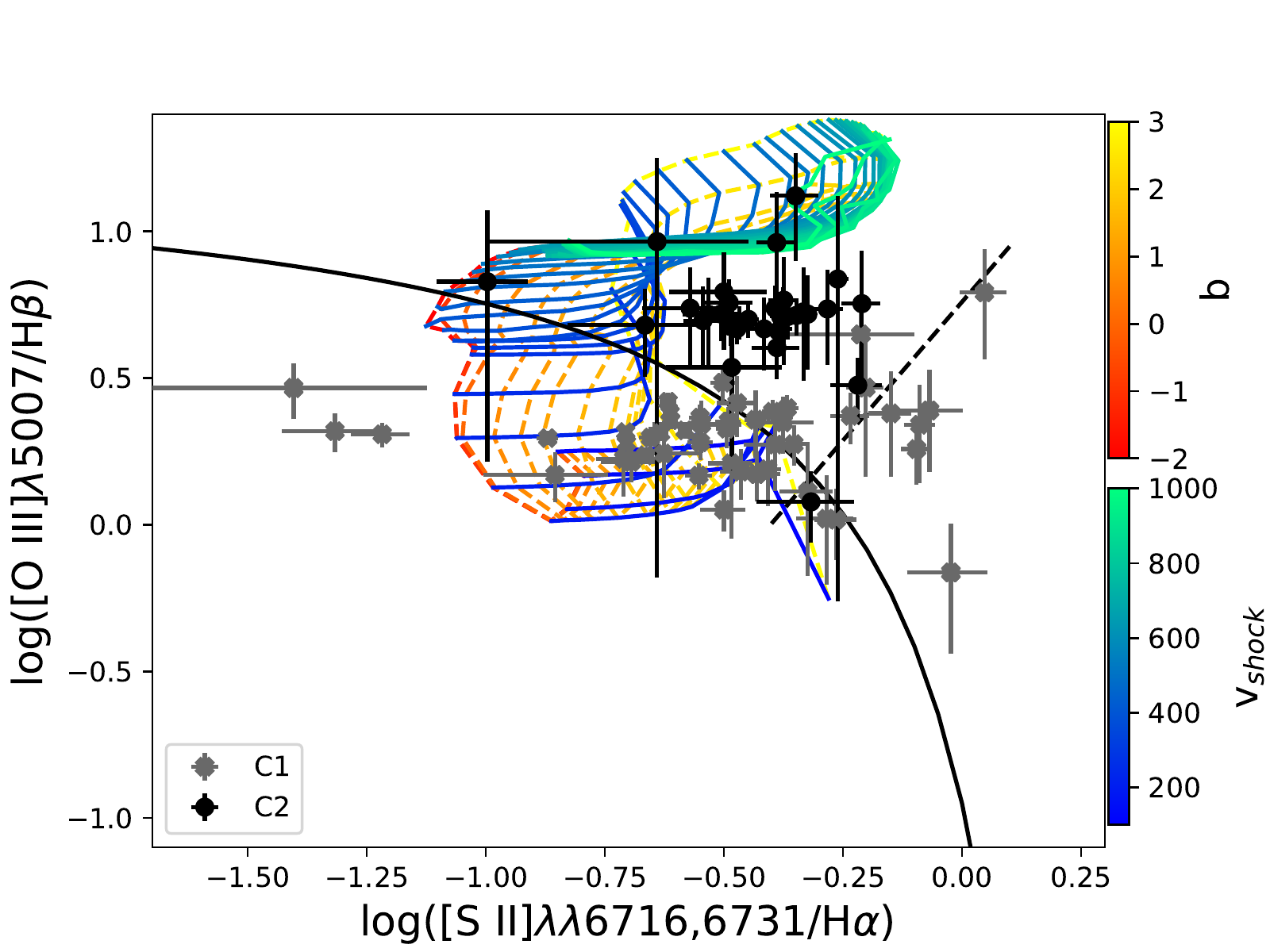}
\end{minipage}
\caption{\oiiihb\ vs \siiha\ for the C2 (black) and C1 (gray) components of \tb, compared with AGN (left), shock (middle) and shock$+$precursor (right) models. The grids of the AGN models are color-coded by the power-law indices and ionization parameters of the AGN, and those of the shock and shock$+$precursor models are color-coded by the values of magnetic parameter $b$ and shock velocity v$_{shock}$. See Section \ref{53} for more details on these model parameters. In all three panels, the black, solid lines are the theoretical line separating AGN (above right) and star-forming galaxies (below left) from \citet{Kewley2001}. The black, dashed lines are the theoretical line separating the Seyferts (above left) and LINERs (below right) defined in \citet{Kewley2006}. 
\label{fig:lrmodel1}}
\end{figure*}

\begin{figure*}[!htb]
\begin{minipage}[t]{0.33\textwidth}
\includegraphics[width=\textwidth]{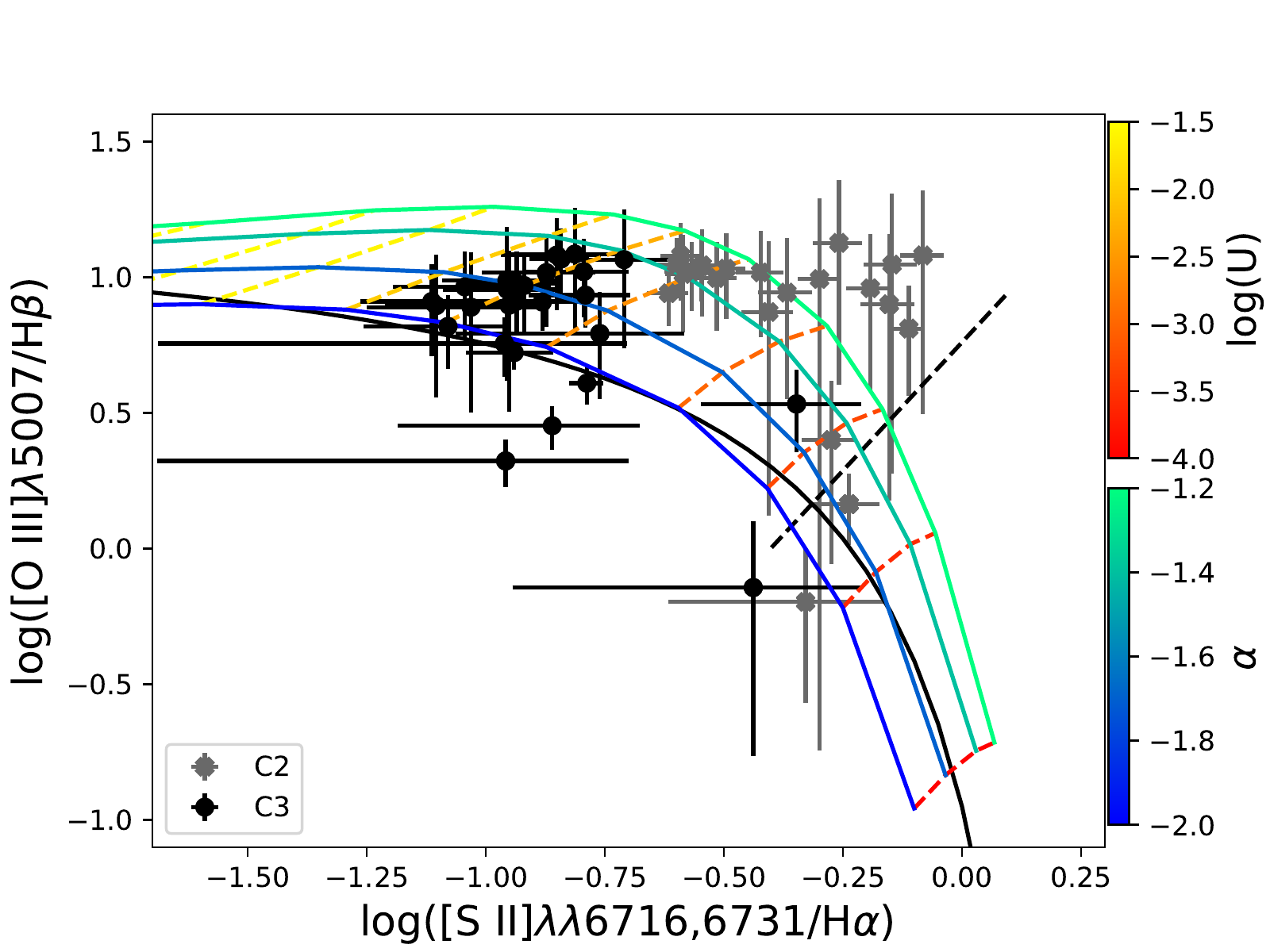}
\end{minipage}
\begin{minipage}[t]{0.33\textwidth}
\includegraphics[width=\textwidth]{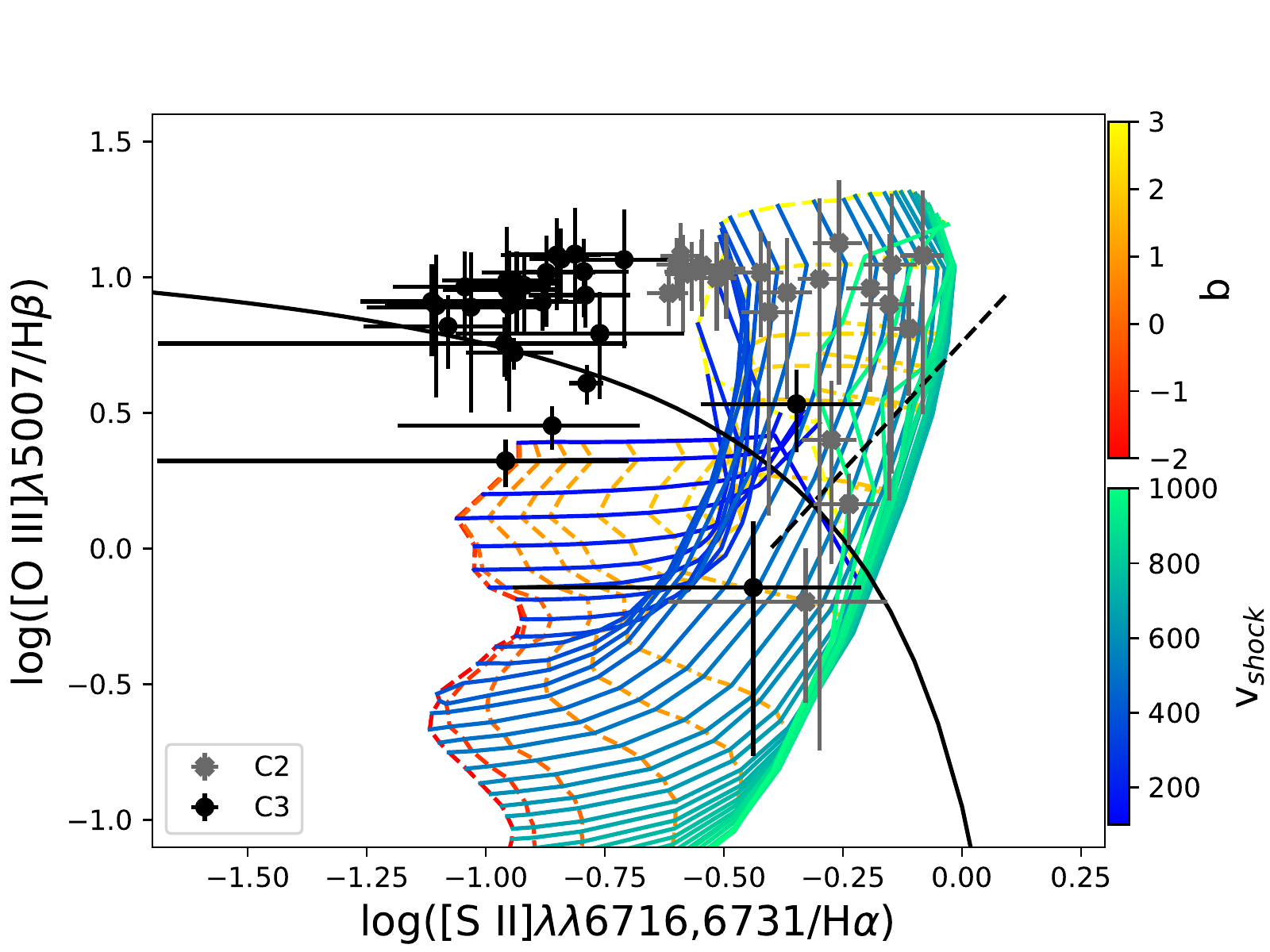}
\end{minipage}
\begin{minipage}[t]{0.33\textwidth}
\includegraphics[width=\textwidth]{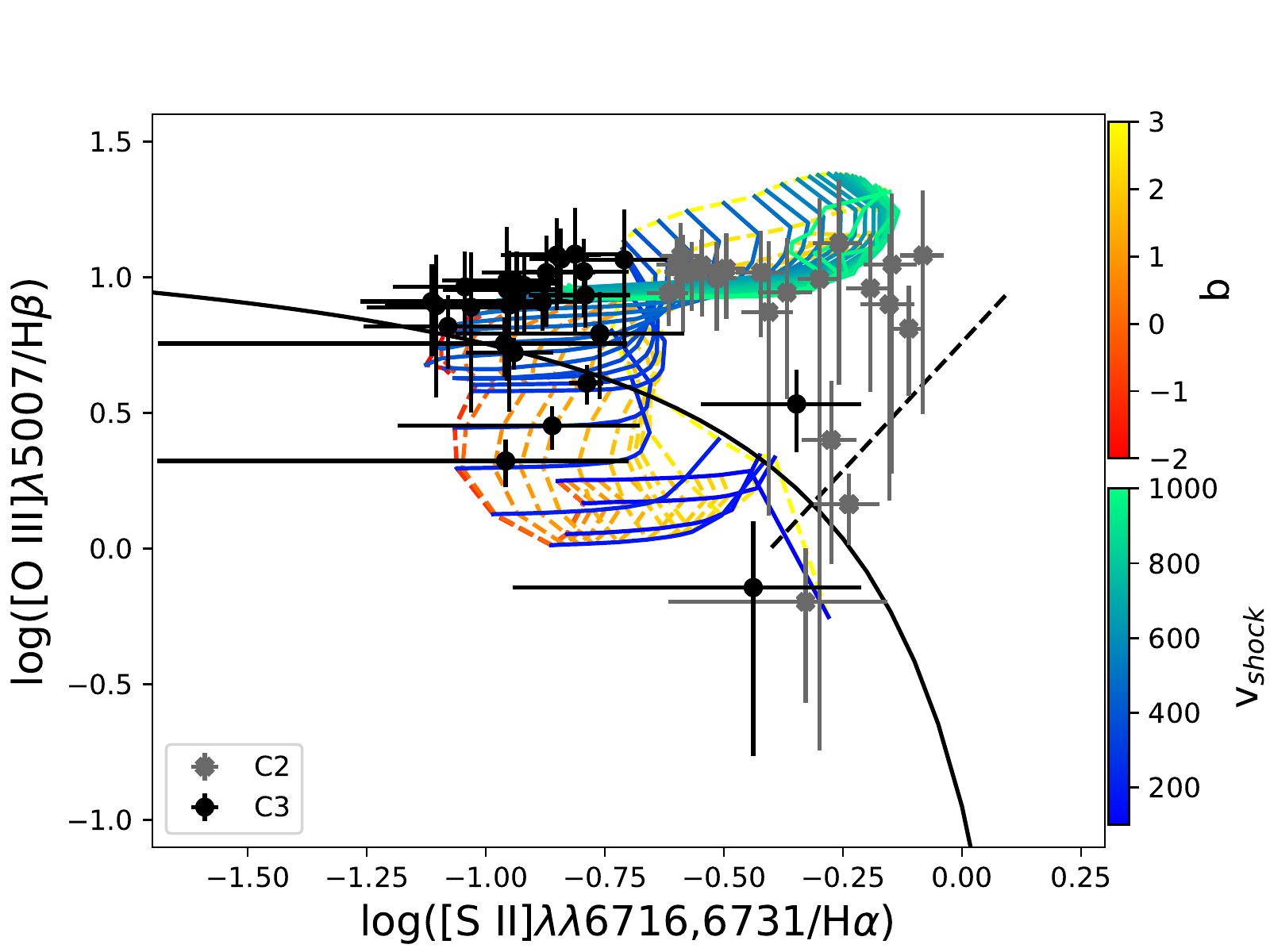}
\end{minipage}
\caption{ Same as Fig.\ \ref{fig:lrmodel1} but for the C3 (black) and C2 (gray) components of \ta.
\label{fig:lrmodel2}}
\end{figure*}

\begin{figure}[!htb]   
\epsscale{2.3}
\plottwo{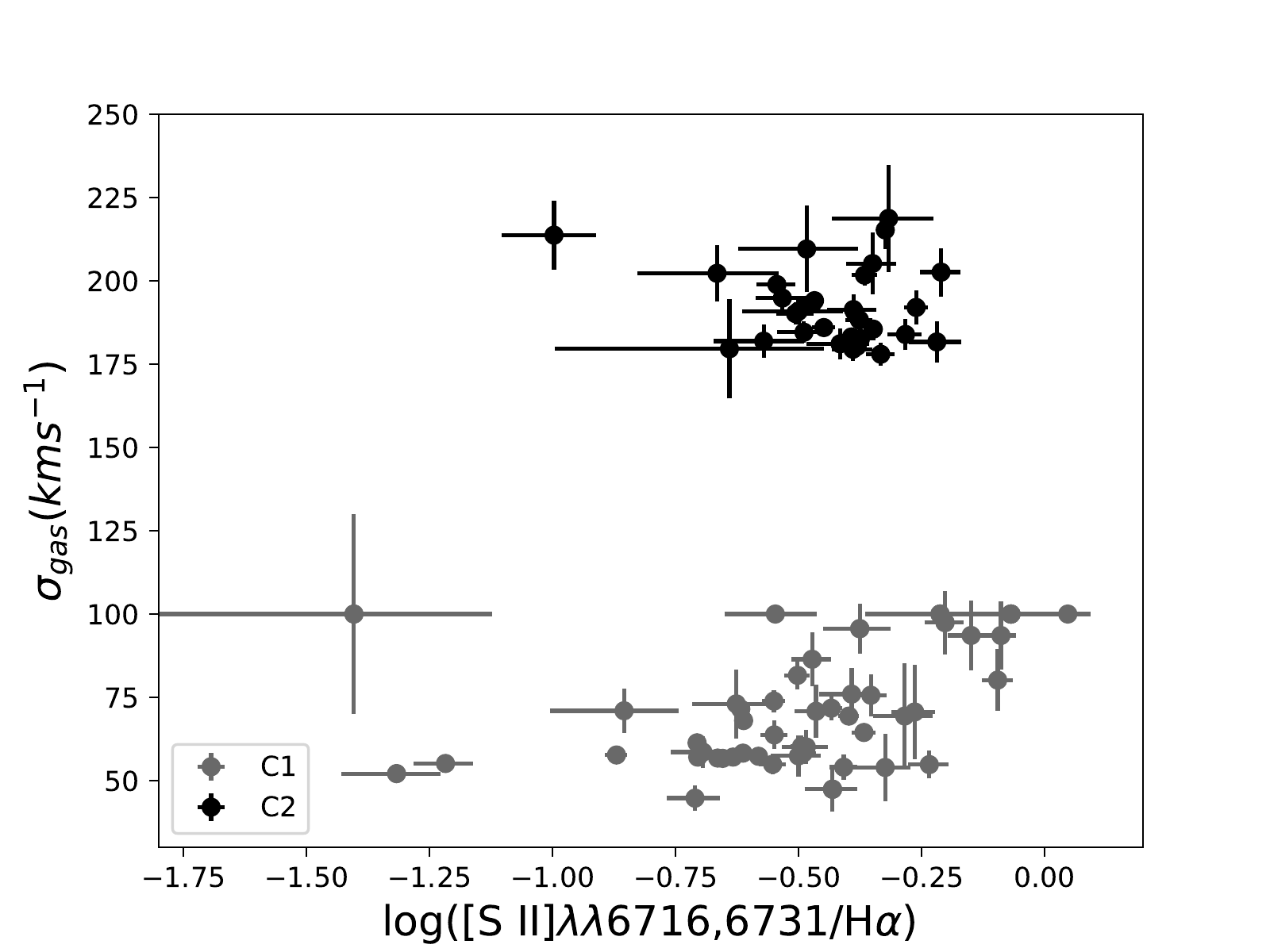}{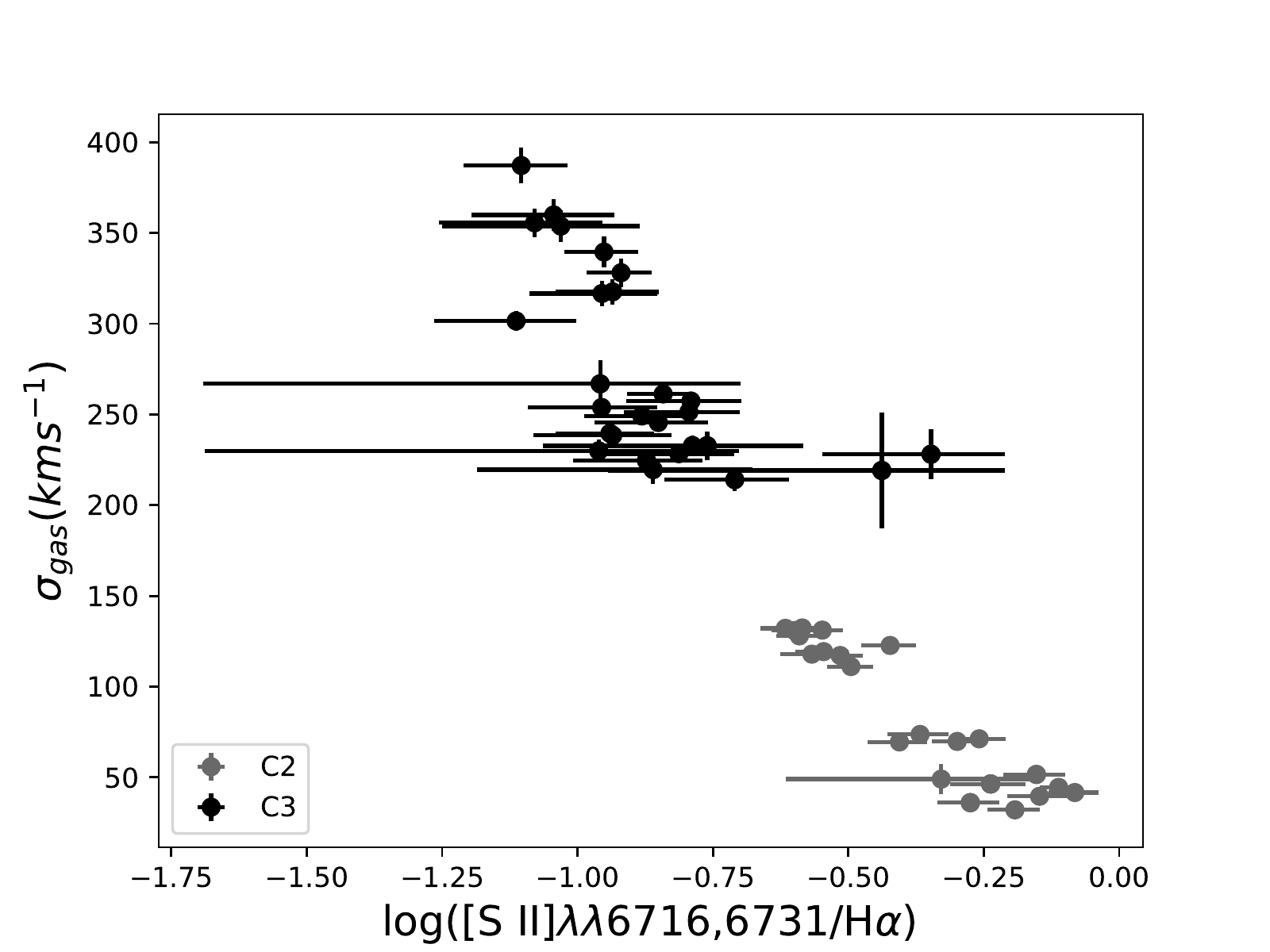}
\caption{\siiha\ ratios versus gas velocity dispersions for the outflow components in \tb\ (top) and \ta\ (bottom) based on the GMOS data.}
\label{fig:lrvel}
\end{figure}

\begin{figure*}[!htb]
\begin{minipage}[t]{0.33\textwidth}
\includegraphics[width=\textwidth]{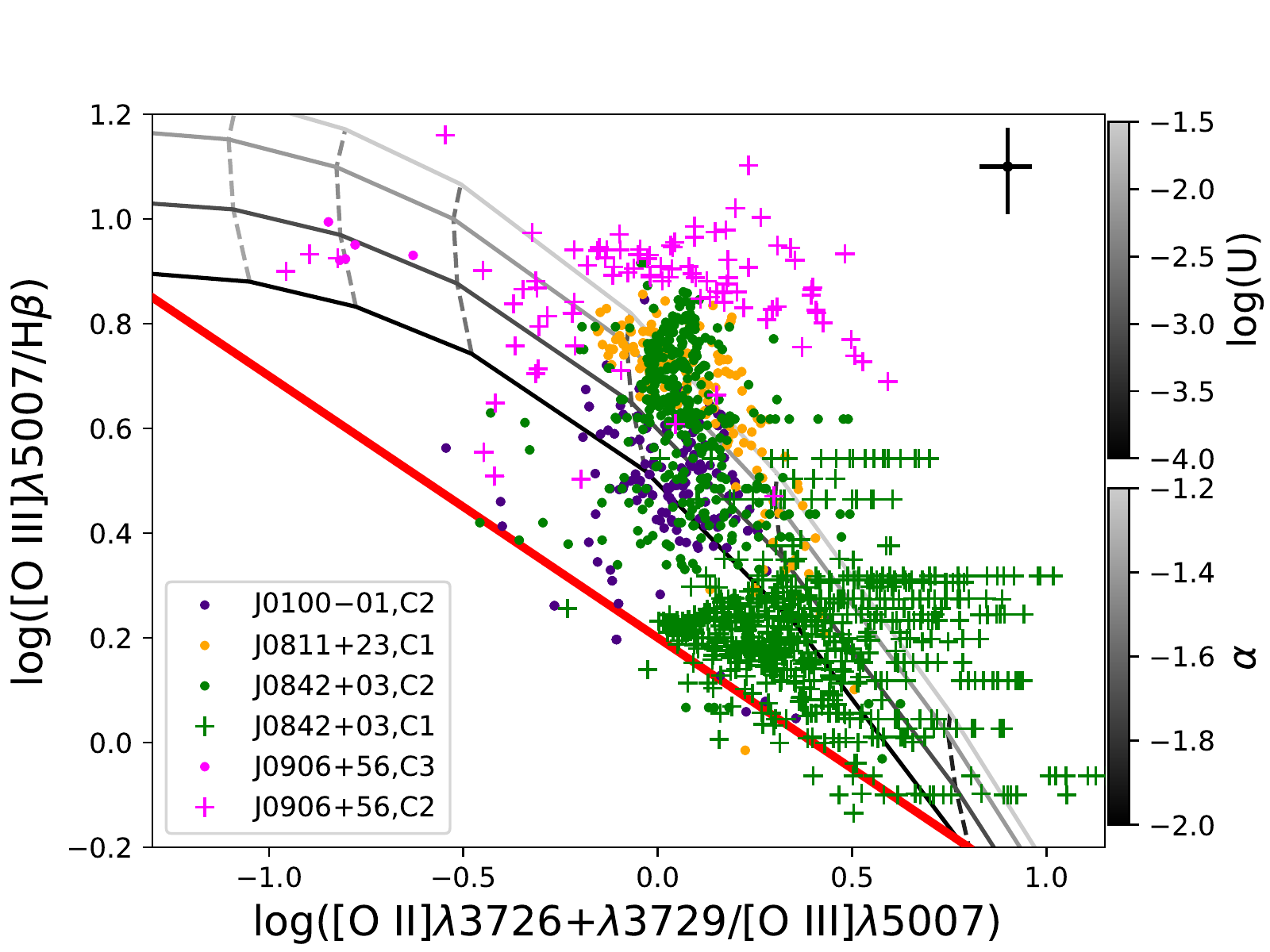}
\end{minipage}
\begin{minipage}[t]{0.33\textwidth}
\includegraphics[width=\textwidth]{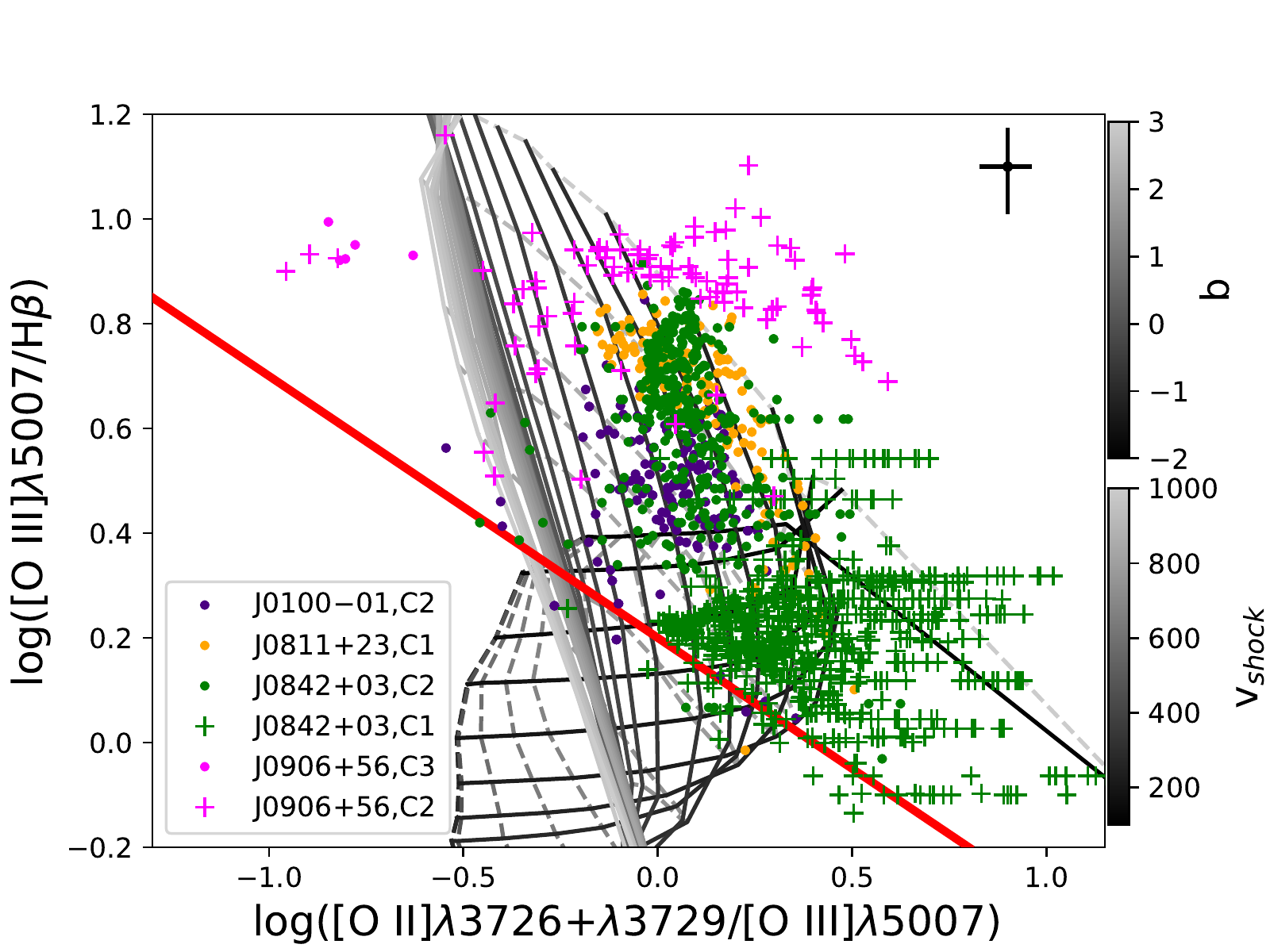}
\end{minipage}
\begin{minipage}[t]{0.33\textwidth}
\includegraphics[width=\textwidth]{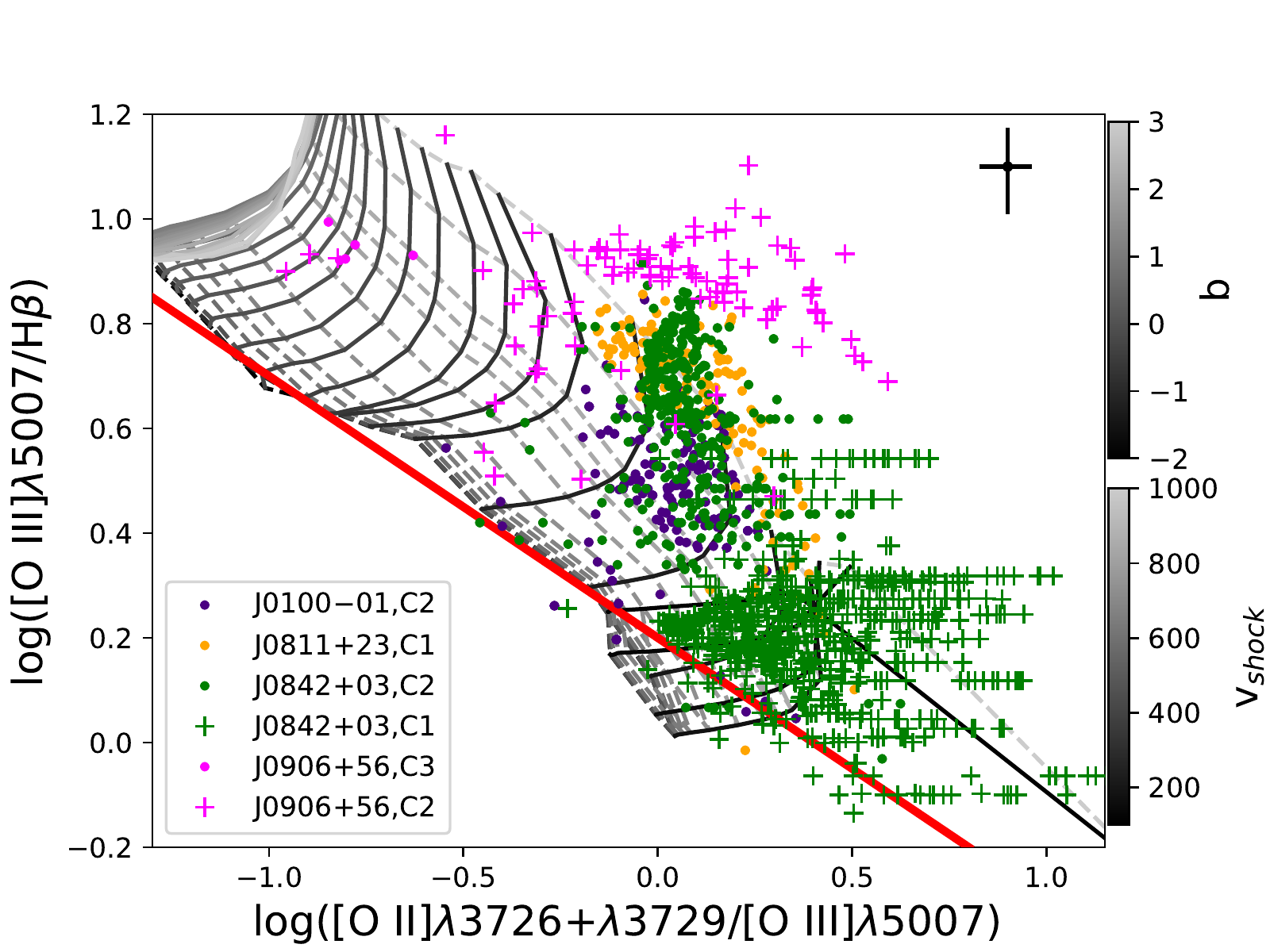}
\end{minipage}
\begin{minipage}[t]{0.33\textwidth}
\includegraphics[width=\textwidth]{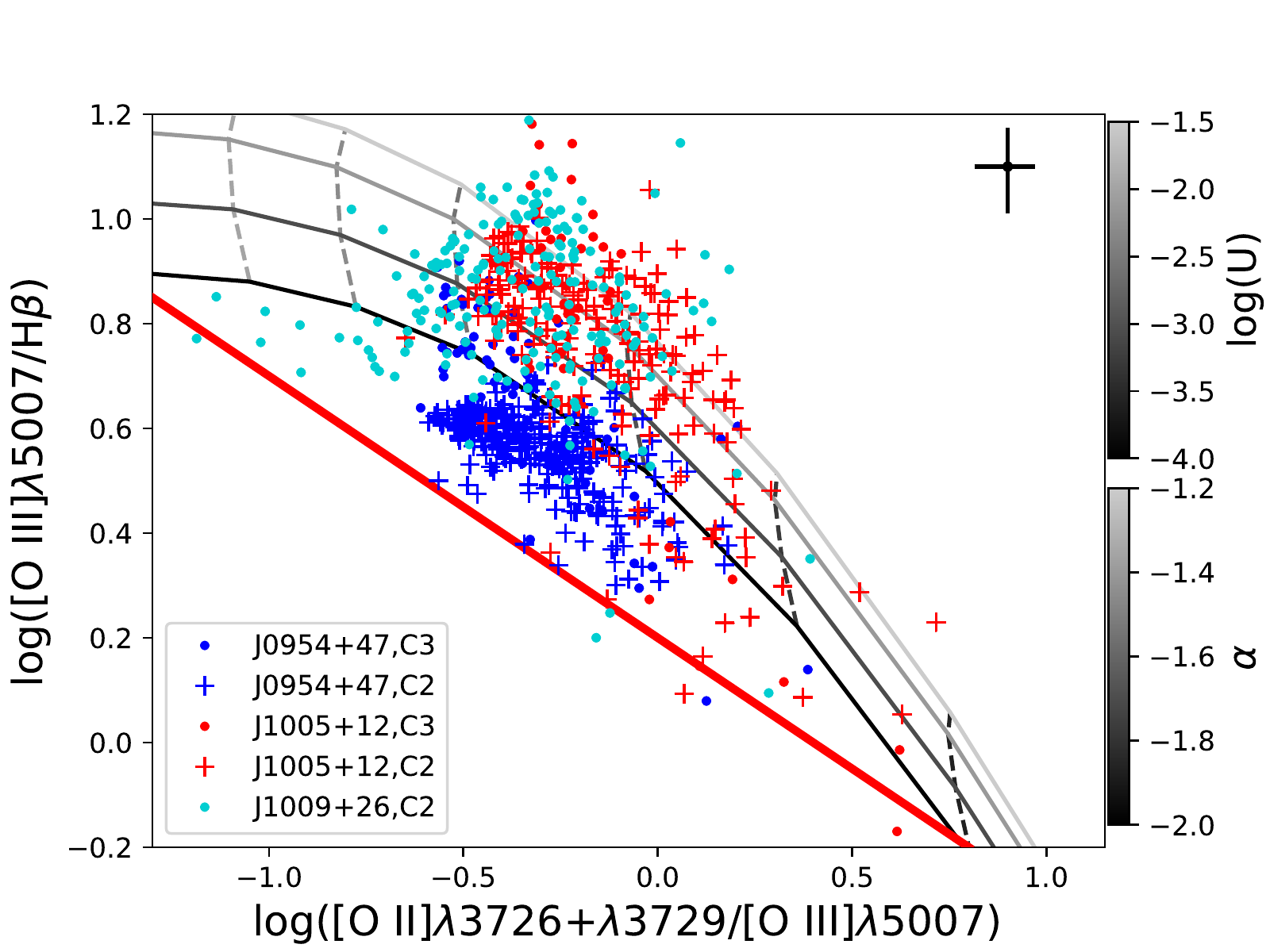}
\end{minipage}
\begin{minipage}[t]{0.33\textwidth}
\includegraphics[width=\textwidth]{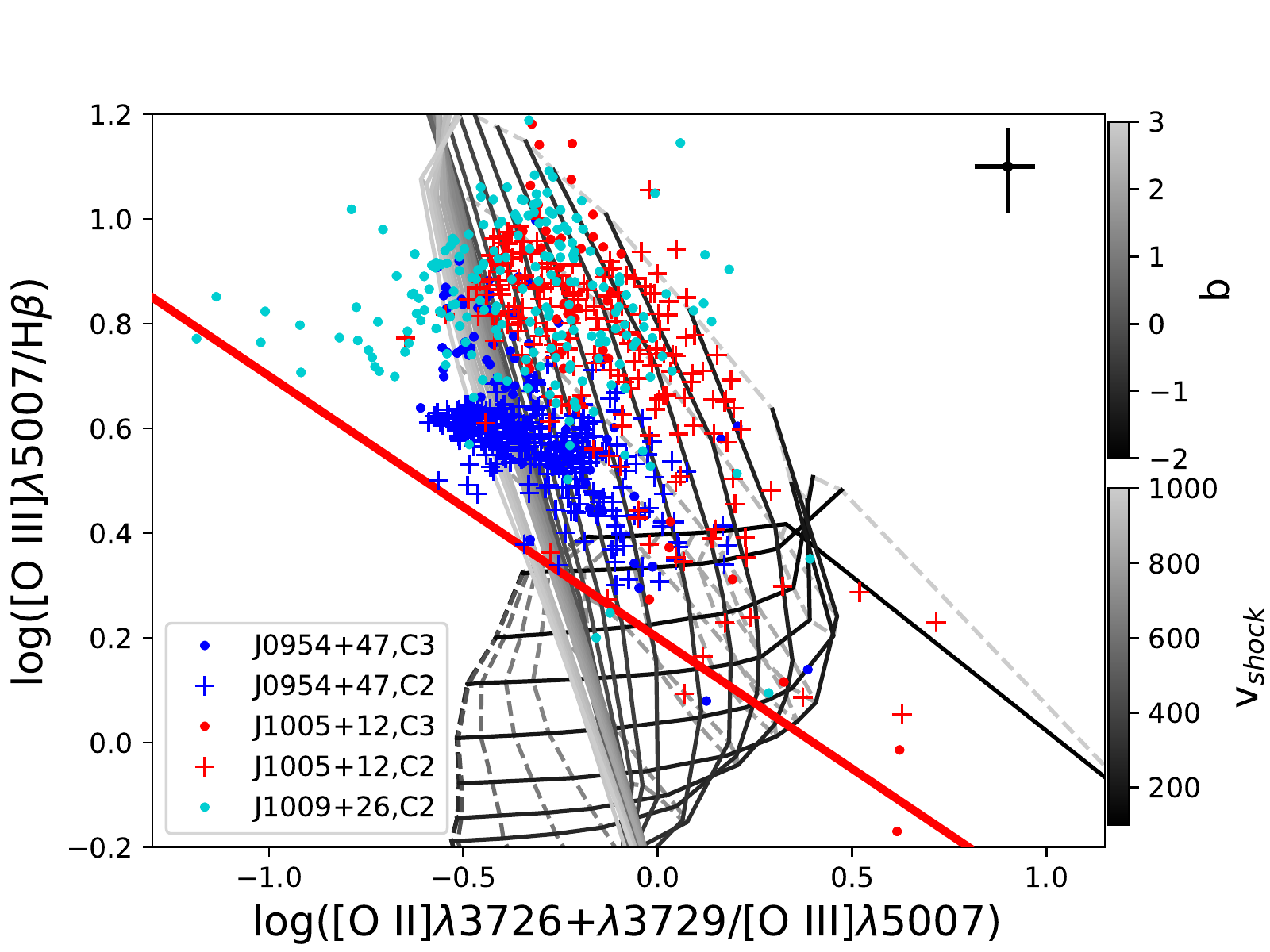}
\end{minipage}
\begin{minipage}[t]{0.33\textwidth}
\includegraphics[width=\textwidth]{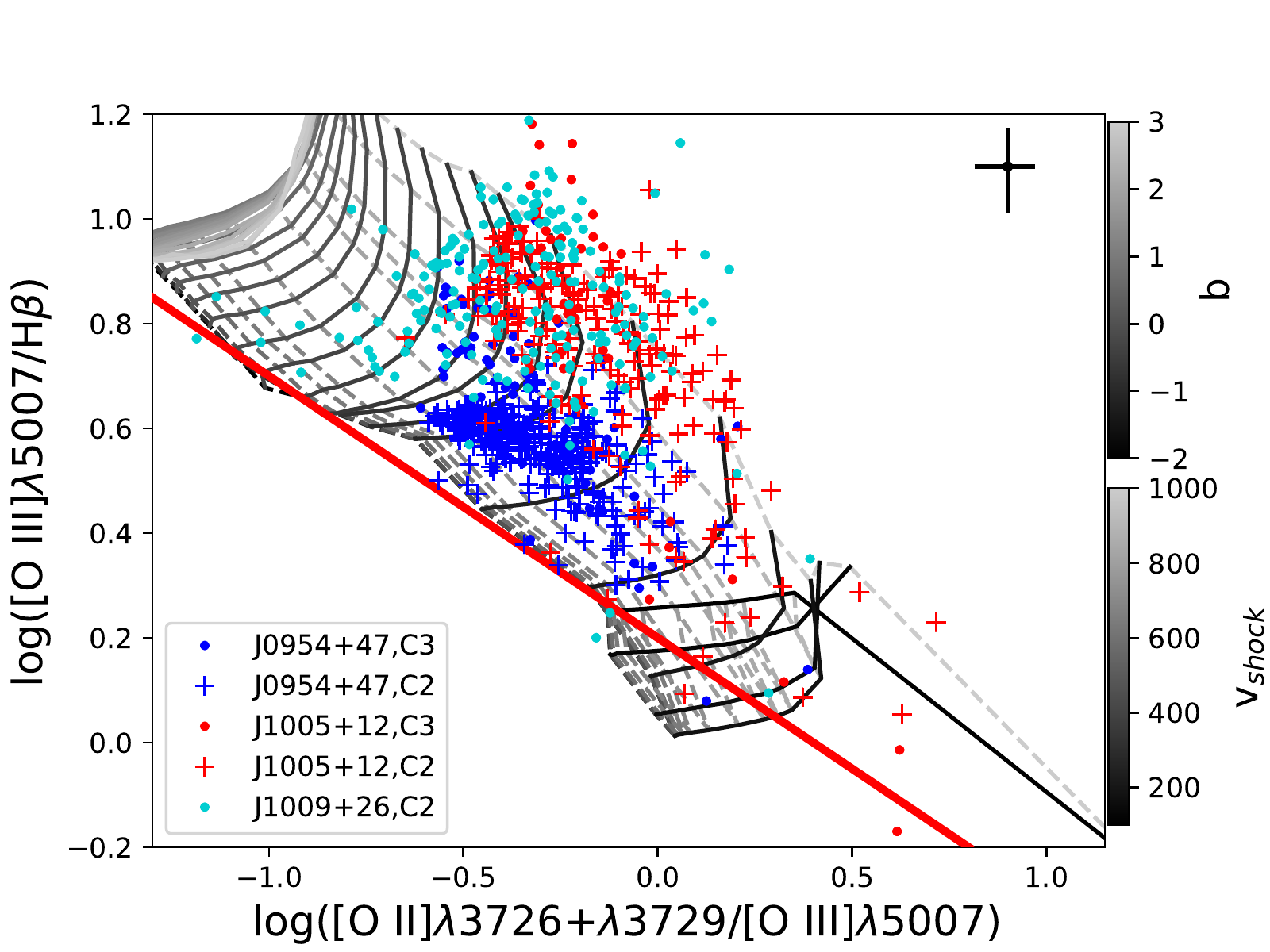}
\end{minipage}
\caption{\oiioiiihb\ for the outflow components of all seven targets (1st row: C2 component in \te, C1 component in \tg, C2 and C1 components in \tb, as well as C3 and C2 components in \ta; 2nd row: C3 and C2 components in \tc\ and \td, as well as C2 component in \tf) based on the KCWI data, compared with AGN (left column), shock (middle column) and shock$+$precursor (right column) models (gray-scale model grids). The median values of the errors of the data points are noted by the black crosses on the top right corners. For the AGN models, the constant power-law indices of the AGN are drawn in solid lines and the constant ionization parameters are drawn in dashed lines. For the shock and shock$+$precursor models, the constant magnetic parameters $b$ are drawn in dashed lines, and the constant shock velocities v$_{shock}$ are drawn in solid lines. The red, solid lines represent the approximate upper boundary of line ratios that can be generated by star-forming activity, based on the Starburst99 models with continuous star formation history from \citet[][]{Levesque2010}. See Section \ref{53} for more details on the model parameters.}
\label{fig:lrmodelkcwi}
\end{figure*}

\begin{figure}[!htb]   
\epsscale{2.3}
\plottwo{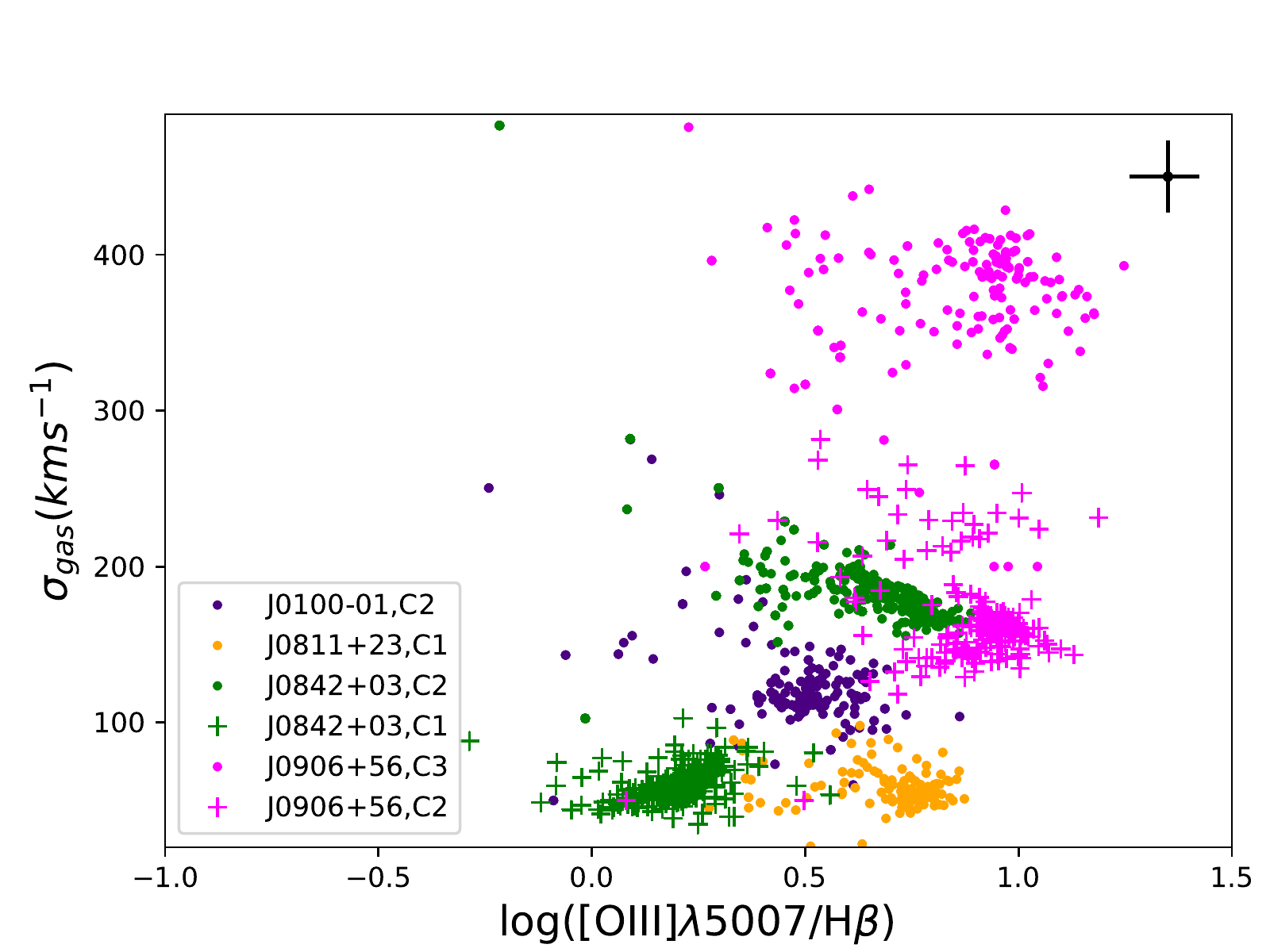}{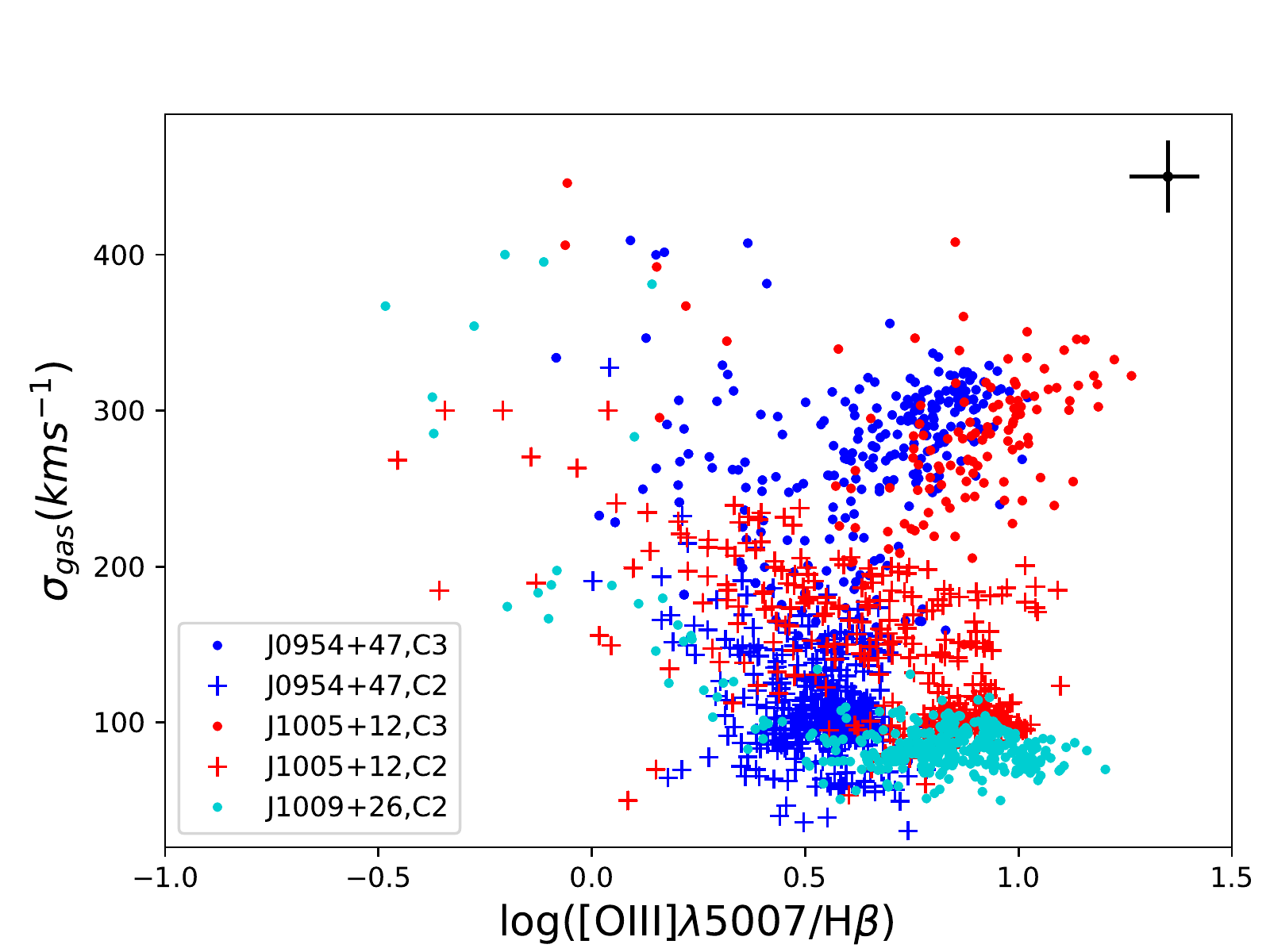}
\caption{\oiiihb\ ratios versus gas velocity dispersions for the outflowing gas in all seven targets (The results are split into two panels for a better view of the data points) with detected outflows based on the KCWI data. The median values of the errors of the data points are shown as the black crosses in the top-right corners.}
\label{fig:lrvelkcwi}
\end{figure}

\subsection{Outflow Ionization: AGN or Shocks?} \label{53}

The line ratio maps and spatially resolved BPT and VO87 diagrams for \tb\ and \ta\ (Figs.\ \ref{fig:J0842bpt1} $-$ \ref{fig:J0842bpt2} and Figs.\ \ref{fig:J0906bpt2} $-$ \ref{fig:J0906bpt3} in Appendix \ref{4}, respectively) suggest that the outflows in our targets are largely photoionized by the AGN. Here we examine further the evidence that supports this statement. In particular, we examine the possibility that fast shocks caused by the interaction of the outflows with the surrounding ISM may contribute, or even dominate, the heating and ionization of the outflowing gas. Shock excitation is a telltale sign of fast starburst-driven winds \citep{VeilleuxRupke2002,Sharp2010}, and has also been suspected in a few AGN-driven outflows \citep[e.g.][]{Hinkle2019}.

 First, we compare the BPT and VO87 line ratios measured in the clear outflow components, C2 and C1 components in \tb\, and C3 and C2 components in \ta, to those of typical AGN models \citep{Groves2004} and shock models \citep{Allen2008} extracted from the ITERA library \citep{ITERA}. For the AGN models, the free parameters are the gas number density, the metallicity, the photon index of the AGN continuum, $\alpha$, and the ionization parameter $U$, where $U \equiv n_{\rm ion}/n_e$, where $n_{\rm ion}$ is the density of ionizing photons and $n_e$ is the electron density. We find that the line ratios probed by our data are not sensitive to the gas number density in the range (100--1000 cm$^{-3}$) relevant to our targets. We further compared the AGN models with metallicity of 0.5 $\times$ solar and solar to the data, and conclude that the one with solar metallicity is a better match to the data. Therefore, the gas number density and metallicity of the AGN model grids are fixed at 1000 cm$^{-3}$ and solar values in our following model comparison, respectively. For the shock models, we consider two types of models, one where only the ionization from the shock itself is considered (called shock model hereafter), and one where the ionization is caused by both the shock and the precursor region ahead of the shock front (called shock$+$precursor model hereafter). The free parameters for both sets of models are the pre-shock particle number density $n$, the metallicity, the shock velocity v$_{shock}$, and the magnetic parameter $b \equiv log[B/n^{\frac{1}{2}} / (1\ {\mu}G\ cm^{\frac{2}{3}}) ]$ (where $B$ is the transverse magnetic field). We have fixed the pre-shock particle number density $n$ to 1000 cm$^{-3}$ and the metallicity to solar value, which follows the same set-up as that for the AGN models. The full extent of the line ratio predictions from the shock and shock$+$precursor models with other density and metallicity settings is mostly covered by the model grids we adopt here, and they are thus omitted from the discussion below.

The results for the \oiiihb\ vs \siiha\ diagram are shown in Fig.\ \ref{fig:lrmodel1} for \tb\ and Fig.\ \ref{fig:lrmodel2} for \ta, where the comparison with the AGN, shocks, and shock$+$precursor models are displayed in the left, middle, and right panels, respectively. The results for the other two VO87 diagnostic diagrams, \oiiihb\ vs \niiha\ and \oiiihb\ vs \oiha, are in general similar to those from the \oiiihb\ vs \siiha\ diagram in terms of how well the data and the models match with each other. They are thus omitted in the following discussion.

For the C2 component of \tb, the AGN models match the observed line ratios with $-$3.5 $\lesssim$ log(U) $\lesssim$ $-$2 and $-$2 $\lesssim$ $\alpha$ $\lesssim$ $-$1.2. The shock models can reproduce the majority of the observed line ratios with relatively large $b$ parameters ($\gtrsim$1.5) and small shock velocities ($\lesssim$700 \kms). As for the shock$+$precursor models, either the observed \oiiihb\ ratios or the \siiha\ ratios are systematically lower than the model predictions, by $\sim$0.3 dex on average. For the C1 component, most of the data points lie in, or close to the region for the star-forming galaxies in the diagram. This is consistent with their systematically lower \oiiihb\ ratios compared to the AGN models. However, the shock and shock$+$precursor models are apparently better matches to the line ratios of the C1 component.

For \ta, the observed \oiiihb\ and \siiha\ ratios can be mostly reproduced by AGN models with ionization parameters in the range of $-$3 $\lesssim$ log(U) $\lesssim$ $-$1 and photon indices in the full range provided by the model grids ($-$2 $<$ $\alpha$ $<$ $-$1.2). However, either the observed \oiiihb\ ratios or the \siiha\ ratios are systematically larger than the predictions of shock models by at least $\sim$0.3 dex, contrary to the case for \tb. This discrepancy becomes larger as the shock velocity increases. Once the ionization from the precursor region is considered, the model predictions match the observed line ratios almost as well as the AGN models, although the data have few constraints on the shock velocity and the $b$ parameters. As for the C2 component, the AGN models still match the data relatively well, except that $\sim$1/3 of the data points show slightly higher \oiiihb\ ratios. The shock models with relatively high $b$ parameter ($\gtrsim$1) are also a good match to the data. Finally, the shock$+$precursor models have some trouble explaining $\sim$1/2 of the data points with lower \oiiihb\ ratios. 

Overall, the AGN models more easily reproduce the observed \oiiihb\ and \siiha\ ratios of the C2 component in \tb\ and C3 component in \ta. The shock models generate line ratios consistent with observations for \tb\ but not for \ta, while the shock$+$precursor models match the observations for \ta\ but not for \tb. As for the C1 component in \tb, the AGN models are a worse match to the data, which agrees with the expectation that it is contaminated by emission from the host galaxy, as discussed in Appendix \ref{42}. Nevertheless, the AGN and shock models can both explain the line ratios of the C2 component in \ta\ apparently.

Second, as shown in Fig. \ref{fig:lrvel}, there is no positive correlation between the emission line widths ($\sigma_{gas}$) and the \siiha\ line ratios for the individual outflow components of targets \tb\ and \ta, contrary to theoretical predictions \citep[e.g.,][]{Allen2008} and what is usually found in systems where shocks are the dominant source of ionization \citep[e.g.][]{Veilleux1995,Allen1999,Sharp2010,Rich2011,Rich2012,Rich2014,Ho2014}.
This conclusion still holds even when we consider the two outflow components together in each target. Overall, these results suggest that shock ionization is not important in \tb\ and \ta. The outflowing gas in these two objects thus appears to be primarily photoionized by the AGN.

For the other targets, where only KCWI data are available, the \niiab, \siiab, \oi, and \ha\ emission lines are not covered by the data, so we cannot directly compare the results with model predictions in the BPT and VO87 diagrams. Instead, we compare the KCWI data-based line ratios of the outflow components with model predictions in the \oiioiiihb\ diagrams as shown in Fig. \ref{fig:lrmodelkcwi}. The emission line fluxes are extinction corrected in the same way as stated in Section \ref{34}. The same AGN, shock, and shock$+$precursor models as those shown in Fig. \ref{fig:lrmodel1} and Fig. \ref{fig:lrmodel2} are adopted in this analysis. Additionally, we plot the approximate upper boundary of line ratios predicted by a set of star-forming galaxy models from \citet[][]{Levesque2010} as a red solid line in Fig. \ref{fig:lrmodelkcwi}. Excluding the outflow components with possible contribution from non-outflowing gas (i.e., C2 components in \ta, \tc, and \td, as well as C1 component in \tb), the results suggest that: (1) the star-forming models cannot reproduce the observed line ratios of the outflowing gas in the targets, therefore indicating that massive young stars are not the dominant ionization source in the outflowing gas; (2) the predictions from the shock models can match the observed line ratios of the outflowing gas relatively well, although the models may not be able to explain the observed data with the highest \oiiihb\ ratios and lowest \oiioiii\ ratios; (3) the AGN and shock$+$precursor models can explain the observed line ratios equally well and are both slightly better matches to the observations than the shock models. Moreover, the C2 component of \ta\ and the C1 component of \tb\ have lower \oiioiii\ ratios than the predictions of all three model sets in general, and the C2 component of \tc\ have lower \oiiihb\ ratios than those of the AGN models. These results are consistent with our conclusions in Appendix \ref{42}, \ref{41}, and \ref{43} that these outflow components are partially contaminated by emission from non-outflowing gas.

Next, we have examined the \oiiihb\ line ratios vs the emission line widths ($\sigma_{gas}$) based on the KCWI data for all seven targets with detected outflows in Fig.\ \ref{fig:lrvelkcwi}. To the first order, one would expect a positive correlation between the \oiiihb\ line ratios and gas velocity dispersions \citep[e.g., see Fig. 16 \& 17 in][]{Allen2008}. However, no such clear correlation is seen in our data, which is a similar conclusion to that derived from the \siiha\ ratios. In addition,  for the C2 components in both \te\ and \tf, their observed line widths are significantly smaller (by $\sim$300--400 \kms\ on average) than the shock velocities predicted by the shock and shock$+$precursor models shown in the middle and right columns of Fig. \ref{fig:lrmodelkcwi}. This is apparently contradictory to the expectation that the emission line velocity dispersion reflects the shock velocity when the shocks dominate the ionization of the gas. These results again suggest that shock ionization is not important in our targets. 

Overall, our analysis indicates that AGN is most likely the dominant source of ionization for the outflows in our targets.

\subsection{Electron Densities of the Outflows} \label{54}

The electron density, n$_e$, in the ionized gas may be derived from the \siia/\siib\ ratios or \oiia/\oiib\ ratios, following well-established calibrations \citep[e.g.,][]{Sanders2016}.

For the two targets in our sample with GMOS observations, the spatially-resolved electron density maps derived from the flux ratios of the total \siiab\ line emission show possible radial trends of decreasing electron density outwards, but the errors on $n_e$ are too large to draw robust ($>$5$\sigma$) conclusions. For the other targets, the electron density maps derived from the \oiia/\oiib\ ratios from the KCWI data are even more noisy, which again prevent us from determining the radial trend of the electron densities. Consequently, the electron densities in individual velocity components cannot be measured reliably based on the spatially-resolved maps in these systems. 

To check further the possible difference of [S~II]-based electron densities among different velocity components, we then turn to use the spectra spatially integrated over the whole GMOS data cubes for targets \tb\ and \ta, and the Keck/LRIS spectra for the other targets\footnote{Notice that for the Keck/LRIS data, the emission line profiles are fit with two Gaussian components as described in \citet{ManzanoKing2019}, and here the outflow components in \te, \tg, \tc, \td, and \tf\ refer to the broad components from their best fits}. However, for most of our targets, the measured electron densities of the outflow components still show large uncertainties and thus no useful information of the electron density contrast among individual velocity components can be obtained from our data. The only exceptions are \tb\ and \td, where no clear differences in electron densities are seen among individual velocity components. In the discussion below, we thus adopt the electron densities measured from the \siia/\siib\ ratios based on the total line flux in each object as the electron densities for the outflowing gas (see Table \ref{tab:energetics}).

\subsection{Dust Extinction of the Outflows} \label{55}

From the GMOS data of \tb\ and \ta, we find that the clearly outflowing line-emitting material (the C2 component in \tb\ and the C3 component in \ta) has \hahb\ ratios that are higher than the intrinsic values of typical H~II regions or AGN narrow line region \citep[2.87 and 3.1, respectively;][]{Osterbrock2006}, suggesting dust extinction affects the line emission of the outflows in these objects. Adopting the extinction curve from \citet{ccm89} with $R_V$ = 3.1, the derived extinction values, $A_V$, measured from the spectra integrated over the whole data cube, are on the order of 1 mag. For comparison, the other velocity components in these two targets show slightly smaller $A_V$ by $\sim$0.2 magnitude on average. A more detailed look at the spatially-resolved $A_V$ maps of the outflow components reveals possible radial trends of decreasing $A_V$ at larger radii in both targets. As for the other targets observed with KCWI, the outflow components in \hg\ are in general too faint to allow us to draw robust conclusions.

\subsection{Comparison with the Keck/LRIS Data} \label{56}

The fast outflows in our targets were initially discovered by \citet{ManzanoKing2019} based on Keck/LRIS long-slit data. The properties of the outflows measured from these long-slit data are in broad agreement with those reported here.

The column (10) in Table \ref{tab:obs} lists the 5-$\sigma$ detection limits of an \oiii\ emission line with FWHM of 1000 \kms\ in the GMOS and KCWI data. Excluding the shallower observation of \te, these detection limits are in general comparable to those of the Keck/LRIS data, which are in the range of $\sim$1--3$\times$10$^{-17}$ \flxarc. 

In \te, \tb, \ta, \tc, and \td\ (GMOS data and KCWI data with small slicer setup), the kinematic properties of the outflows (\vwu\ and \wba) measured from these three data sets are similar, but the better spectral resolutions of the GMOS and KCWI IFS data compared with the LRIS data\footnote{Recall that FWHM $\simeq$ 100 \kms\ at 4610 \AA\  for GMOS,  $\simeq$ 80 \kms\ at 4550 \AA\ for the small-slicer setup of KCWI, and $\simeq$ 190 \kms\ for Keck/LRIS \citep{ManzanoKing2019}.} reveal more details in the shapes of the emission line profiles in \ta, \tc, and \td, where three Gaussian components are required to adequately describe the line profiles. The spatial extents of the outflows are broadly consistent with each other after taking into account the sensitivity of the various data sets. 

In \tg\ and \tf\ (KCWI data with the medium and small slicer setup, respectively), blueshifted \oiii\ velocity components are detected in both the Keck/LRIS and KCWI data sets, although they are narrower (by a factor of $\sim$3 on average) and show smaller blueshifts (by a factor of $\sim$4 on average) in the KCWI data when compared to those in the Keck/LRIS data. As for \tx\ (KCWI data with medium slicer setup), a very faint ($\sim$2$\times$10$^{-17}$ \flxarc), broad (\wba\ $\simeq$ 1600 \kms), and redshifted (\vwu\ $\simeq$ 150 \kms) velocity component is reported in the Keck/LRIS data, but it is not detected in the KCWI data. The origin of this apparent discrepancy is not clear although the slightly coarser spectral resolution of LRIS might make it more capable of detecting such a broad feature.

\section{Discussion} \label{6}

\subsection{Energetics of the Outflows} \label{61}

The ionized gas mass of the outflows can be calculated based on either the \oiiib\ line luminosity or the Balmer line (\ha\ or \hb) luminosity of the outflowing, line-emitting gas. We have compared the ionized gas mass of the outflows based on these emission lines, and find that the [O~III]-based values are systematically smaller than the \ha\ or H$\beta$-based values by $\sim$0.2 dex on average, assuming solar metallicity and following equation (29) in \citet{Veilleux2020} (If we assume instead a 0.5$\times$solar metallicity, the average difference increases to $\sim$0.5 dex). This difference may be caused by the uncertainties on the ionization fraction correction (which is assumed to be unity in the previous calculation) and gas-phase metallicity that is assumed in the [O~III]-based ionized gas mass. In order to avoid introducing such uncertainties into our results, the best global fits (Section\ \ref{322}) to the \ha\ (GMOS data) and \hb\ (KCWI data) line emission are thus used to calculate the energetics of the outflows in the following discussion. From \citet{Osterbrock2006} and assuming case B recombination with $T$ = 10$^4$ K, we have 

\begin{eqnarray}
M_{\rm out}= 4.48~\msun \left(\frac{L_{H\alpha,corr}}{10^{35}~{\rm erg~s}^{-1}}\right) \left(\frac{<n_e>}{100\ {\rm cm}^{-3}}\right)^{-1}
\end{eqnarray}
where $L_{H\alpha,corr}$ is the extinction-corrected \ha\ luminosity using the measured Balmer decrement from the total emission line fluxes of the spatially-integrated spectra and adopting an intrinsic \ha/\hb\ ratio of 2.87, appropriate for Case B recombination  \citep[][]{Osterbrock2006}, and the \citet{ccm89} extinction curve with $R_V$ = 3.1. For the KCWI data sets, where \ha\ was not observed, we instead use the extinction-corrected \hb\ luminosity $L_{H\beta,corr}$ and then convert it to $L_{H\alpha,corr}$ using $L_{H\alpha,corr}$ = 2.87 $L_{H\beta,corr}$ as above. 

The calculations of the mass, momentum, and kinetic energy outflow rates depend on the spatial extent of the outflows. As discussed in Section \ref{52}, while the outflows in \te, \tg, \tb, and \tf\ are spatially resolved in the IFS data, our analysis of the IFS data on \ta, \tc, and \td\ does not provide a conclusive outflow size in these objects. For the later, we thus calculate the energetics of the outflows in both scenarios, one where the outflows are spatially resolved and one where they are not.

As presented in Section \ref{51}, while the outflows are mainly traced by the broadest/most blueshifted velocity components  (C3 in \ta, \tc, \td, C2 in \te, \tb, \tf, and C1 in \tg) in the seven targets with detected outflows, the C2 components in \ta, \tc, and \td, as well as the C1 component in \tb\ may also trace significant portion of the outflowing gas in these systems. In the following calculations of the outflow energetics, we thus consider not only the primary outflow components of each target (C3 in \ta, \tc, \td, C2 in \te, \tb, \tf, and C1 in \tg), but also the C2 components in \ta, \tc, and \td, as well as the C1 component in \tb, recording their results separately.

\subsubsection{Spatially Resolved Outflows} \label{611}

We begin with the scenario where the detected outflows are spatially resolved. The mass, momentum, and kinetic energy outflow rates are calculated using a time-averaged, thin-shell, free wind model \citep[e.g.][]{Shih2010,RupkeVeilleux2013b}, where the outflow is spherically-symmetric with a radius $R_{out}$ in 3D space. 

Specifically, the energetics are calculated by summing up quantities over individual spaxels:

\begin{eqnarray}
dM/dt = \sum dm/dt = \sum \frac{m_{\rm out} v_{50,out}\sec\theta}{R_{out}}
\end{eqnarray}
\begin{eqnarray}
dp/dt = \sum (v_{50,out}\sec\theta)dm/dt 
\end{eqnarray}
\begin{eqnarray}
dE/dt = \frac{1}{2}\sum  [(v_{50,out}\sec\theta)^2+3{\sigma_{out}^2}]~dm/dt
\end{eqnarray}
where $m_{\rm out}$, $v_{50,out}$, $\sigma_{out}$ respectively are the ionized gas mass, absolute value of \vwu, and velocity dispersion (= \wba/2.563) measured from the outflow components within individual spaxels.  In these expressions, $\theta = {\rm sin}^{-1}(r_{spaxel}/R_{out})$, the angle between the velocity vector of the outflow in 3D space and the line-of-sight. $R_{out}$, again, is the radius of the spherically-symmetric outflow in 3D space, and is calculated as the maximum extent that the outflow components are detected (S/N of the outflow component of \oiii\ emission $>$ 2) in the sky plane plus half spaxel, converted to an equivalent physical distance. The half spaxel is added artificially since a spherical outflow is formally travelling perpendicular to the line-of-sight at the maximum radius $R_{out}$ (i.e., the $v_{50,out}$ will be 0), and thus no outflow signal can be detected. r$_{spaxel}$ is the projected distance on the sky of a given spaxel with respect to the spaxel with peak outflow flux. In the calculations above, we exclude the spaxels with emission line flux that fall in the lowest 5\% of the full flux range. It should be emphasized that we adopt the [O III]-based $R_{out}$ in the calculation instead of the Balmer-line-based values, which are in general smaller when measured through the fainter \hb\ feature. The mass, momentum, and kinetic energy outflow rates scale as $R_{out}^{-1}$ in the above equations and would thus be higher if the \hb-based $R_{out}$ were used in the calculations.

The electron densities used in the above equations are measured from the \siia/\siib\ ratios, following the conversion presented in \citet{Sanders2016}. As discussed in details in Section \ref{54}, the \siia/\siib\ ratios are calculated using the total line fluxes from the spatially-integrated GMOS spectra or the Keck/LRIS spectra for the other targets without GMOS observations (see Table \ref{tab:energetics}). Neither the \siia/\siib\ ratios of individual spaxels nor the \siia/\siib\ ratios of the outflow components could be used due to their large uncertainties.

We multiply the energetics by a factor of two to account for the far side of the outflow that is blocked by the galaxy, except for the C2 component of \ta, which is purely redshifted and likely represents the back side of the outflow traced by the C3 component (see discussion in Appendix\ \ref{411}). The results of the calculations are listed in Table \ref{tab:energetics}.

It is important to point out that the geometries of the outflows in \te\ and \tf\ may deviate significantly from the spherically-symmetric wind model adopted in the calculations, given the apparent biconical morphologies of the outflows on the sky plane. Nevertheless, if we assume a biconical geometry \citep[e.g., bipolar super-bubble as adopted in][]{RupkeVeilleux2013b} for the outflows in these targets, the estimated change in the mass, momentum and kinetic energy outflow rates are comparable to the errors listed in Table \ref{tab:energetics}. This may also be true for the C2 components in \tc\ and \td, if their apparent biconical/asymmetric morphologies on the sky plane arise from the geometry of the outflowing gas.

\begin{deluxetable*}{ccccccccccccc}[!htb]
\tablecolumns{13}
\tabletypesize{\tiny}
\tablecaption{Energetics of the Outflows \label{tab:energetics}}
\tablehead{
\colhead{Name} & \colhead{Comp.} & \colhead{Data Set} & \colhead{$n_e$ (cm$^{-3}$)}  &   \colhead{log($M$/\msun)}  & \colhead{R$_{out}$(kpc)}   & \colhead{R$_{out,ur}$(kpc)}    & \multicolumn{2}{c}{log[(d$M$/d$t$)/(\msunyr)]} & \multicolumn{2}{c}{log[(d$E$/d$t$)/(erg s$^{-1}$)]}   & \multicolumn{2}{c}{log[($c$ d$p$/d$t$)]/(\lsun)]} \\
& & & & & & & resolved & unresolved & resolved & unresolved & resolved & unresolved \\
(1) & (2) & (3) & (4) & (5) & (6) & (7) & (8) & (9) & (10) & (11) & (12) & (13)
}
\startdata
\hline
\te                  & C2                  & KCWI                  & 60$\pm{50}$                    & 7.3$^{+0.3}_{-0.8}$ & 3.1 & ... & $-$0.5$^{+0.3}_{-0.8}$ & ...       & 40.8$^{+0.3}_{-0.8}$ & ...        & 9.5$^{+0.3}_{-0.8}$ & ...      \\ 
\\
\tg                  & C1                  & KCWI                  & 590$\pm{160}$                  & 4.8$^{+0.1}_{-0.1}$ & 0.9 & ... & $-$2.5$^{+0.1}_{-0.1}$ & ...       & 37.1$^{+0.1}_{-0.1}$ & ...        & 6.8$^{+0.1}_{-0.1}$ & ...     \\
\\
\tb & C2 & GMOS                  & \multirow{4}{*}{470$\pm{150}$} & 5.4$^{+0.1}_{-0.2}$ & 0.8 & ... & $-$1.4$^{+0.1}_{-0.2}$ & ...       & 39.3$^{+0.1}_{-0.2}$ & ...        & 8.5$^{+0.1}_{-0.2}$ & ...      \\
    & C1 & GMOS                  &                                & 5.4$^{+0.2}_{-0.2}$ & 0.9 & ... & $-$1.6$^{+0.1}_{-0.2}$ & ...       & 38.1$^{+0.1}_{-0.1}$ & ...        & 8.0$^{+0.1}_{-0.2}$ & ...      \\
                                    &  C2                   & KCWI                  &                                & 5.9$^{+0.1}_{-0.2}$ & 1.6 & ... & $-$1.2$^{+0.1}_{-0.2}$ & ...       & 39.4$^{+0.1}_{-0.2}$ & ...        & 8.6$^{+0.1}_{-0.2}$ & ...      \\ 
                                    &  C1                   & KCWI                  &                                & 6.0$^{+0.1}_{-0.2}$ & 1.6 & ... & $-$1.6$^{+0.2}_{-0.3}$ & ...       & 38.1$^{+0.1}_{-0.1}$ & ...        & 7.8$^{+0.1}_{-0.1}$ & ...      \\ 
 \\
  \ta & C3                  & GMOS & \multirow{4}{*}{570$\pm{360}$} & 5.8$^{+0.2}_{-0.4}$ & 1.1 & 0.3 & $-$1.8$^{+0.2}_{-0.4}$ & $>-$0.9 & 39.2$^{+0.2}_{-0.4}$ & $>$40.2  & 7.5$^{+0.2}_{-0.4}$ & $>$8.4 \\
                                      & C2                  &   GMOS  &                   &      5.4$^{+0.2}_{-0.4}$ &  1.2 & 0.3 & $-$2.1$^{+0.2}_{-0.4}$ & $>-$1.6 & 37.8$^{+0.2}_{-0.4}$ & $>$38.7  & 7.1$^{+0.2}_{-0.4}$ & $>$7.6 \\
                                    & C3                  & KCWI &                              & 5.9$^{+0.2}_{-0.4}$ & 2.1 & 0.4 & $-$1.5$^{+0.2}_{-0.4}$ & $>-$0.9 & 39.9$^{+0.2}_{-0.4}$ & $>$40.3  & 8.3$^{+0.2}_{-0.4}$ & $>$8.7 \\
                                     & C2                  &  KCWI                     &                                & 6.1$^{+0.2}_{-0.4}$ & 2.2 & 0.4 & $-$1.4$^{+0.2}_{-0.4}$ & $>-$0.7 & 39.1$^{+0.2}_{-0.4}$ & $>$39.7  & 8.2$^{+0.2}_{-0.4}$ & $>$8.7 \\
 \\
\tc & C3                  & KCWI & \multirow{2}{*}{470$\pm{80}$}  & 5.8$^{+0.1}_{-0.1}$ & 1.6 & 0.4 & $-$1.5$^{+0.1}_{-0.1}$ & $>-$1.0 & 39.6$^{+0.1}_{-0.1}$ & $>$39.9  & 8.2$^{+0.1}_{-0.1}$ & $>$8.4 \\ 
                                    & C2                  & KCWI                      &                                & 6.3$^{+0.1}_{-0.1}$ & 1.8 & ... & $-$2.1$^{+0.1}_{-0.1}$ & ...       & 38.9$^{+0.1}_{-0.1}$ & ...        & 7.9$^{+0.1}_{-0.1}$ & ...      \\ 
\\
\td & C3                  & KCWI & \multirow{2}{*}{450$\pm{100}$} & 5.2$^{+0.1}_{-0.1}$ & 0.3 & 0.1 & $-$1.2$^{+0.1}_{-0.1}$ & $>-$0.6 & 40.1$^{+0.1}_{-0.1}$ & $>$40.4 & 9.0$^{+0.1}_{-0.1}$ & $>$9.2 \\
                                    & C2                  &  KCWI                     &                                & 5.6$^{+0.1}_{-0.1}$ &  0.7 & ... & $-$1.7$^{+0.1}_{-0.3}$ & ...       & 38.8$^{+0.1}_{-0.1}$ & ...        & 7.8$^{+0.1}_{-0.1}$ & ...      \\
\\
\tf                  & C2                  & KCWI                  & 150$\pm{60}$                   & 5.5$^{+0.2}_{-0.2}$ & 0.8 & ... & $-$2.0$^{+0.2}_{-0.2}$  & ...       & 38.2$^{+0.2}_{-0.2}$ & ...        & 7.5$^{+0.2}_{-0.2}$ & ...      
\enddata
\tablecomments{Column (1): Short name of the target; Column (2): Individual outflow components from the best fits; Column (3): Instrument used for the observations; Column (4): Electron density measured from the \siiab\ line ratio based on the total line flux from the spatially-integrated, GMOS spectra or keck/LRIS spectra (see Section \ref{54}); Column (5): Ionized gas mass of the corresponding outflow component; Column (6): Outflow radius adopted in the calculation of mass, momentum and kinetic energy outflow rates when the outflows are spatially resolved (Column (8), (10), and (12), respectively); Column (7): Outflow radius adopted in the calculation of mass, momentum and kinetic energy outflow rate when the outflow is spatially unresolved (Column (9), (11), and (13), respectively); (8): Ionized gas mass outflow rate of the corresponding outflow component when the outflow is spatially resolved;  Column (9): Same as in Column (8) but with the assumption that the outflow is spatially unresolved; Column (10): Ionized gas kinetic energy outflow rate of the corresponding outflow component when the outflow is spatially resolved; Column (11): Same as in Column (10) but with the assumption that the outflow is spatially unresolved; Column (12): Ionized gas momentum outflow rate of the corresponding velocity component when the outflow is spatially resolved; Column (13): Same as in Column (12) but with the assumption that the outflow is spatially unresolved.}
\end{deluxetable*}


\subsubsection{Spatially Unresolved Outflows} \label{612}

If instead the outflows are unresolved by the IFS data, the total mass of the outflowing gas remains unchanged, but the time-averaged mass, momentum, and kinetic energy outflow rates are affected since they depend inversely on the size of the outflows. As discussed above, the C3 components of \ta, \tc, and \td\, and the C2 component of \ta\ may be spatially unresolved. In this scenario, we adopt $\frac{1}{2}$~$\times$~FWHM(PSF) as a conservative upper limit to the true outflow radius $R_{out,ur}$, and get:

\begin{eqnarray}
dM/dt = \frac{M_{\rm out,tot} v_{out,tot}}{R_{out,ur}}
\end{eqnarray}
\begin{eqnarray}
dp/dt = v_{out,tot} dM/dt
\end{eqnarray}
\begin{eqnarray}
dE/dt = \frac{1}{2} (v_{50,out}^2+3{\sigma_{out}^2}) dM/dt
\end{eqnarray}
Here, $M_{\rm out,tot}$ is the total mass of the outflowing gas, and $R_{out,ur}$ is the upper limit on the radius of the outflow. The quantities $v_{50,out}$ and $\sigma_{out}$ are the median values of \vwu\ and $\sigma$ (= \wba/2.563) of the outflow components measured across the data cube (see Table \ref{tab:kinematics}). The adopted electron densities are the same as those in the spatially resolved scenario. The lower limits on the outflow rates obtained under these assumptions are listed in Table \ref{tab:energetics}.

\begin{figure}[!htb]   
\epsscale{1.1}
\plotone{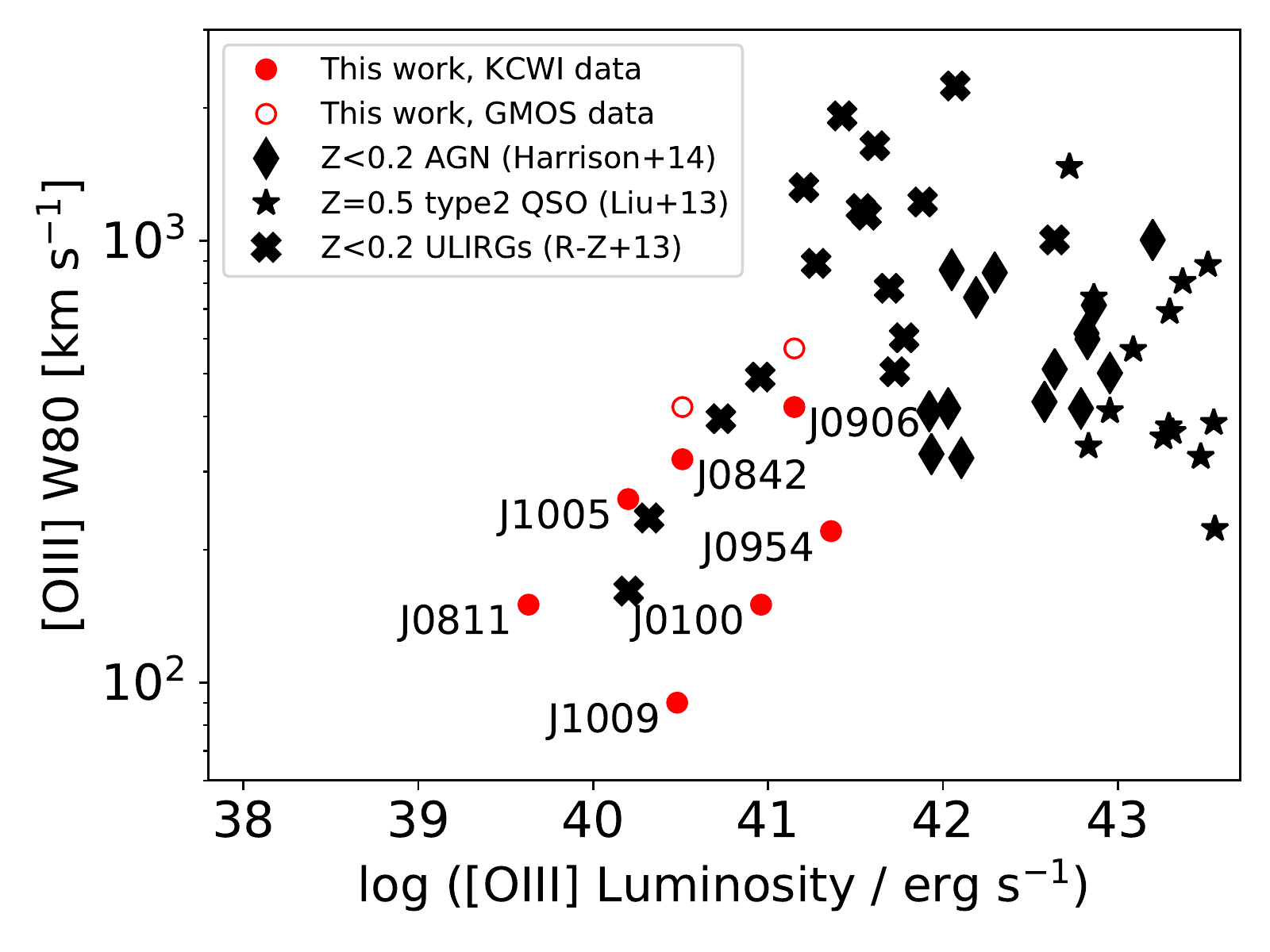}
\caption{\oiii\ line widths \wba\ vs \oiii\ luminosities for the seven targets with detected outflows (red filled circles indicate the KCWI data and red open circles indicate the GMOS data of \tb\ and \ta) as well as more luminous AGN and ULIRGs taken from the literature \citep[black symbols;][]{Liu2013a,Liu2013b,Harrison2014,RZ2013}, as indicated in the legend. All measurements refer to the total, spatially-integrated \oiii\ line emission from each object. The typical errors of the measurements are similar to the size of the data points.}
\label{fig:fenxiw80}
\end{figure}


\begin{figure*}[!htb]   
\epsscale{1.1}
\plottwo{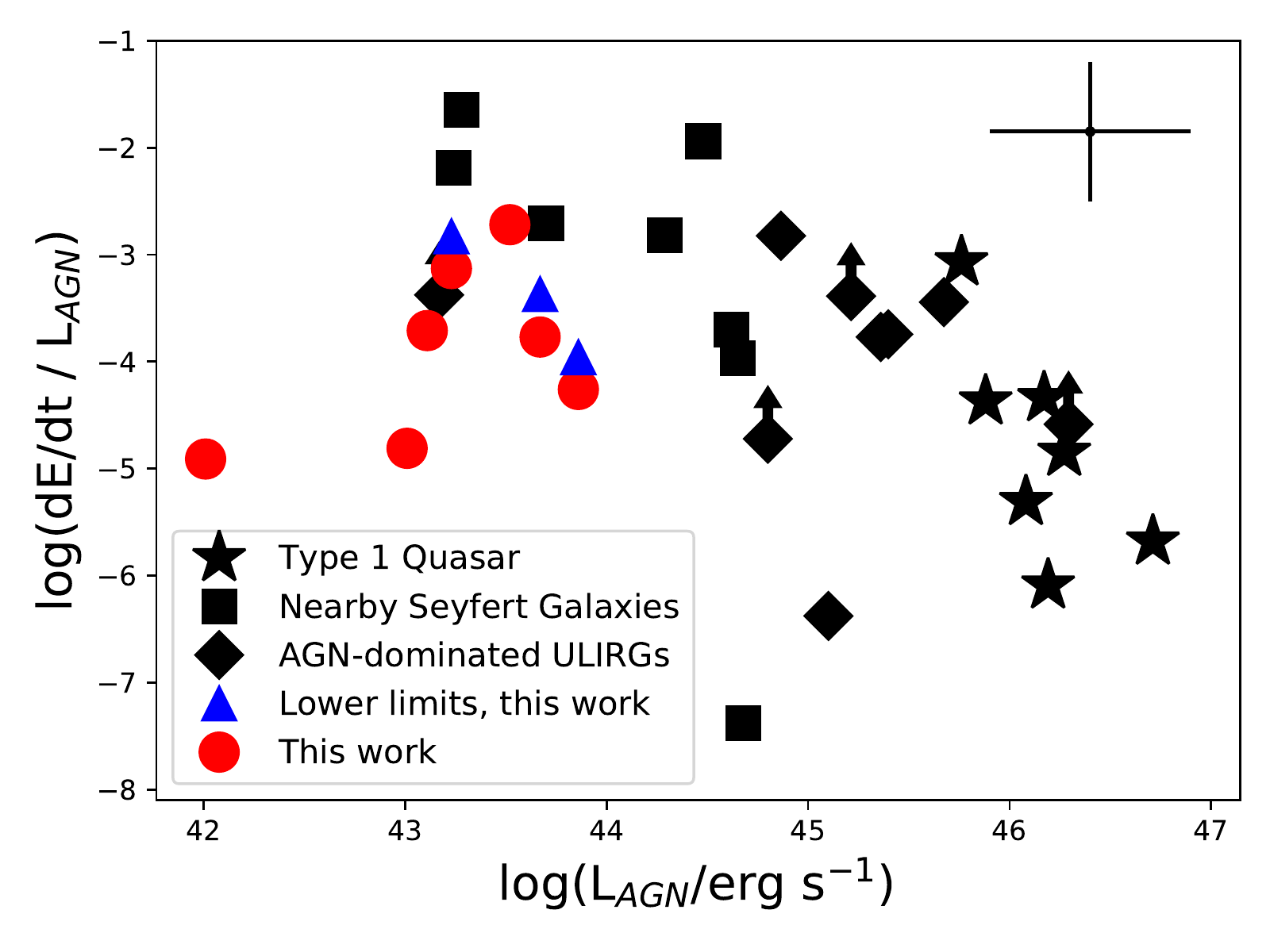}{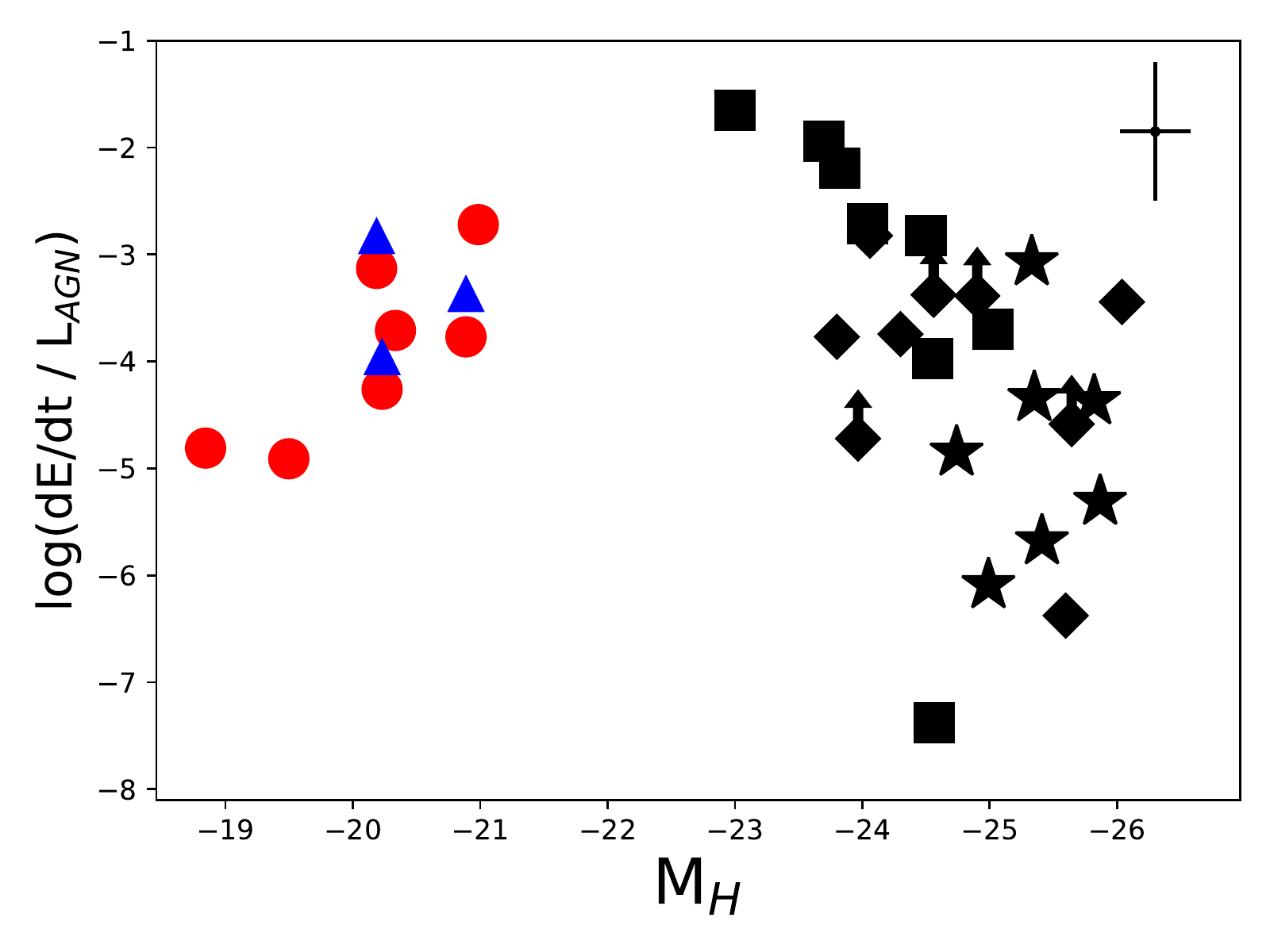}
\caption{Ratios of the kinetic energy outflow rates, based on the KCWI data, to the AGN bolometric luminosities as a function of (left) the AGN bolometric luminosties and (right) H-band absolute magnitudes, for the seven targets with detected outflows in our sample (red circles) and lower limits (blue triangles) if the outflows in \ta, \tc, and \td\ are spatially unresolved (see Section \ref{52} and \ref{612}). Here we have neglected the kinetic energy outflow rates calculated from the C2 components in \ta, \tc, \td\ and the C1 component in \tb\ as their contributions are modest. Also plotted as a comparison are the values from a sample of $z <$ 0.3 but more powerful type 1 quasars and nearby Seyfert galaxies from or collected by \citet{Rupke2017}, as well as a sample of $z <$ 0.15, AGN-dominated ULIRGs from \citet{Rose2018}. The absolute H-band magnitudes shown in the right panel are derived from the 2MASS \citep{2MASS} H-band magnitudes taken from the IRSA/2MASS archive, except for those of the Type 1 quasars and three of the ULRIGs from \citet{Rose2018}, which are the AGN-subtracted, host-only H-band magnitudes quoted from \citet{veilleux2006,veilleux2009a}. The estimated typical errors of the data points are noted as black crosses in the upper-right corners of both panels. 
}
\label{fig:ke}
\end{figure*}


\subsection{Comparison with More Luminous AGN} \label{62}

The most direct measure of the magnitude of an outflow is its velocity. Various definitions have been used in literature to represent outflow velocities \citep[e.g., see a brief summary in Sec. 3.1 in][]{Veilleux2020}. \wba\ of the overall spatially integrated emission line profiles have been used as surrogates for characteristic outflow velocities in many studies \citep[e.g.][]{Liu2013a,Liu2013b,RZ2013,Harrison2014,nadiajenny2014}. In Fig. \ref{fig:fenxiw80}, the values of \wba\ derived from the \oiii\ line emission integrated over our data cubes and the \oiii\ luminosities (\loiii) of our targets are compared with published values in low-z AGN and/or Ultraluminous Infrared Galaxies (ULIRGs) with strong outflows. Remarkably, four of our targets (\tb, \ta, \tc, and \td) have \wba\ that are comparable to those of AGN with \loiii\ that are two orders of magnitude larger than those of our targets. However, in general, the data points suggest a positive correlation between \oiii\ \wba\ and luminosities, spanning 4 orders of magnitude in \loiii\ and 1.5 orders of magnitude in \wba.  This correlation simply implies that more powerful AGN provide more energy to drive faster outflows. 

A more physically meaningful, albeit also more model-dependent, estimate of the importance of an outflow is the kinetic energy outflow rate. In Fig. \ref{fig:ke}, the kinetic energy outflow rates of our targets (based on the KCWI data), normalized by their AGN luminosities (see Table \ref{tab:targets}), are compared with those of low-z Seyferts and type 1 quasars studied in \citet{Rupke2017}, as well as those of the z $<$ 0.15, AGN-dominated ULIRGs from \citet{Rose2018}. The results for the C2 components of \ta, \tc, and \td, as well as the C1 component of \tb\ are omitted in this analysis due to their relatively modest contribution, as they have on average $\sim$1 dex smaller dE/dt than those of either the C3 or C2 components of these targets. The values shown in this figure assume the spatially-resolved scenario by default (red filled circles; see Section\ \ref{611}) for all of our sources. For \ta, \tc, and \td, we also show the lower limits obtained by assuming that the outflow components are spatially unresolved (blue filled triangles). The measurements of \tb\ and \ta\ based on the GMOS data are also omitted as they have dE/dt smaller than (but close to) those based on the KCWI data. Compared with our targets, those Seyferts and quasars have both more powerful AGN (with higher median AGN luminosity by $\sim$1 to 3 orders of magnitude) and more massive host galaxies (with brighter median H-band absolute magnitudes\footnote{The absolute H-band magnitudes of our targets and all other sources  are derived from the 2MASS \citep{2MASS} H-band magnitudes taken from the IRSA/2MASS archive \irsaurl, except for those of the Type 1 quasars and three of the ULRIGs from \citet{Rose2018}, which are the AGN-subtracted, host-only H-band magnitudes quoted from \citet{veilleux2006,veilleux2009a}. While the H-band magnitudes of the Seyfert galaxies are not AGN-subtracted, the contribution from the AGN is probably not substantial: the H-band magnitudes of the Seyfert galaxies are close to the QSO-subtracted ones of the Type 1 quasars, which is consistent with the fact that the stellar velocity dispersions of the two samples are comparable when they are measured or recorded in \citet{Rupke2017}.} by $\sim$4 to 5 mag.). Nevertheless, our targets have ratios of kinetic energy outflow rates to AGN luminosities that are comparable to those measured in the more luminous AGN. This result adds support to the idea that the outflows in the dwarf galaxies are scaled-down versions of the outflows in the more luminous AGN and are fundamentally driven by the same AGN processes. We examine this issue in more detail in Section \ref{63}. 


\subsection{What Drives these Outflows: AGN or Starbursts?} \label{63}

The results from the previous sections favor AGN-related processes as the main driver of the detected outflows.  First, the velocities of the outflows detected in our dwarf galaxies are often large. The maximum \wba\ of outflow components in six targets exceed 600 \kms, including three that exceed 1000 \kms. If we adopt the definition of bulk outflow velocities V$_{out}=$\wba$/$1.3 as in some studies \citep[e.g.][where they assume spherically-symmetric or wide-angle bi-cone outflows]{Liu2013b,Harrison2014}, six out of the seven targets with detected outflows have outflow velocities $\gtrsim$500 \kms. To put these numbers into perspective, a velocity of 500 \kms\ is equivalent to an energy of 1 keV per particle, and is difficult to achieve with stellar processes \citep{Fabian2012araa}. The high velocities of the outflows seen in most of our targets thus suggest that AGN plays an important role in driving these outflows.

Second, as shown in Fig.\ \ref{fig:ke} and discussed in Section \ref{62}, the AGN are also powerful enough to drive the outflows in our targets. The ratios of kinetic energy outflow rates to bolometric AGN luminosities of our targets are in the range of $\sim$1$\times$10$^{-5}$ -- 2$\times$10$^{-3}$. These ratios are far less than unity, and are within the range of values seen in other more luminous AGN, suggesting that the AGN are more than capable of driving these outflows. 

The lower limits of the ionized gas mass entrainment efficiency $\eta$, defined as the ratio of ionized gas mass outflow rate over the star formation rate, are in the range of $\sim$0.1 -- 0.8, with a median of $\sim$0.3 (the range and median are $\sim$0.1 -- 0.6 and $\sim$0.2, respectively, if we exclude the contributions from the C2 components in \ta, \tc, and \td, and from the C1 component in \tb). Note that these are lower limits since our adopted SFR are upper limits (see Section\ \ref{35}). This is comparable to the average value ($\sim$0.19) measured for the neutral outflows in low-redshift, AGN/starburts-composite ULIRGs \citep{Rupke2005c}. In the more luminous AGN, apparently higher $\eta$\ are reported in the literature. For example, $\eta$ $\simeq$ 6 $-$ 20 are reported for a sample of $z <$ 0.2 luminous type 2 AGN \citep{Harrison2014}. Meanwhile, much lower $\eta$\ values, with a median of 0.8, are reported for a sample of type 1 quasars at z$<$ 0.3 in \citet{Rupke2017} once the quasar emission is subtracted and both the neutral and ionized phases of the outflows are considered. In their sample, the median value of $\eta$ drops further to 0.03 when the ionized phase alone is considered. In short, the $\eta$ measured in our targets fall in the wide range seen in various studies of outflows in more luminous AGN. In addition, if the outflows in \ta, \tc, and \td\ are spatially unresolved, then the lower limits of $\eta$ can be as high as $\sim$3, uncomfortably high for starburst-driven outflows in the low-z universe \citep[e.g.][where $\eta < 1$ in general]{Arribas2014}. This is even more so if we also consider the possible contribution from the C2 components to the outflow energetics in these targets. 

There is also circumstantial evidence against starburst driving of these outflows. Given the upper limits of SFR estimated from the \oiiab\ emission, all of the galaxies in our sample lie either slightly, or significantly, below the main sequence of star-forming galaxies in the low-z universe \citep[e.g.][]{Brinchmann2004}, while the star formation-driven outflows are observed much more frequently in galaxies above the star formation main sequence \citep[e.g.][]{Heckman2015,Roberts-Borsani2020}.

More quantitatively, we can examine if stellar processes are physically capable of driving the observed outflows.  The typical kinetic energy output rate from core collapse supernovae is $\sim$7$\times10^{41}(\alpha_{SN}/0.02)(\dot{M_{\star}}/\msun \ yr^{-1})$ \citep{Veilleux2005,Veilleux2020}. Adopting the SFR upper limits of our targets (Table \ref{tab:targets}), and assuming a constant supernovae rate of $\alpha_{SN}=0.02$, the expected maximum kinetic energy output rates from core-collapse supernovae in our targets are in the range of $\sim$7$\times$10$^{39}$ -- 5$\times$10$^{41}$ erg $s^{-1}$, with a median of $\sim$2$\times$10$^{41}$ erg $s^{-1}$. These are $\sim$6 -- 720 times larger than the kinetic energy outflow rates based on the scenario that the outflows are spatially resolved. Stellar processes thus cannot be overlooked as a potential source of energy for these outflows.

However, it should be pointed out that we have only considered the warm ionized phase of the outflowing gas and adopted the energetics calculated in the spatially resolved scenario. If the outflows in \ta, \tc, and \td\ are spatially unresolved, the kinetic energy outflow rates may be comparable to, if not larger than, the kinetic energy output from the stellar process as estimated above. This argument is slightly stronger if we consider the contribution from the C2 components to the outflow energetics in these targets, too. Additionally, it is possible that a significant fraction of the energy is carried in a hot, thin gas phase instead, which has been predicted by recent simulations \citep[e.g.][]{Koudmani2019,Koudmani2020}. 

Overall, the outflows in our targets are likely driven by AGN, but we cannot formally rule out the possibility that star formation activity may also help in launching the outflows, as is often the case among low-z ULIRGs and luminous AGN \citep[e.g.][]{RupkeVeilleux2013b,Harrison2014,Fluetsch2019}.  More stringent constraints on the star formation rates of our targets need to be obtained before we can draw a more robust conclusion about the role of stellar processes in these outflows. 

\subsection{Does the Outflowing Gas Escape the Galaxies?} \label{64}

To help us evaluate the impact of these outflows on their host galaxies, it is interesting to examine the question of whether some of the outflowing gas is able to escape the host galaxy. This requires comparing the kinematics of the outflows with the local escape velocity, $v_{\rm esc}(r) = \sqrt{2[\Phi(\infty) - \Phi(r)]}$, where $\Phi(r)$ and $\Phi(\infty)$ are the values of the gravitational potential at $r$ and $r = \infty$, respectively, in the case of a spherically-symmetric galaxy. 

One may estimate the escape velocity in terms of observed quantities, like the circular velocity $v_{\rm circ}$ of the galaxy, by assuming a simple density profile such as that of a singular isothermal sphere. A conservative estimate of the escape velocity in that case gives $v_{\rm esc} \simeq 3v_{\rm circ}$ \citep{Veilleux2020}. Our IFS data do not probe the flat portion of the rotation curve, so we adopt the maximum of the measured stellar velocities (v$_\star$) and velocity dispersions ($\sigma_\star$) to calculate the lower limits of the circular velocities in our targets, where $v_{\rm circ} = \sqrt{v_\star^2+2\sigma_\star^2}$ \citep[e.g., See Section 2.4 of ][]{Veilleux2020}. We have not applied any deprojection corrections to the circular velocities and outflow velocities, given that the 3D morphologies of the outflows are poorly constrained.

Alternatively, the escape velocity may be derived by assuming a NFW dark matter density profile \citep{NFW} and a total halo mass determined from abundance matching \citep{Moster2013}, which has been done in \citet{ManzanoKing2019}. Since the escape velocity always peaks at the center, it can serve as a conservative upper limit to the escape velocity throughout the galaxy. For our targets, the escape velocities at $r$ = 0 obtained through this approach are larger by $\sim$50\% on average than those based on the empirical circular velocities above. We adopt the more conservative $r =$ 0, NFW-based escape velocities in the remainder of our discussion. 

\begin{deluxetable}{ccc}[!htb]
\tabletypesize{\normalsize}
\tablecaption{Outflow Escape Fractions\label{tab:escape}}
\tablehead{
\colhead{Target} & \colhead{V$_{esc}$ [\kms]} & \colhead{$f_{esc}$} \\
\colhead{(1)} & \colhead{(2)} & \colhead{(3)}
}
\startdata
\te & 320 & 1\%   \\
\tg & 260 & 0.1\% \\
\tb & 300 & 6\%   \\
\ta & 300 & 6\%   \\
\tc & 320 & 1\%   \\
\td & 380 & 1\%   \\
\tf & 240 & 0.3\% 
\enddata
\tablecomments{Column (1): Short name of the target; Column (2): Escape velocity at the center of each galaxy assuming a NFW density profile, rounded to the nearest 10 \kms; Column (3): Escape fraction of the \oiii\ line emitting gas, based on flux rather than mass. This number does not take into account possible density contrasts between the outflowing and quiescent gas components in these systems and projection effects; see Section\ \ref{64} for more details.}
\end{deluxetable}

For all of the targets, we next define the escape fraction ($f_{\rm esc}$) as the ratio of  \oiii\ flux with absolute velocities larger than the escape velocity summed up across the data cube, to the total emission line flux in the whole data cube. Notice here that the escape fraction is defined as a flux ratio rather than a mass ratio, so it does not take into account possible density contrasts between the outflowing and quiescent (non-outflowing) gas components \citep[e.g.][]{Hinkle2019,Fluetsch2019,Fluetsch2020}, which may affect the luminosity-to-mass conversion factor. In addition, the values of $f_{\rm esc}$ obtained here are conservatively low since we have not applied deprojection corrections to the gas velocities in the outflows. Some fraction of the escaping gas may not be accounted for here if the velocities of this gas, projected along our line of sight, fall below $v_{\rm esc}$.

The results from our IFS data are summarized in Table \ref{tab:escape}. The escape fractions range from 0.1\% to 6\%. Taking into account that the escape velocities are likely overestimated for the reasons mentioned earlier and that the outflow velocities are potentially underestimated due to projection effects, this suggests that at least some small portion of the outflowing gas may travel a long way from the centers and help contribute to the metal enrichment of the circumgalactic medium in these dwarf galaxies \citep[as reported in a number of studies; e.g.][]{Bordoloi2014}.

\section{Conclusions} \label{7}

In this paper, we report the results from an integral field spectroscopic study with Gemini/GMOS and Keck/KCWI of the warm ionized gas in a sample of 8 low-redshift (0.01 $\lesssim$ z $\lesssim$ 0.05) dwarf galaxies with known AGN and suspected outflows. The main results are summarized as follows:

\begin{itemize}

\item 

Warm ionized outflows are detected in 7 out of the 8 targets. The IFS data in most targets reveal broad, blueshifted velocity components tracing rapid outflows (\vwu\ down to $\sim$$-$240 \kms\ and \wba\ up to $\sim$1200 \kms) and narrow components tracing the rotation of the host galaxies. In \ta, \tc, and \td, the multi-Gaussian fits require a third velocity component with intermediate line widths, which probably traces portion of the outflowing gas and/or turbulent gas. In \tg\ and \tb, the narrow components are in general blueshifted and may trace the outflows or a mixture of outflowing and rotating gas in these systems. 

\item

The two-dimensional velocity structures and radial profiles of the outflowing kinematic components indicate that the outflows are spatially resolved by the IFS data in at least four cases (\tb, \te, \tf, \tg), with the emission extending up to $\sim$3 kpc from the galactic centers. In \te\ and \tf, the outflowing kinematic components show apparent biconical morphologies in projection. Additionally, clear non-radial velocity gradients/structures are also seen in those components of \tg\ and \tb. In \ta, \tc, and \td, the kinematic components that have intermediate line widths and probably trace part of the outflows are also spatially resolved.  However, the fast outflows traced by the kinematic components with the broadest line widths in these targets are not clearly spatially resolved. An attempt at deconvolving the data cubes gives inconclusive results.

\item
The clearly outflowing gas in all of the targets have line ratios that are consistent with AGN photoionization. A general lack of positive correlation between the gas kinematics and the \siiha\ or \oiiihb\ line ratios, and inconsistencies between the observed line ratios and the predictions from shock models, indicate that shocks likely do not play a major role in heating and ionizing the outflowing gas in these systems.

\item
Assuming a simple thin-shell, free wind model, the warm, ionized gas mass outflow rates of our targets range from $\sim$3$\times$10$^{-3}$ to $\sim$3$\times$10$^{-1}$ \msunyr, and the kinetic energy outflow rates range from $\sim$1$\times$10$^{37}$ erg s$^{-1}$ to $\sim$6$\times$10$^{40}$ erg s$^{-1}$ (excluding the contribution from the velocity components that likely trace portion of the outflows in targets \tb, \ta, \tc, and \td). In \ta, \tc, and \td, where the outflows may be spatially unresolved, the lower limits of the mass outflow rates and kinetic energy outflow rates are $\sim$2--10 times higher than those obtained in the scenario where they are spatially resolved.

\item

The overall emission line widths measured from the spatially-integrated spectra of our targets, together with the results from samples of more luminous AGN studied in the recent literature, show a positive trend with increasing \oiii\ luminosities. When normalized by the bolometric AGN luminosities, the kinetic energy outflow rates of these outflows are comparable to those of more luminous AGN in massive systems. The outflows in these dwarf galaxies act as scaled-down versions of those in more luminous AGN, in shallower potential wells.

\item
The outflows are likely driven by the central AGN, since i) the outflows are faster than typical outflows driven by stellar processes; ii) the AGN is powerful enough to drive the outflows given the efficiency of other low-redshift AGN; (iii) the lower limits of the ionized gas mass entrainment efficiency (i.e. mass outflow rates to SFR $\simeq$ 0.1--0.8, based on the upper limits on SFR estimated from the \oiiab\ emission) fall in the wide range seen in various studies of outflows in more luminous AGN, and may be uncomfortably high (with lower limits up to $\sim$3) for starburst-driven outflows in the low-z universe if the outflows are spatially unresolved in targets \ta, \tc, and \td; (iv) the dwarf galaxies of our sample all lie either slightly or significantly below the main sequence of star-forming galaxies, whereas starburst-driven outflows typically take place in star-forming galaxies above that main sequence.  However, we cannot formally rule out, based on energetic arguments, the possibility that the star formation activity in these galaxies also partially contributes to driving these outflows.

\item
A small but non-negligible fraction (at least 0.1\%--6\%) of the outflowing ionized gas in our targets has velocities large enough to escape from the host galaxies, if no additional drag force is present. These outflows may thus contribute to the enrichment of the circumgalactic medium in dwarf galaxies.

\end{itemize}

If such AGN-driven outflows are also present in dwarf galaxies at high redshifts, they will increase the porosity of these dwarf galaxies and thus their contribution to the reionization of the universe \citep[e.g.][]{Silk2017}. They may also help explain the current core-cusp controversy regarding the dark matter distribution in dwarf galaxies \citep[e.g.][]{Maccio2020}. A proper treatment of such AGN feedback will need to be included in seed black hole formation models \citep[e.g.][]{Mezcua2019a}.

\clearpage
\acknowledgements

We thank the anonymous referee for thoughtful and constructive comments that improved this paper. S.V. and W.L. acknowledge partial support for this work provided by NASA through grants {\em HST} GO-15662.001A and GO-15915.001A from the Space Telescope Science Institute, which is operated by AURA, Inc., under NASA contract NAS 5-26555. It also made use of NASA's Astrophysics Data System Abstract Service and the NASA/IPAC Extragalactic Database (NED), which is operated by the Jet Propulsion Laboratory, California Institute of Technology, under contract with the National Aeronautics and Space Administration. G.C. acknowledges partial support from the National Science Foundation, under grant number AST 1817233. Additional support was provided by NASA through a grant from the Space Telescope Science Institute (Program AR- 14582.001-A), which is operated by the Association of Universities for Research in Astronomy, Incorporated, under NASA contract NAS5-26555. 

A significant part of the data presented herein were obtained at the W. M. Keck Observatory, which is operated as a scientific partnership among the California Institute of Technology, the University of California, and the National Aeronautics and Space Administration. The Observatory was made possible by the generous financial support of the W. M. Keck Foundation. The authors wish to recognize and acknowledge the very significant cultural role and reverence that the summit of Maunakea has always had within the indigenous Hawaiian community. We are most fortunate to have the opportunity to conduct observations from this mountain. The rest of the observations were obtained at the international Gemini Observatory, a program of NSF’s NOIRLab, which is managed by the Association of Universities for Research in Astronomy (AURA) under a cooperative agreement with the National Science Foundation. on behalf of the Gemini Observatory partnership: the National Science Foundation (United States), National Research Council (Canada), Agencia Nacional de Investigaci\'{o}n y Desarrollo (Chile), Ministerio de Ciencia, Tecnolog\'{i}a e Innovaci\'{o}n (Argentina), Minist\'{e}rio da Ci\^{e}ncia, Tecnologia, Inova\c{c}\~{o}es e Comunica\c{c}\~{o}es (Brazil), and Korea Astronomy and Space Science Institute (Republic of Korea). The data were processed using the Gemini IRAF package. This publication makes use of data products from the Two Micron All Sky Survey, which is a joint project of the University of Massachusetts and the Infrared Processing and Analysis Center/California Institute of Technology, funded by the National Aeronautics and Space Administration and the National Science Foundation.

\software{Astropy (\url{http://dx.doi.org/10.1051/0004-6361/201322068}), Gemini IRAF package \citep{Gemini}, IFSRED \citep{ifsred}, IFSFIT \citep{ifsfit}, KCWI data reduction pipeline (\kcwiurl) MPFIT\citep{mpfit}, pPXF \citep{ppxf}, Kinemetry \citep{Kinemetry}.}


\bibliography{dwarfoutflow}

\clearpage
\appendix

\begin{figure}[!htb] 
\plotone{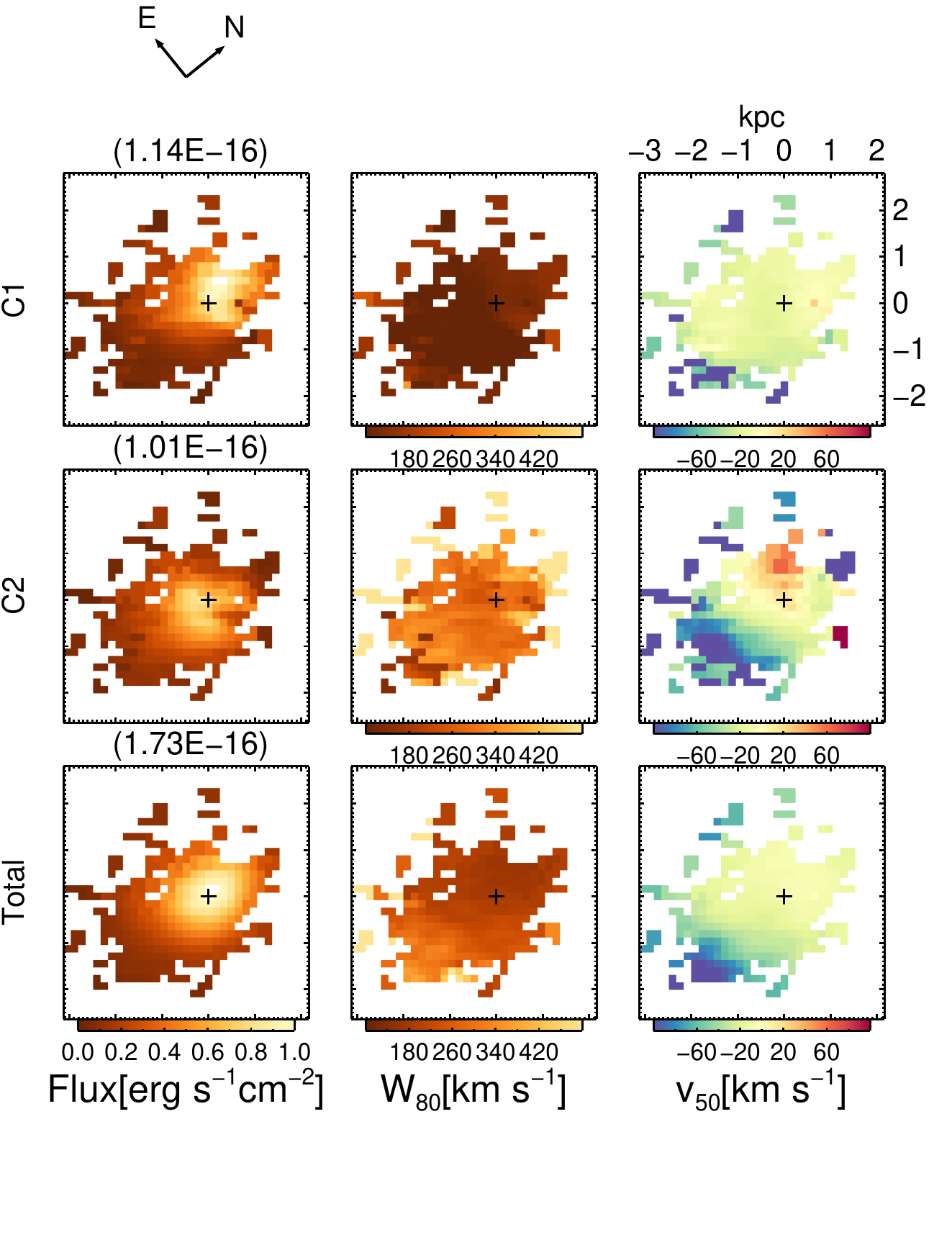}
\caption{Voronoi-binned maps of the \oiii\ flux and kinematics in \te\ based on the KCWI data. The orientation of the maps is indicated by the compass at the top of the figure. Maps of the properties of the individual velocity components derived from the multi-Gaussian fits (C1, C2), and those of the overall emission profiles (Total), are shown from top to bottom. The flux maps are shown in the leftmost column, where each map is normalized to the maximum flux value in the map, which is listed in cgs units above each panel. The line widths \wba\ and velocities \vwu\ are shown in the middle and rightmost columns, respectively. In each panel, the black cross indicates the spaxel where the peak of the total \oiii\ emission line flux falls. The coordinates of the panels are in kpc.}
\label{fig:o3map1}
\end{figure}

\begin{figure}[!htb]   
\epsscale{0.5}
\plotone{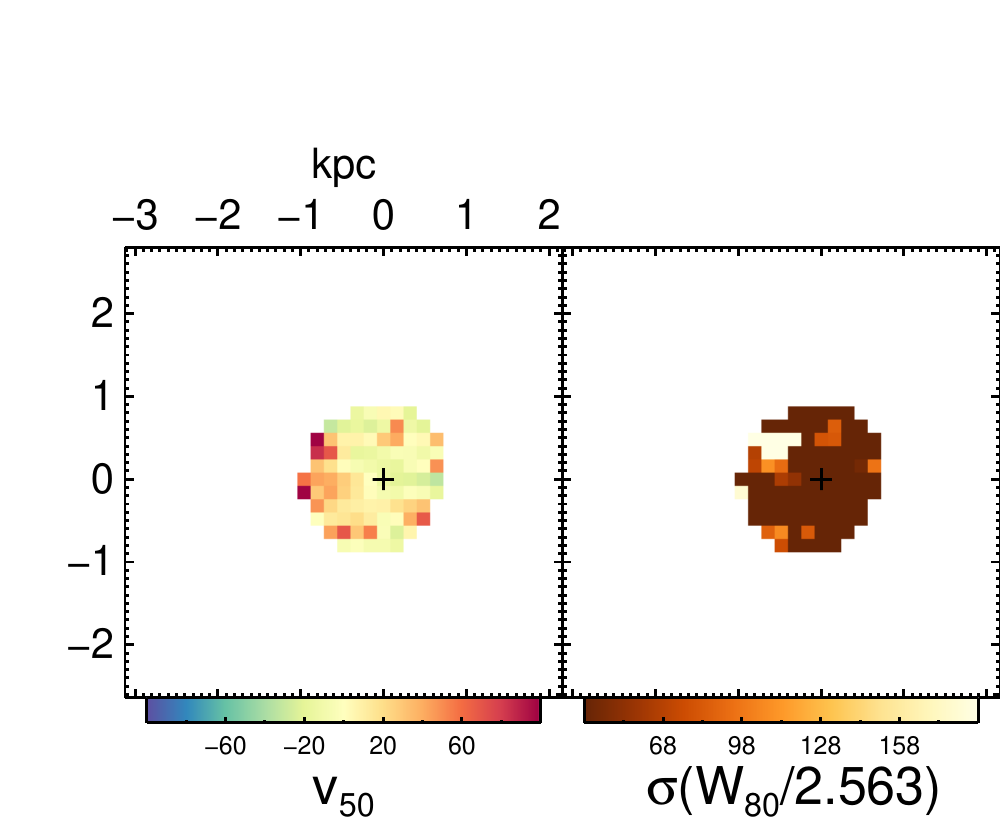}
\caption{Maps of the stellar median velocity (left) and velocity dispersion (right) in \te\ on the same spatial scale as Fig. \ref{fig:o3map1}. The map of the median velocity is drawn on the same color scale as the \vwu\ maps of \oiii\ (Fig.\ \ref{fig:o3map1}). The map of the velocity dispersion is also drawn on the same color scale as the \wba\ maps of \oiii, namely the same color represents the same line width in all maps. For a Gaussian profile, the conversion between \wba\ and velocity dispersion $\sigma$ is \wba\ $=$ 2.563 $\sigma$. In each panel, the black cross indicates the spaxel where the peak of the total \oiii\ emission line flux is located.}
\label{fig:stelmap1}
\end{figure}

\begin{figure}
\epsscale{0.5}
\plotone{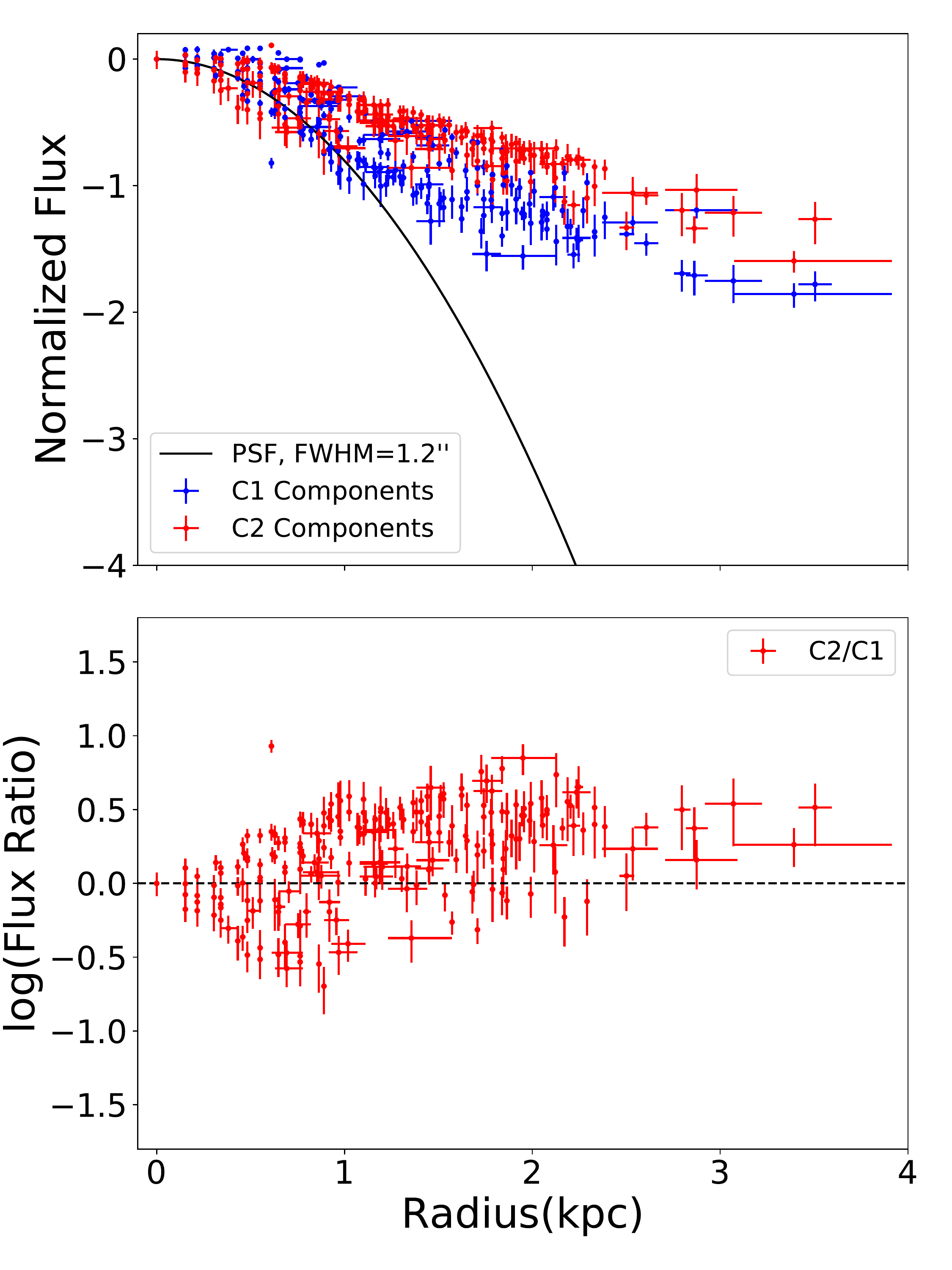}
\caption{Top panel: Radial profiles of the \oiii\ fluxes from the two velocity components of \te. For each component, the fluxes are normalized to those of the spaxel with peak emission line flux (i.e., the spaxel indicated by the black cross in Fig. \ref{fig:o3map1}). The PSF profile is derived from fits to the spectrophotometric standard stars of the IFS observations using single Gaussian profiles (see Section \ref{22} for more details). Lower panel: C2/C1 flux ratios on a logarithmic scale as a function of distance from the spaxel with peak emission line flux. In both panels, the data points beyond the maximal spatial extent of the outflow component (C2 in this target) are omitted. In addition, the error bars in radius (x-axis) are set either to zero for single spaxels or to reflect the radial coverage of the spatial bin.}
\label{fig:radial1}
\end{figure}

\section{Results: Individual Objects} \label{4}

The detailed results from our analysis are presented in this Appendix. For each object, we show the maps of the \oiii\ flux and kinematics, globally and for each velocity component, a map of the stellar kinematics, and the radial profiles of the line fluxes from individual velocity components. In addition, the line ratio maps and spatially resolved BPT and VO87 diagrams are also shown for \ta\ and \tb. In all cases, the systematic velocities of our targets are determined from the stellar velocities measured from the spectra integrated over the whole data cubes. 
In the few objects where the broader velocity component shows kinematic characteristics that are apparently similar to those of a rotating gas disk, we have also attempted to fit the velocity field with \textit{Kinemetry} \citep{Kinemetry}, a software based on a generalized harmonic expansion method of the two-dimensional velocity field.

\subsection{\te} \label{45}

Fig. \ref{fig:o3map1}--\ref{fig:radial1} present the KCWI maps of the \oiii\ flux and kinematics, the map of the stellar kinematics, and the \oiii\ flux radial profiles of the individual velocity components of target \te.

\subsubsection{Maps of the \oiii\ Flux and Kinematics} \label{451}

Two velocity components (C1 and C2) are sufficient to describe the \oiii\ line profiles in this galaxy. The spatial distribution of the \oiii\ flux is not symmetric with respect to the galaxy center (Fig.\ \ref{fig:o3map1}). More flux is present in the southern portion of the galaxy than in the north. This is especially true when considering the C2 component (discussed in more detail below).

The \oiii\ line profiles show a mild velocity gradient similar to that of the C1 component, and both of them appear to be systematically slightly blueshifted by $\sim$20 \kms\ with respect to the stellar velocities. Note, however, that the stellar velocities are only measured reliably in the inner kpc of this galaxy (Fig.\ \ref{fig:stelmap1}), so the amplitude and position angle of the stellar velocity gradient is uncertain. The line widths \wba\ of the \oiii\ line profiles are generally narrow except in the southwestern portion of the galaxy, where \wba\ reach $\sim$440 \kms.

The C1 component shows a mild velocity gradient with \vwu\ ranging from $\sim$$-$60 \kms\ to 0 \kms and a median of $\sim$$-$20 \kms. The C1 line widths are in general narrow (median \wba\ $\simeq$ 120 \kms), consistent with the idea that the C1 component is made of quiescent gas rotating in the galaxy.

The flux asymmetry is more apparent in the C2 component than in the C1 component. The C2 component is significantly blueshifted in the south portion of the galaxy, where \vwu\ reach values of $\sim$$-$240 \kms, well in excess of the stellar velocities measured on smaller scale. A clear gradient in \vwu\ is seen along the N -- S direction (PA$\simeq$10\textdegree), but the most redshifted velocities are $\sim$$+$50 \kms. The line widths of the C2 components are generally large, reaching a maximum value of $\sim$650 \kms. The kinematics of the C2 component may be interpreted as a tilted, biconical outflow, where the near (S) side of the outflow is blueshifted and the far (N) side is redshifted. The redshifted velocities are significantly smaller (in absolute terms) than the blueshifted ones, perhaps an indication that the far side of the outflow is largely blocked by the galaxy. However, without reliable stellar velocities on large scale, it is hard to exclude the possibility that the north portion of the C2 component consists of turbulent, rotating gas within the galaxy.  

To further examine the origin of the \vwu\ gradient seen in the C2 component, we have tested fitting the separate \vwu\ maps of the C2 and C1 components with \textit{Kinemetry} \citep{Kinemetry}. This software fits the two-dimensional map of the line-of-sight velocity distribution of a galaxy by determining the best-fit ellipses along which the profiles of the moments can be extracted and analyzed by means of harmonic expansion. As a product of the fit, the best-fit circular velocity field can be obtained. In practice, we carried out the fits in two steps: (i) we fitted the \vwu\ map with the default setting in \textit{Kinemetry} where the PA and flattening of each ellipse were allowed to vary freely; (2) a second and final fit was applied where the PA and flattening were fixed to the median values measured from step (i). Due to the asymmetry in the flux distribution of the C2 components, we applied the fits described above only to the region within r $\lesssim$ 1.2 kpc where relatively complete ellipses required for the fits can be drawn from the data, and extrapolate the best-fit circular velocity field to the south where the blueshifted emission are mostly seen. For the C1 component, the residual velocities (defined as the difference between the observed \vwu\ and the circular velocities from the best-fit) are consistent with random noise as expected, suggesting that the kinematics of the C1 component can be described as a rotating disk. For the C2 component, on the contrary, we find that there are significant negative residual velocities in the southern portion of the galaxy. This is consistent with our earlier statement that the blueshifted emission on the south side is likely originating from the near side of a biconical outflow.

\subsubsection{Flux Radial Profiles} \label{452}

The \oiii\ flux radial profiles shown in Fig.\ \ref{fig:radial1} confirm that the individual velocity components in \te\ are spatially resolved in the KCWI data. There is a weak trend for the C2/C1 flux ratios to increase radially, further indicating that these components have slightly different flux distributions as stated in Section \ref{451}.

\begin{figure}[!h] 
\plotone{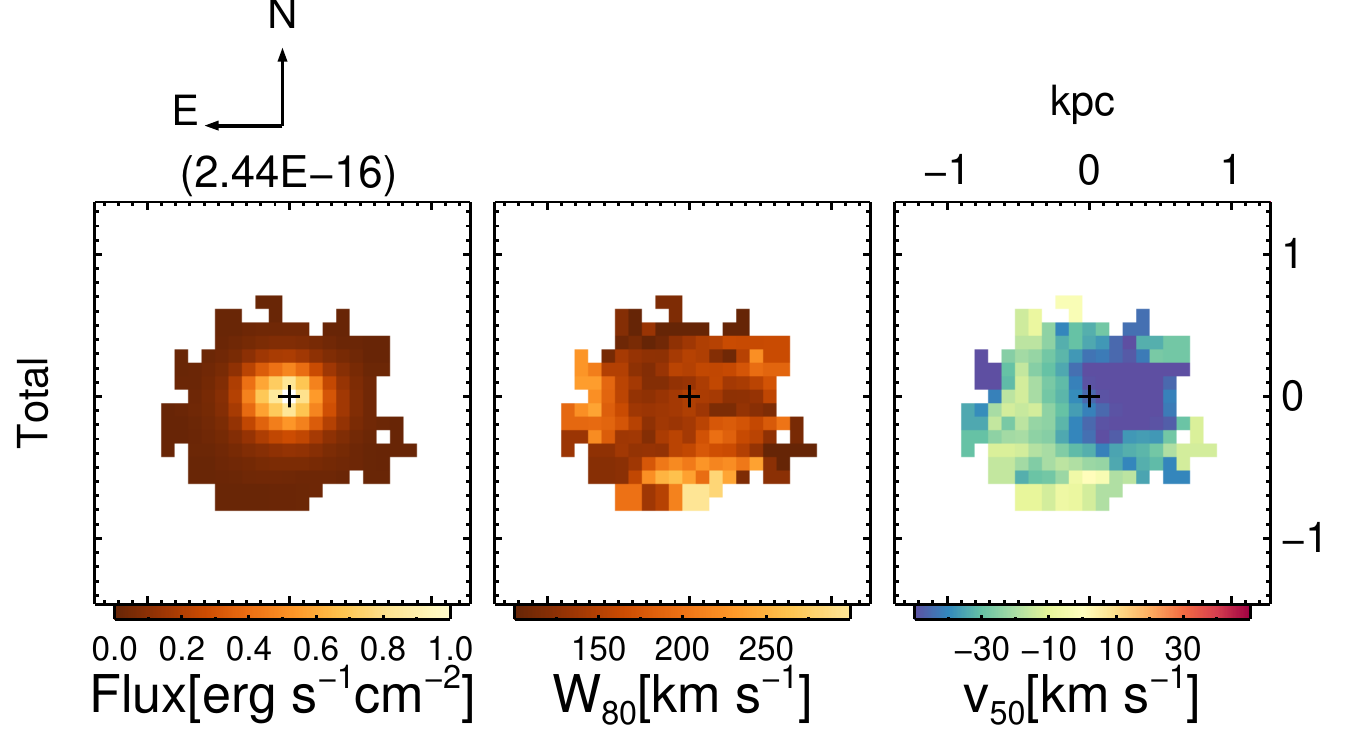}
\caption{Same as Fig.\ \ref{fig:o3map1} but for \tg, where a single Gaussian component is sufficient to fit the \oiii\ line profiles.}
\label{fig:o3map7}
\end{figure}

\begin{figure}[!htb]   
\epsscale{0.6}
\plotone{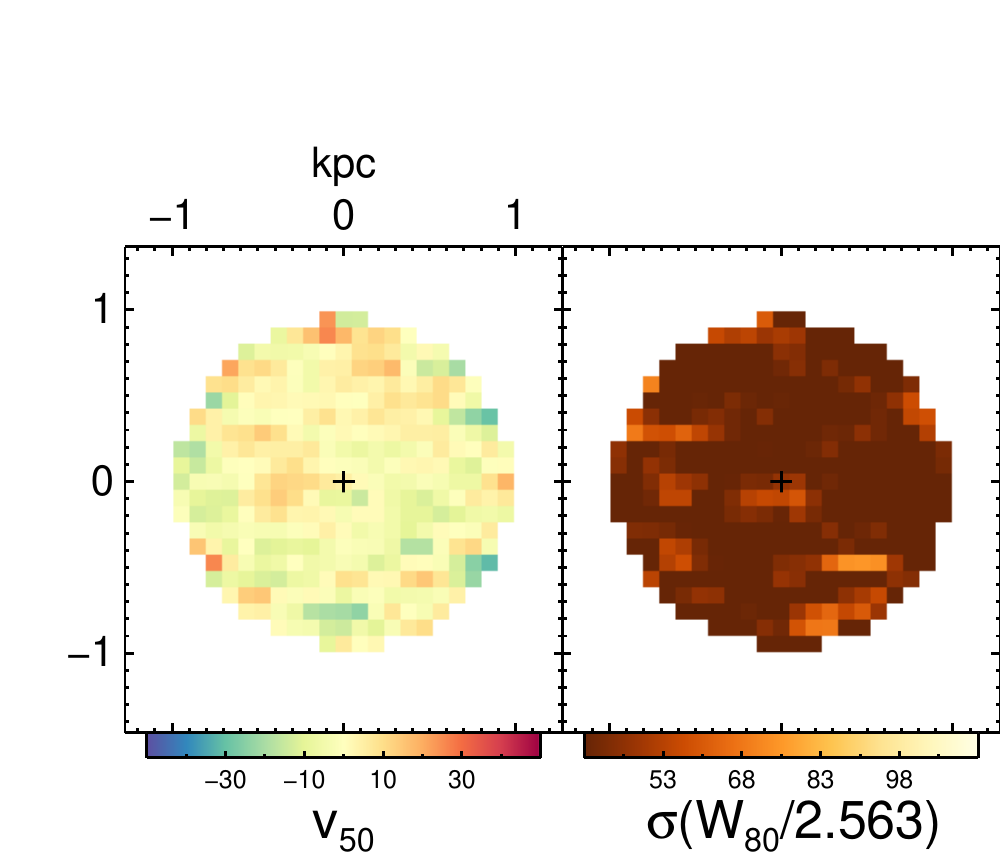}
\caption{Same as Fig.\ \ref{fig:stelmap1} but for \tg.}
\label{fig:stelmap7}
\end{figure}

\begin{figure}[!htb]   
\epsscale{0.5}
\plotone{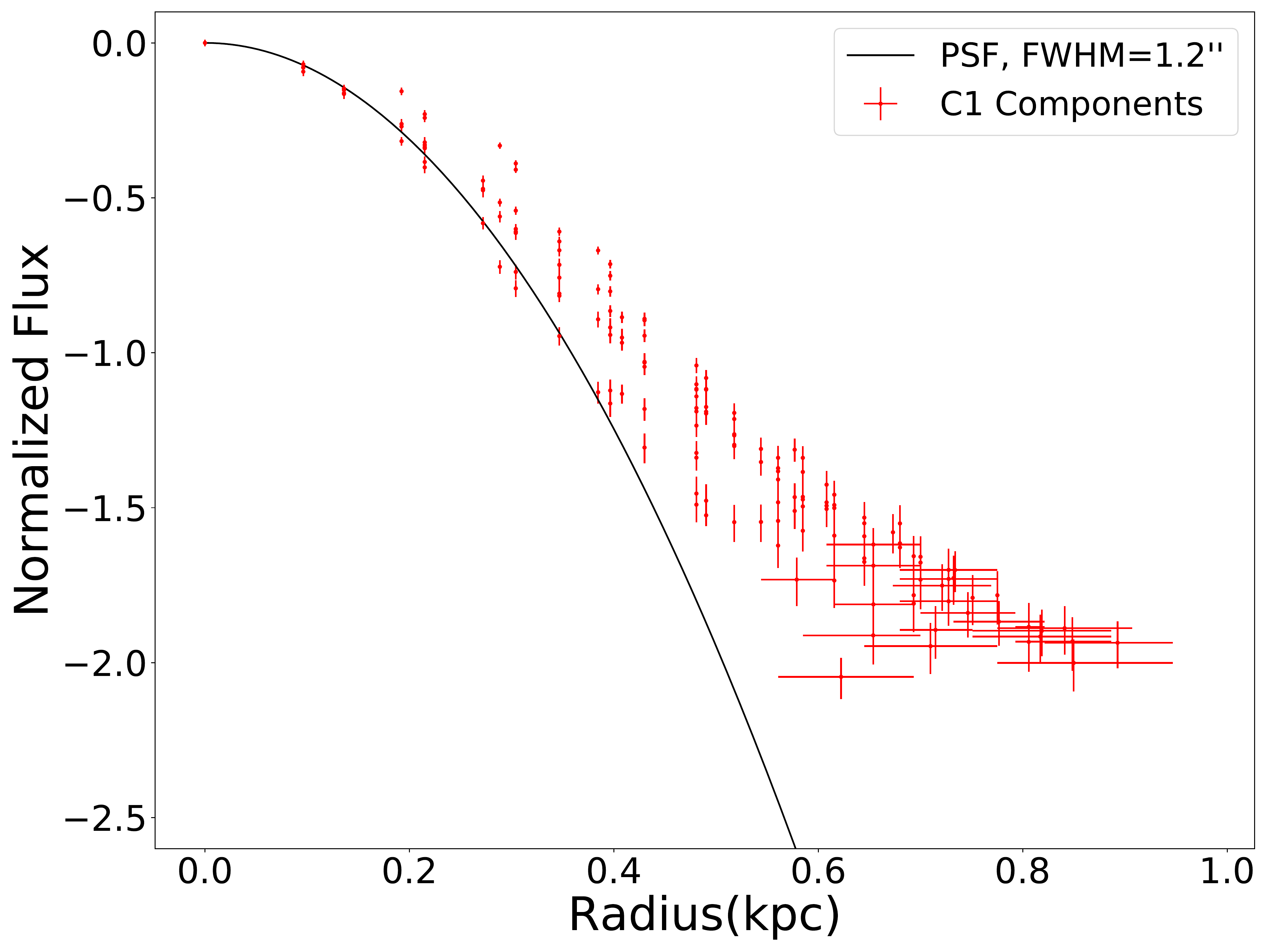}
\caption{Same as the top panel of Fig.\ \ref{fig:radial1} but for \tg. }
\label{fig:radial7}
\end{figure}

\subsection{\tg} \label{47}

The results of our analysis of the KCWI data of \tg\ are presented in  Fig.\ \ref{fig:o3map7}--\ref{fig:radial7}.

\subsubsection{Maps of the \oiii\ Flux and Kinematics} \label{471}
 
A single Gaussian component is sufficient to fit the \oiii\ line profiles in this object. The values of \vwu\ (Fig.\ \ref{fig:o3map7}) are everywhere blueshifted with respect to those of the stellar component (Fig.\ \ref{fig:stelmap7}) and show a gradient from $\sim$$-$60 to $\sim$$-$20 \kms\ along PA $\simeq$ $-$80\textdegree, which is not seen in the stellar velocity field. The line widths of the emission lines are also on average larger than the velocity dispersions of the stellar components (median \wba\ $\sim$140 and 90 \kms, respectively).

These results show that the kinematics of the ionized gas cannot be described by pure rotation. The blueshifted ionized gas likely takes part in a bulk outflow. We speculate that the \vwu\ gradient seen in \oiii\ may be caused by geometrical effects and/or internal velocity gradient in the outflow itself, given that the stellar components show no obvious rotation. 

\subsubsection{Flux Radial Profiles} \label{472}

The \oiii\ flux radial profile (Fig. \ref{fig:radial7}) is clearly more extended than the PSF, consistent with the presence of a clear, spatially resolved velocity gradient in the ionized gas.

\subsection{\tx} \label{48}

\begin{figure}[!htb] 
\plotone{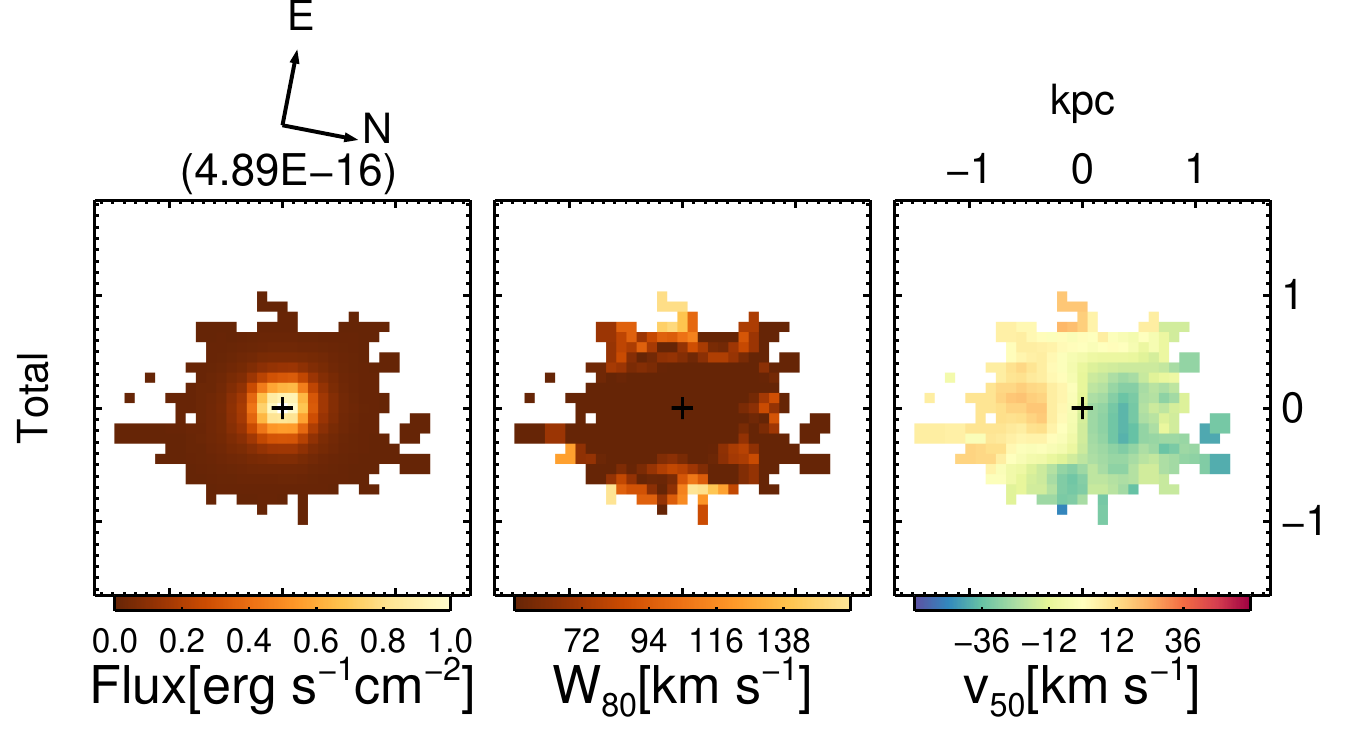}
\caption{Same as Fig.\ \ref{fig:o3map7} but for \tx, where a single Gaussian component is sufficient to fit the \oiii\ line profiles. This is the only object in our sample without a clear sign of outflow.}
\label{fig:o3map8}
\end{figure}

\begin{figure}[!htb]   
\epsscale{0.6}
\plotone{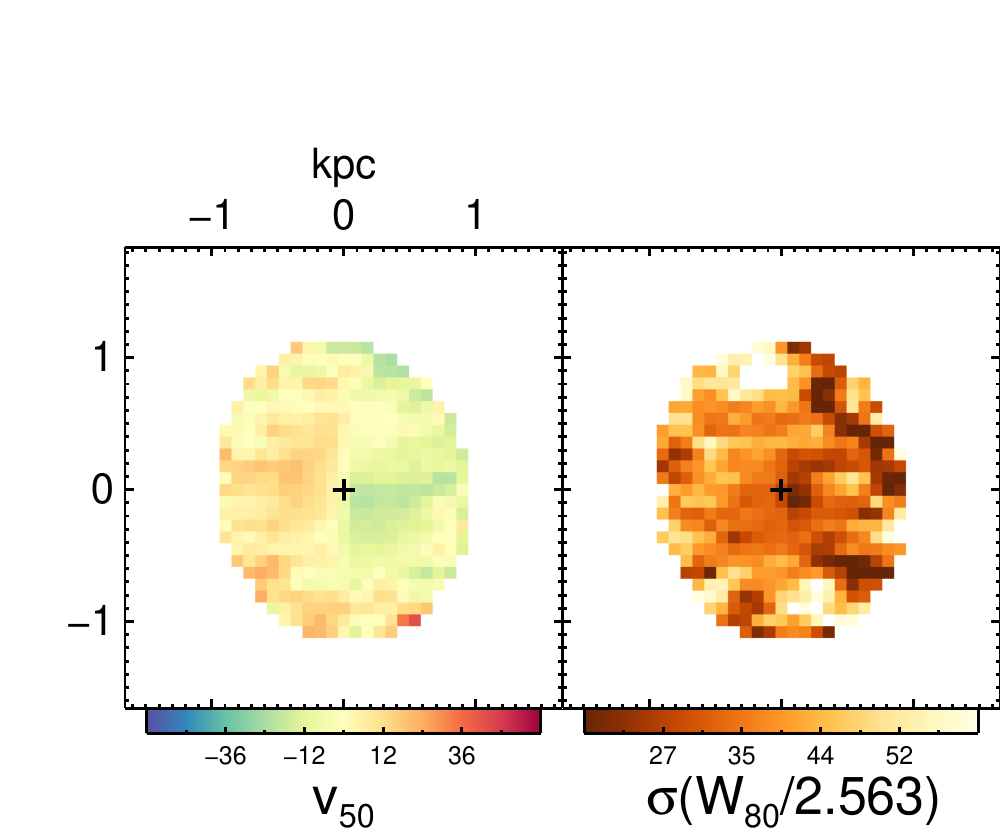}
\caption{Same as Fig.\ \ref{fig:stelmap7} but for \tx. }
\label{fig:stelmap8}
\end{figure}

\begin{figure}[!htb]   
\epsscale{0.5}
\plotone{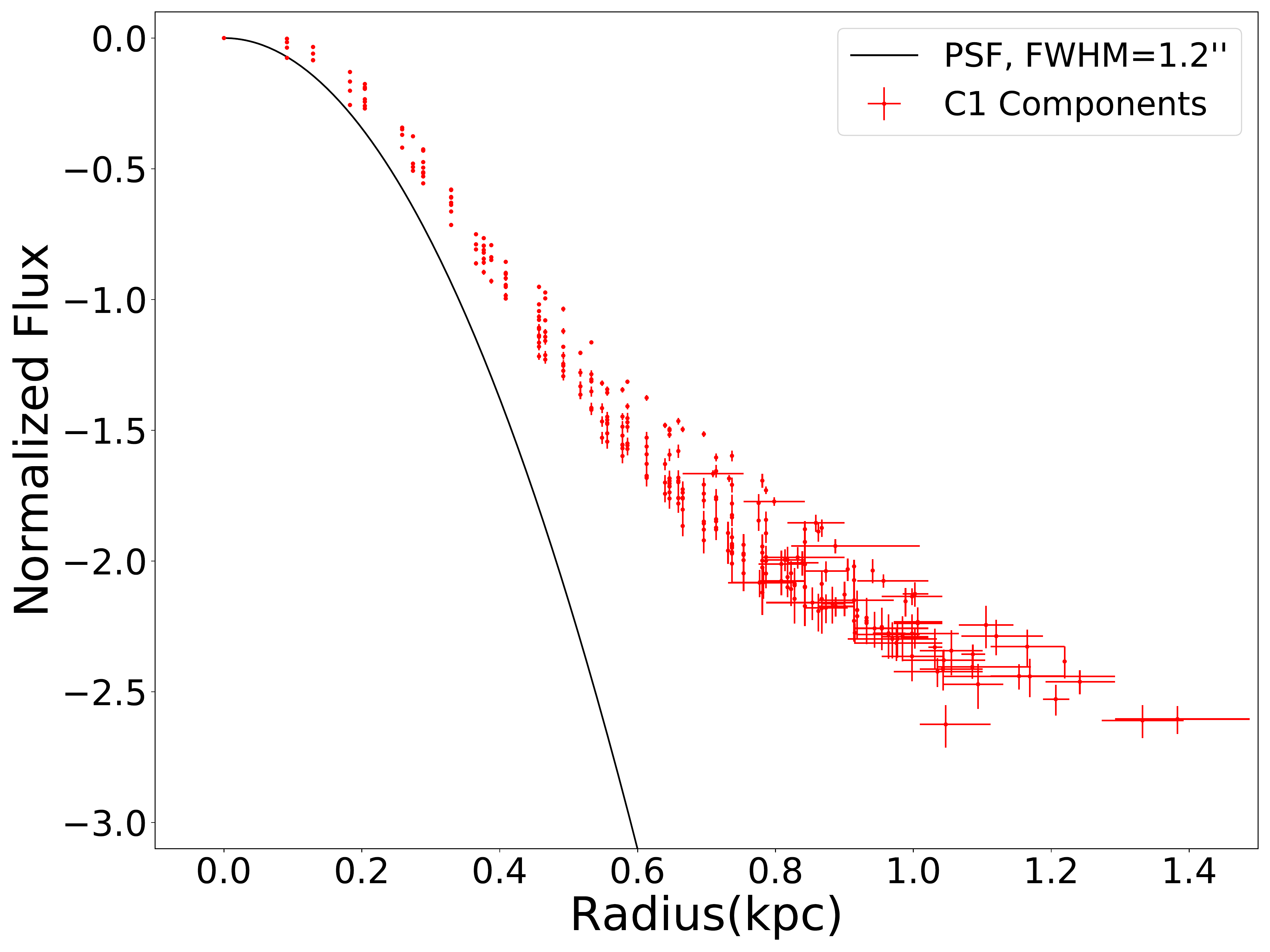}
\caption{Same as Fig.\ \ref{fig:radial7} but for \tx.}
\label{fig:radial8}
\end{figure}

Fig.\ \ref{fig:o3map8}--\ref{fig:radial8} display the results of our analysis of the KCWI data on \tx.

\subsubsection{Maps of the \oiii\ Flux and Kinematics} \label{481}

A single Gaussian component is sufficient to fit the \oiii\ line profiles. The map of \oiii\ \vwu\ (Fig.\ \ref{fig:o3map8}) shows a clear gradient ($-$30 \kms\ to $+$20 \kms) similar to that of the stellar \vwu\ (Fig.\ \ref{fig:stelmap8}). The line widths of the emission lines are smaller than the velocity dispersions of the stellar component. These results suggest that the ionized gas is simply rotating within the galaxy in the same direction as the stars. No clear evidence of outflow is seen in this object.

\subsubsection{Flux Radial Profiles} \label{482}

The \oiii\ flux radial profile (Fig. \ref{fig:radial8}) is clearly more extended than the PSF, which is consistent with the spatially resolved velocity gradients seen in both the ionized gas and underlying stellar population.

\subsection{\tb} \label{42}

\begin{figure*}[!htb]   
\epsscale{1.1}
\begin{minipage}[t]{0.52\textwidth}
\includegraphics[width=\textwidth]{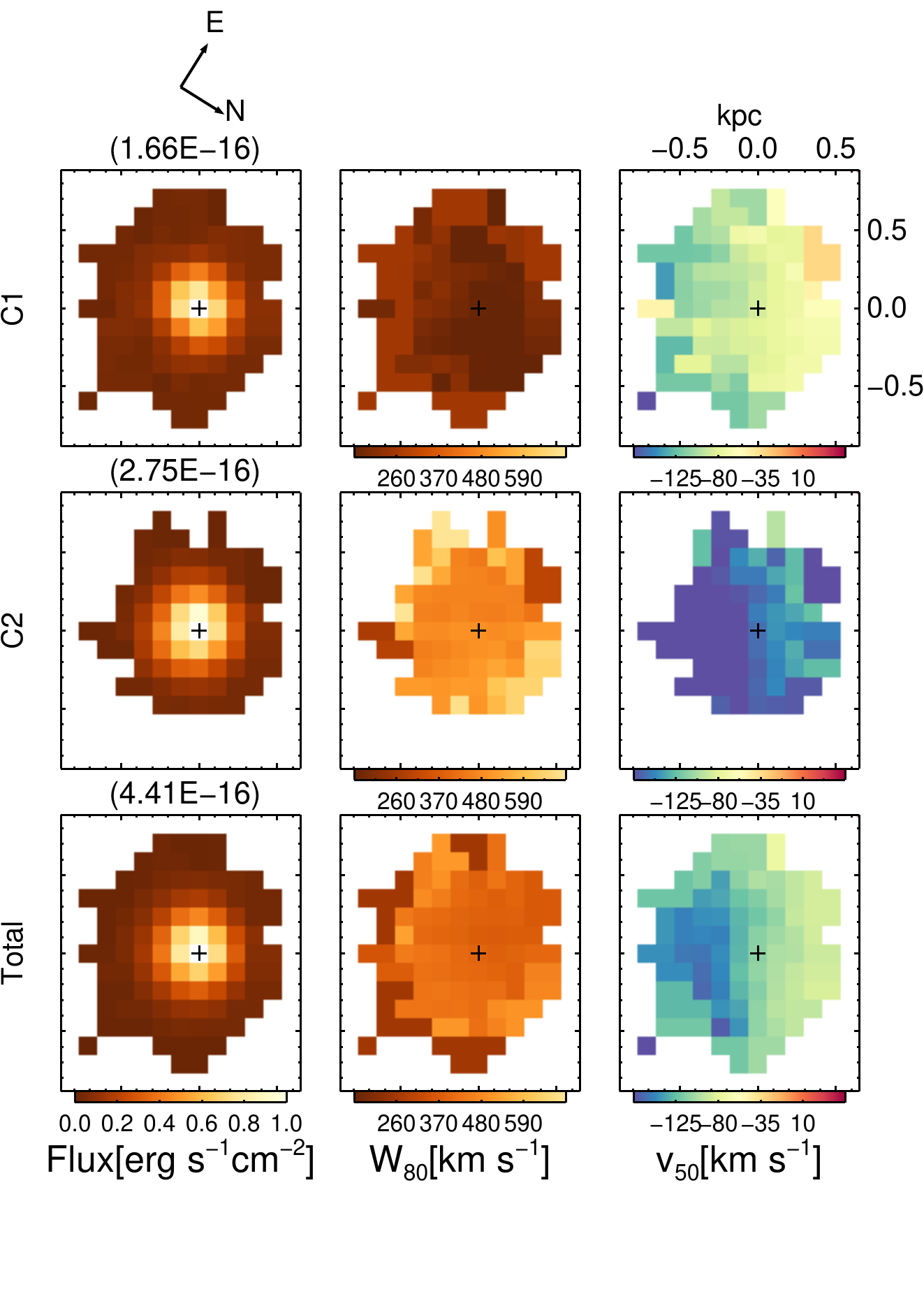}
\end{minipage}
\begin{minipage}[t]{0.5\textwidth}
\includegraphics[width=\textwidth]{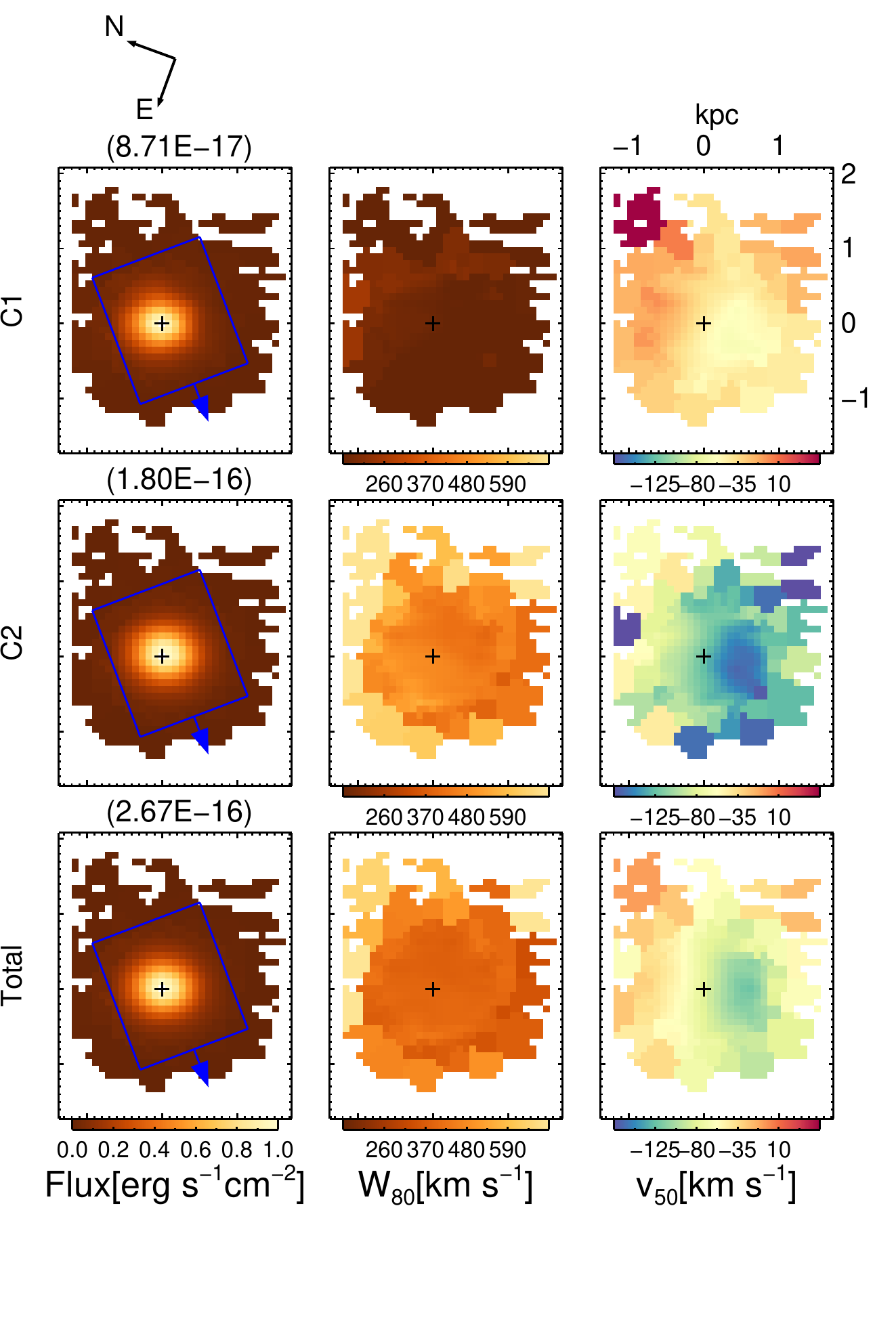}
\end{minipage}
\caption{Same as Fig.\ \ref{fig:o3map1} but for the GMOS data (left) and KCWI data (right) of \tb, where two velocity components C1 and C2 are needed to adequately fit the emission-line profiles. The orientation of the maps are noted at the top of each panel and are different from each other. In the left column of the right panel (KCWI data), the GMOS footprint is overplotted as a blue rectangle with the 0 o'clock direction noted by the blue arrow. Note that the color bars in the right column of each panel (\vwu) are centered on a negative velocity for a better visualization of the velocity gradients in both the C1 and C2 components. }
\label{fig:o3map2}
\end{figure*}

\begin{figure}
\epsscale{0.5}
\plotone{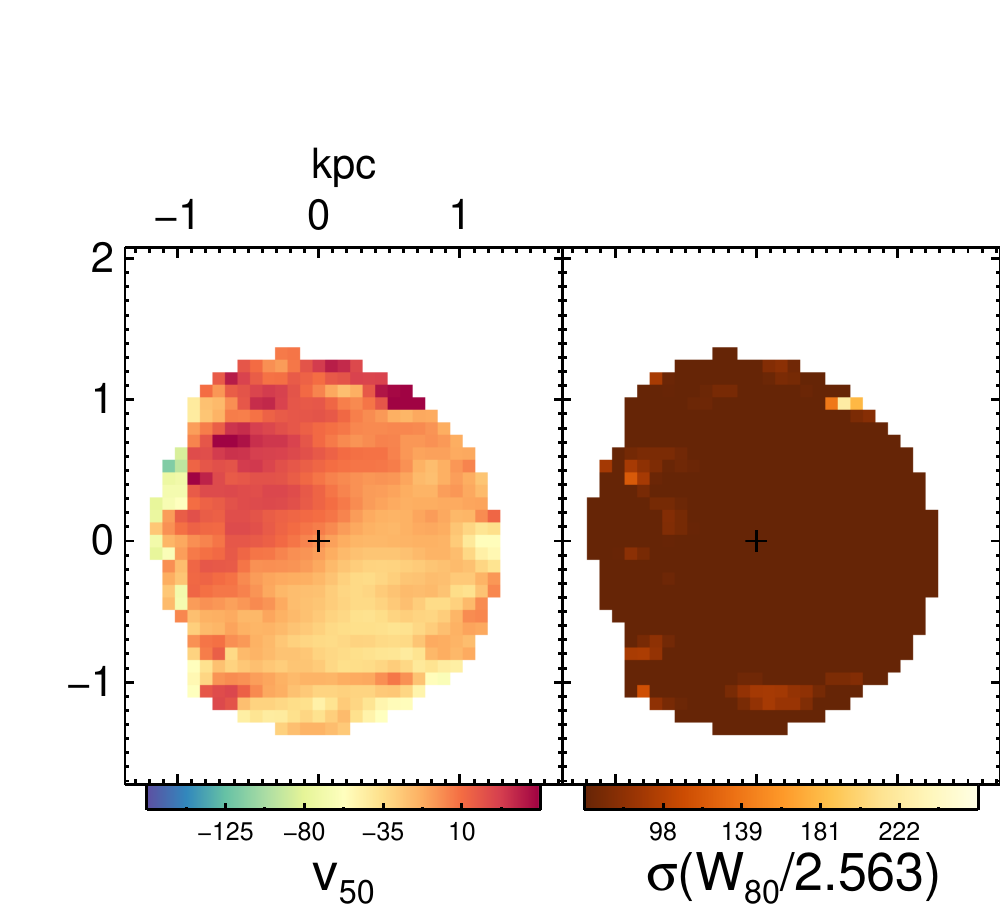}
\caption{Same as Fig.\ \ref{fig:stelmap1} but for \tb.}
\label{fig:stelmap2}
\end{figure}

\begin{figure*}[!htb]   
\plottwo{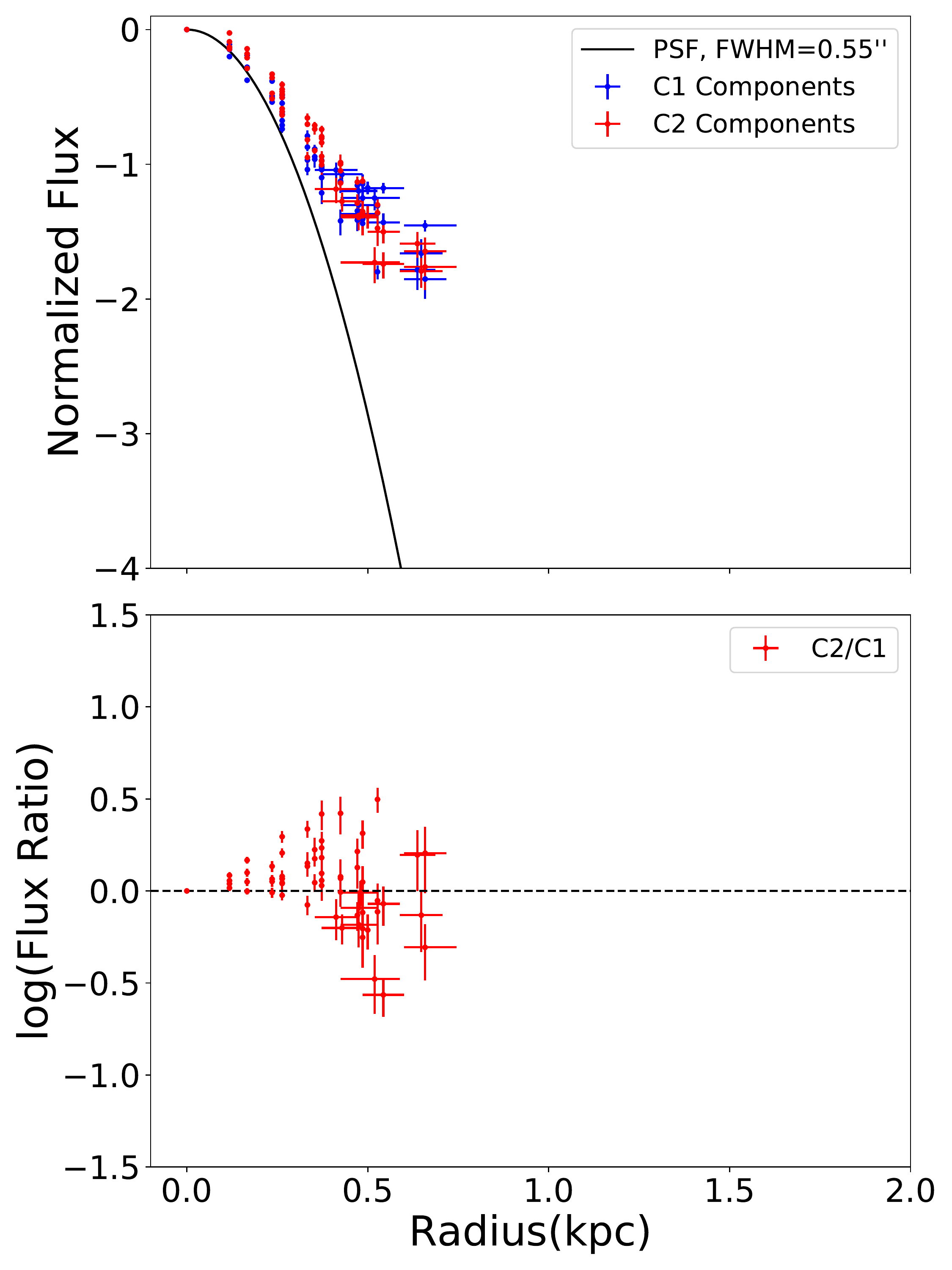}{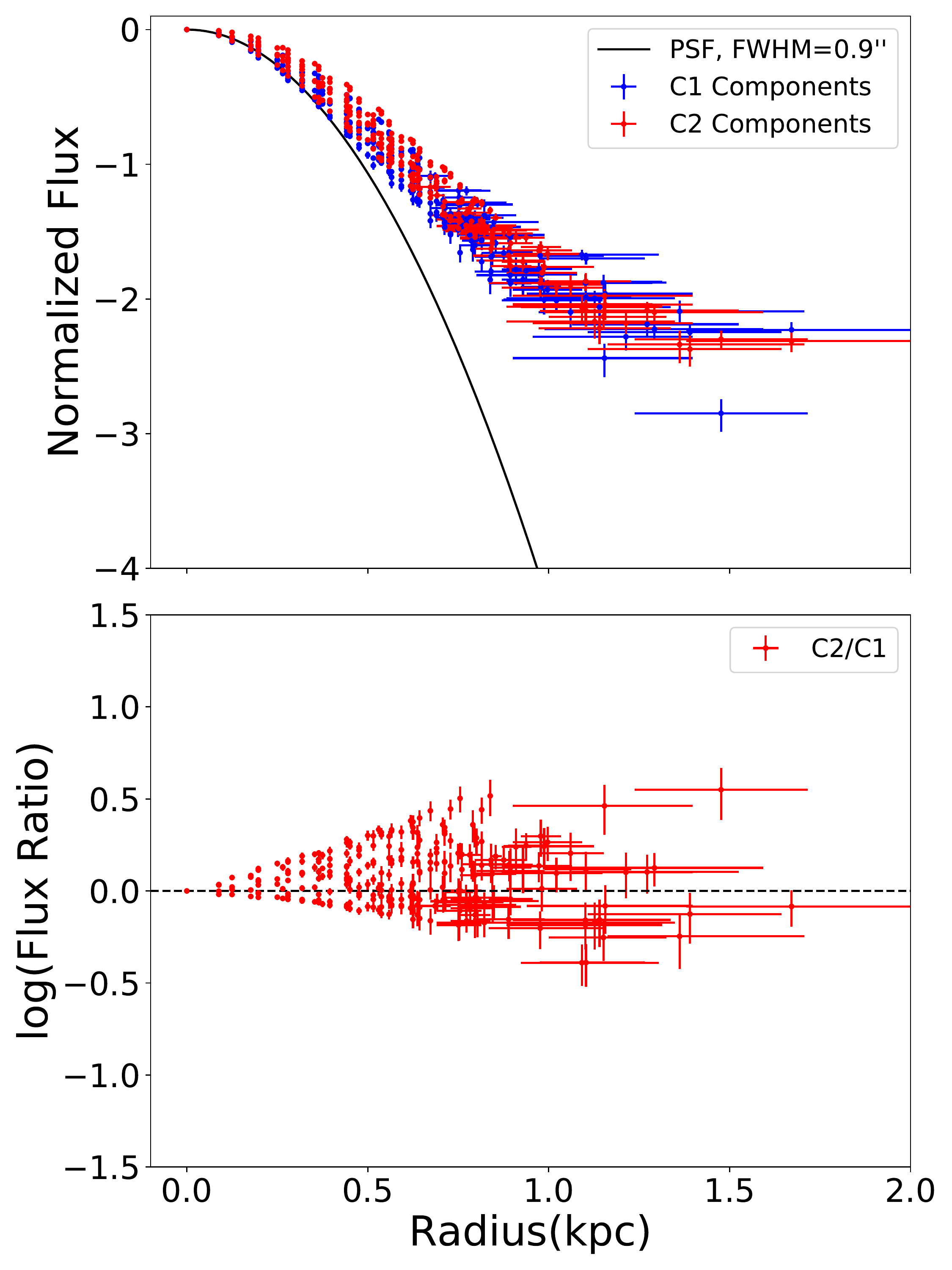}
\caption{Same as Fig.\ \ref{fig:radial1} but for the GMOS data (left) and KCWI data (right) of \tb.}
\label{fig:radial2}
\end{figure*}

\begin{figure}[!htb]   
\epsscale{1.1}
\plotone{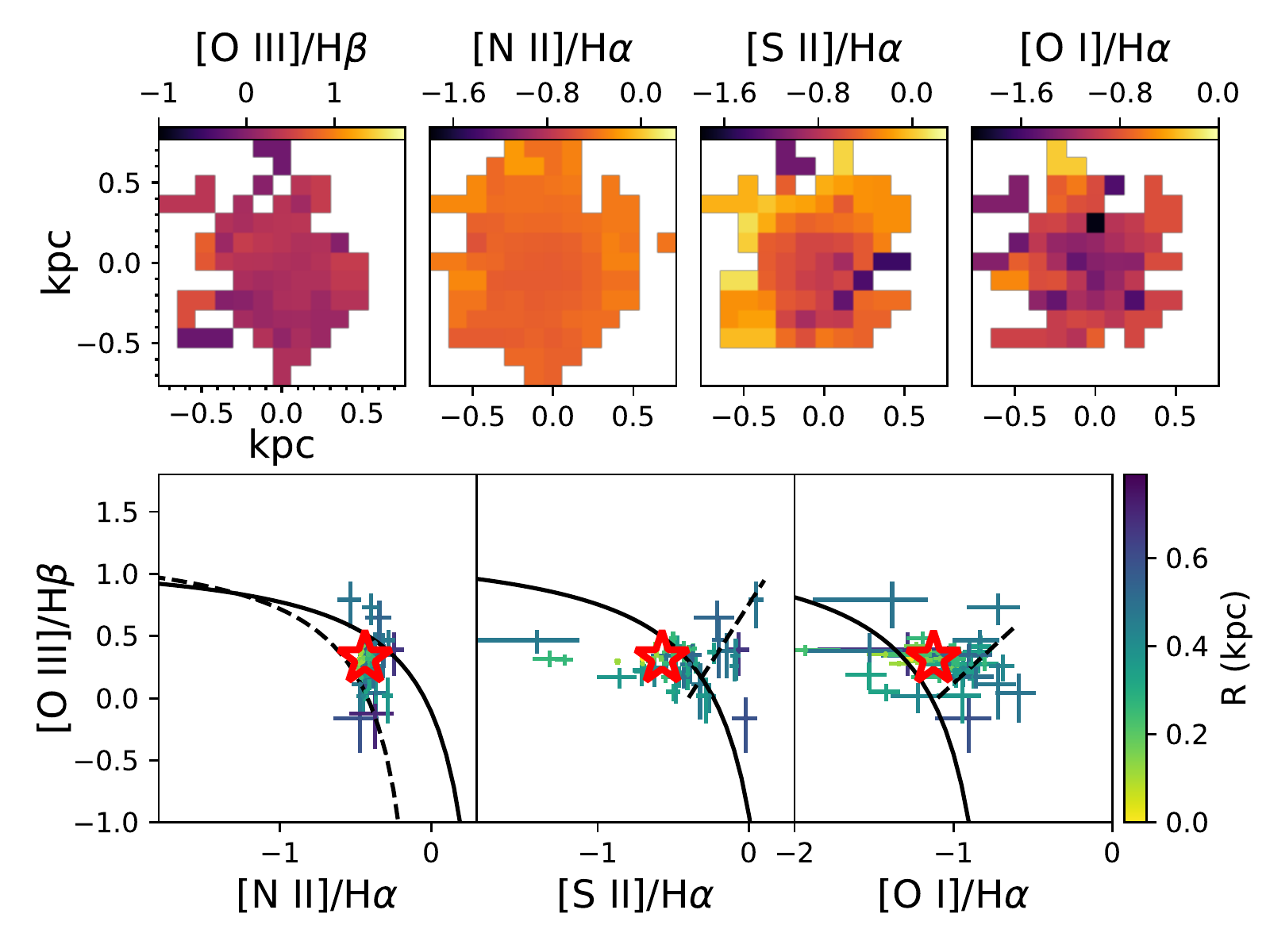}
\caption{Emission line diagnosis for the C1 component of \tb. Top row: Line ratio maps of \oiiihb, \niiha, \siiha\ and \oiha, from left to right. Bottom row: Standard BPT and VO87 diagrams. The data points are color-coded according to their projected distance (in kpc) to the spaxel with peak emission line flux. The large red open star in each panel indicates the line ratios derived from the spatially-integrated spectrum. In all panels, the solid line is the theoretical line separating AGN (above right) and star-forming galaxies (below left) from \citet{Kewley2001}. In the left panel, the dashed line is the empirical line from \citet{Kauffmann2003} showing the same separation. Objects between the dotted and solid lines are classified as composites. In the middle and right panels, the diagonal dashed line is the theoretical line separating Seyferts (above left) and LINERs (below right) from \citet{Kewley2006}.} 
\label{fig:J0842bpt1}
\end{figure}

\begin{figure}[!htb]   
\epsscale{1.1}
\plotone{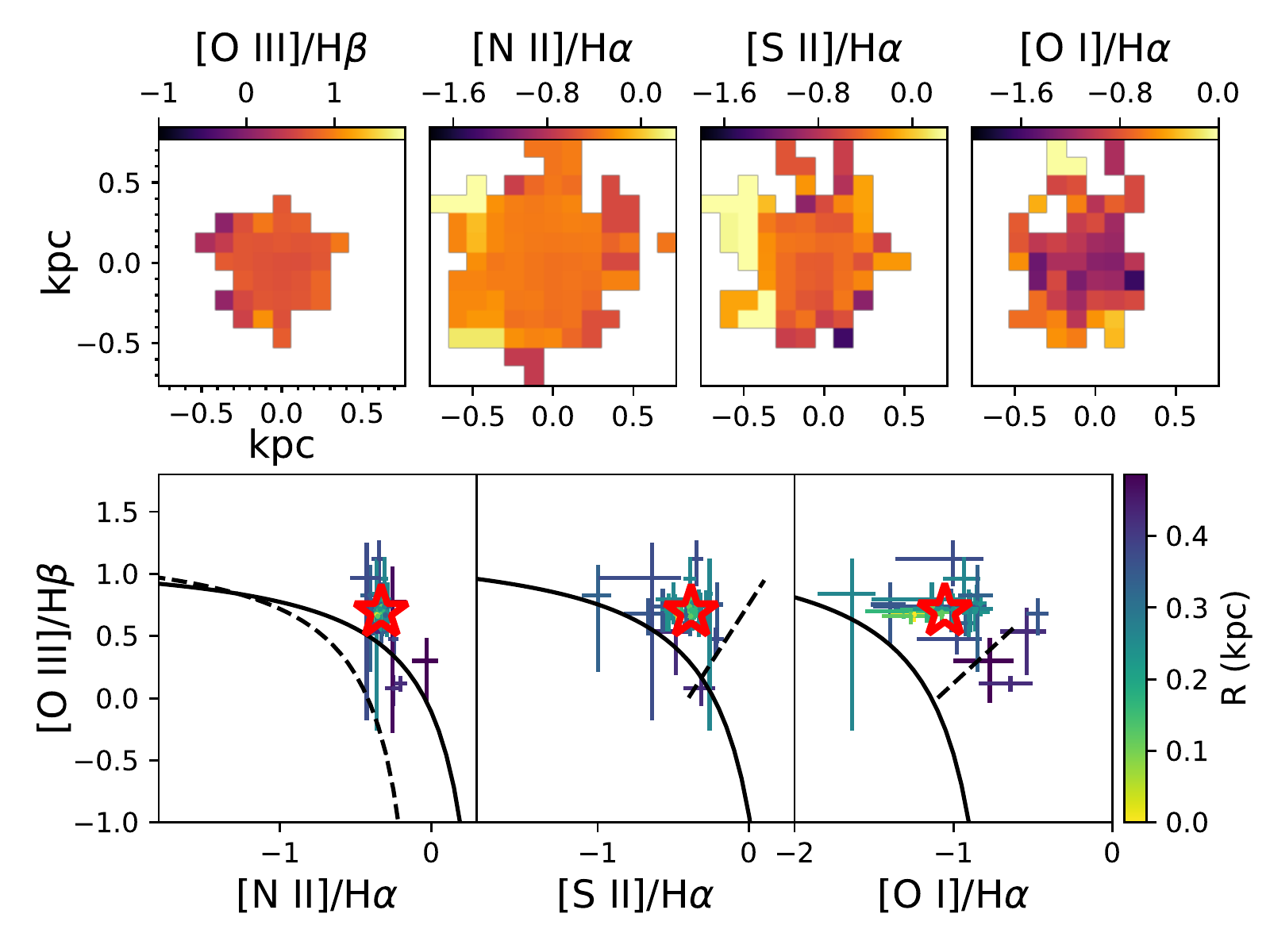}
\caption{Same as Fig.\ \ref{fig:J0842bpt1} but for the component C2.}
\label{fig:J0842bpt2}
\end{figure}

\tb\ was observed with both GMOS and KCWI. The results derived from these two independent data sets agree well with each other. While the GMOS data trace the structures on small spatial scale better than the KCWI data, the KCWI data allow us to probe the fainter emission on the outskirts of the galaxy host.

Fig. \ref{fig:o3map2}--\ref{fig:radial2} present the maps of the \oiii\ flux and kinematics, the map of the stellar kinematics, and the \oiii\ flux radial profiles of the individual velocity components. The line ratio maps and spatially resolved BPT and VO87 diagrams are shown in Fig. \ref{fig:J0842bpt1}--\ref{fig:J0842bpt2}.

\subsubsection{Maps of the \oiii\ Flux and Kinematics} \label{421}

Two Gaussian components (C1 and C2) are enough to describe the \oiii\ emission line profiles in this system (as already shown in Fig.\ \ref{fig:J0842o3fit}). The projected distribution of the \oiii\ emission, both velocity-integrated and in the individual velocity components, is largely symmetric with respect to the galaxy center.

The emission line profiles are in general blueshifted with respect to the systemic velocity (median \vwu\ $\simeq-$110 \kms\ in the GMOS data and $\simeq-$70 \kms in the KCWI data) and show a clear velocity gradient (see the bottom rows in Fig. \ref{fig:o3map2}). The line widths are also much broader than the stellar velocity dispersions (see Fig. \ref{fig:stelmap2}). This indicates that the kinematics of the ionized gas are dominated by non-rotational motion.

The C1 component is on average blueshifted by 80 \kms\ in the GMOS data and 30 \kms\ in the KCWI data, but shows a clear velocity gradient (with \vwu\ ranging from $\sim$$-$110 \kms\ to $\sim$$-$20 \kms\ in the GMOS data, and from $\sim$$-$60 \kms\ to $\sim$$+$10 \kms in the KCWI data) with an orientation of the gradient (PA $\simeq$ 220\textdegree) that is similar to that of the stellar velocity field. The line widths of the C1 component are in general narrow (median \wba\ $\simeq$ 200 \kms), similar to the stellar velocity dispersion. The C1 component likely represents a mixture of both rotating and outflowing gas.

The C2 component is in general significantly blueshifted with respect to the systemic velocity (\vwu\ down to $\sim$$-$220 \kms in the GMOS data, and $\sim$$-$160 \kms\ in the KCWI data) and much broader (\wba\ up to $\sim$650 \kms in the GMOS data and $\sim$750 \kms in the KCWI data) than the C1 component. The C2 component is thus most likely associated with outflowing gas. A clear gradient is seen in the \vwu\ of the C2 component, suggesting that the outflowing gas may have an intrinsic velocity structure and/or an asymmetrical geometry with respect to the line-of-sight. As the orientation of the velocity gradient of the C2 component is similar to that of the C1 component, an alternative explanation for this gradient is that the outflowing gas may have inherited a portion of the angular momentum from the galaxy.

\subsubsection{Flux Radial Profiles} \label{422}

The fluxes from both velocity components are clearly more extended than the PSF (top panels in Fig.\ \ref{fig:radial2}), consistent with the resolved velocity structures seen in the kinematics maps (Fig.\ \ref{fig:o3map2}). There is a slight trend for the C2/C1 flux ratio to increase radially outward in the KCWI data beyond radii of $\sim$0.5 kpc (bottom right panel in Fig.\ \ref{fig:radial2}); this trend is not detected in the GMOS data.

\subsubsection{Ionization Diagnosis} \label{423}
Here we examine the ionization properties of individual velocity components based on the GMOS data with emphasis on the line ratios in the BPT and VO87 diagrams. The KCWI data are not discussed in this context because they do not cover H$\alpha$ and the other important line diagnostics in the red.

The \oiiihb\ ratios of both the C1 and C2 velocity components are roughly constant across the map. The other three line ratios, \niiha, \siiha, and \oiha, are also roughly constant or show only mild radial trends.

The spatially-integrated line ratios of the C2 component fall in the AGN region in all three BPT and VO87 diagrams, while for the C1 component, they lie in the AGN region only in the \oiha\ diagram. Instead, the spatially-integrated line ratios of C1 components are in the composite region of the \niiha\ diagram and in the star-forming region of the \siiha\ diagram. For individual spaxels, the line ratios of the C2 component within r $\simeq$ 0.5 kpc are AGN-like in all three diagrams, while those of the C1 component suggest a significant contribution from star-forming activity, as the majority or a significant fraction of the spaxels are located in the composite regions of the \niiha\ diagram, and in the star-forming region of the \siiha\ diagram.

\subsection{\ta} \label{41}

\begin{figure*}[!htb]   
\epsscale{1.1}
\begin{minipage}[t]{0.5\textwidth}
\includegraphics[width=\textwidth]{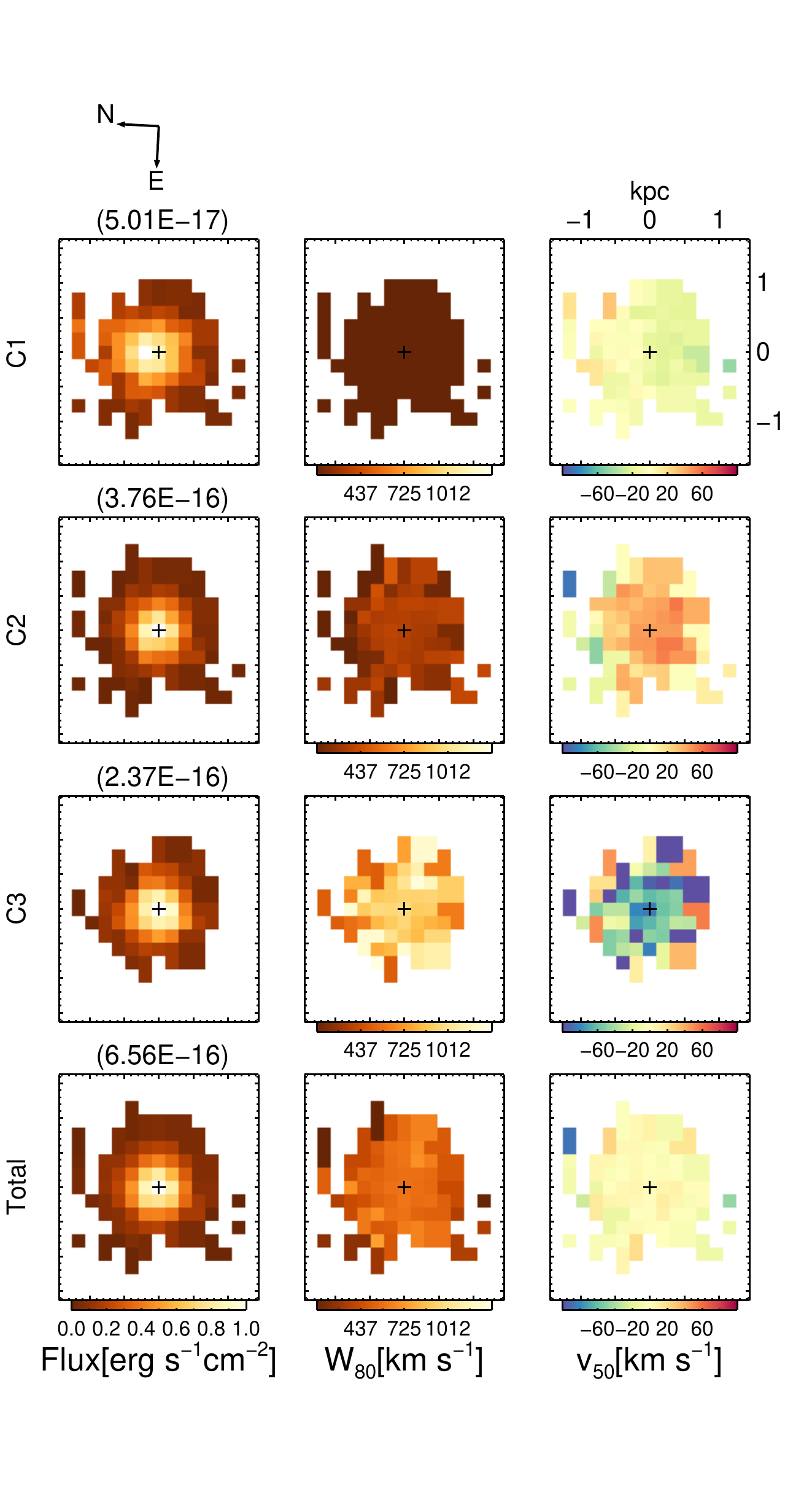}
\end{minipage}
 \begin{minipage}[t]{0.53\textwidth}
\includegraphics[width=\textwidth]{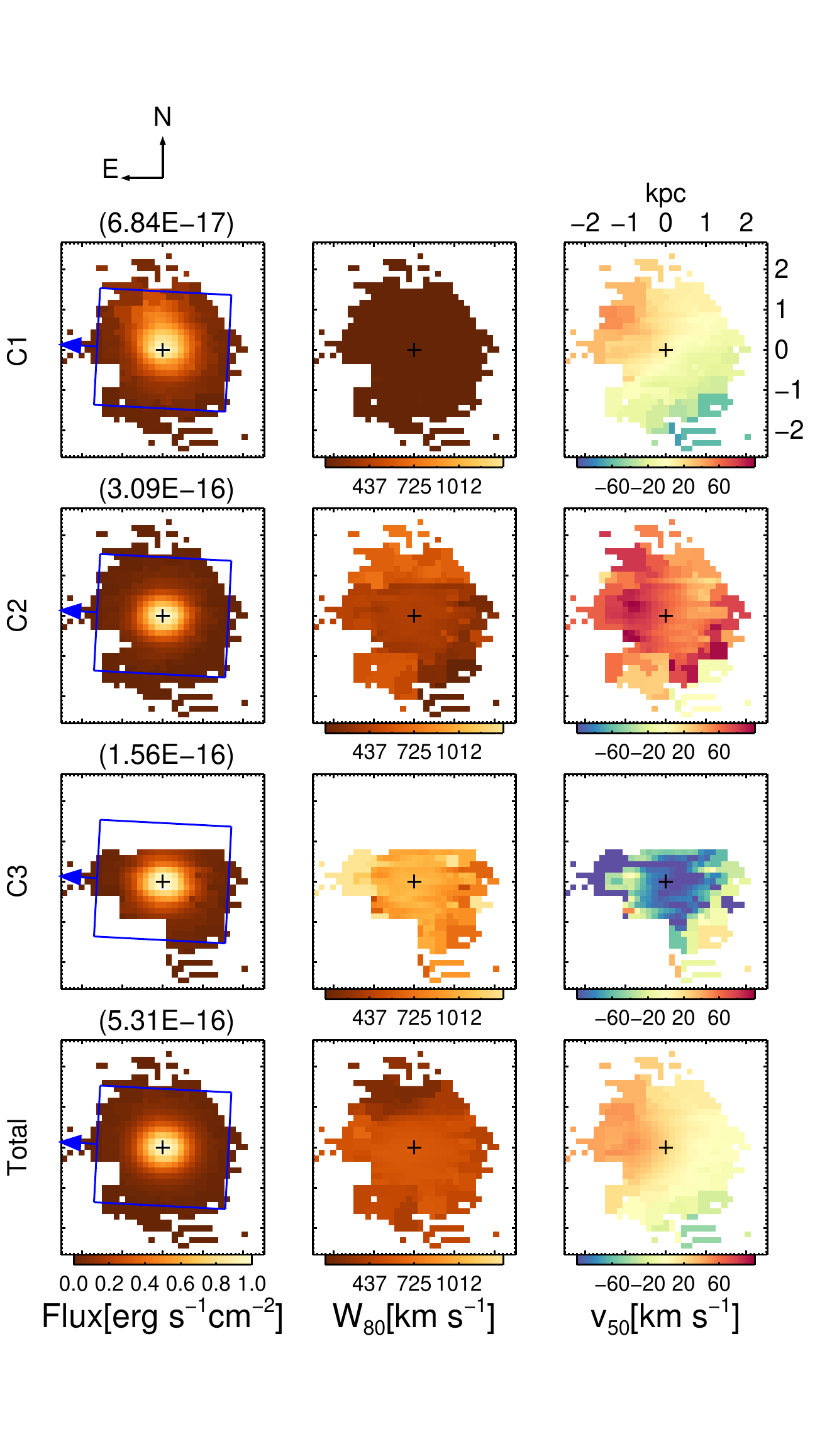}
\end{minipage}
\caption{Same as Fig.\ \ref{fig:o3map2} but for the GMOS data (left) and KCWI data (right) of \ta, where three velocity components C1, C2, and C3 are needed to adequately fit the emission-line profiles.}
\label{fig:o3map5}
\end{figure*}

\begin{figure}[!htb]   
\epsscale{0.5}
\plotone{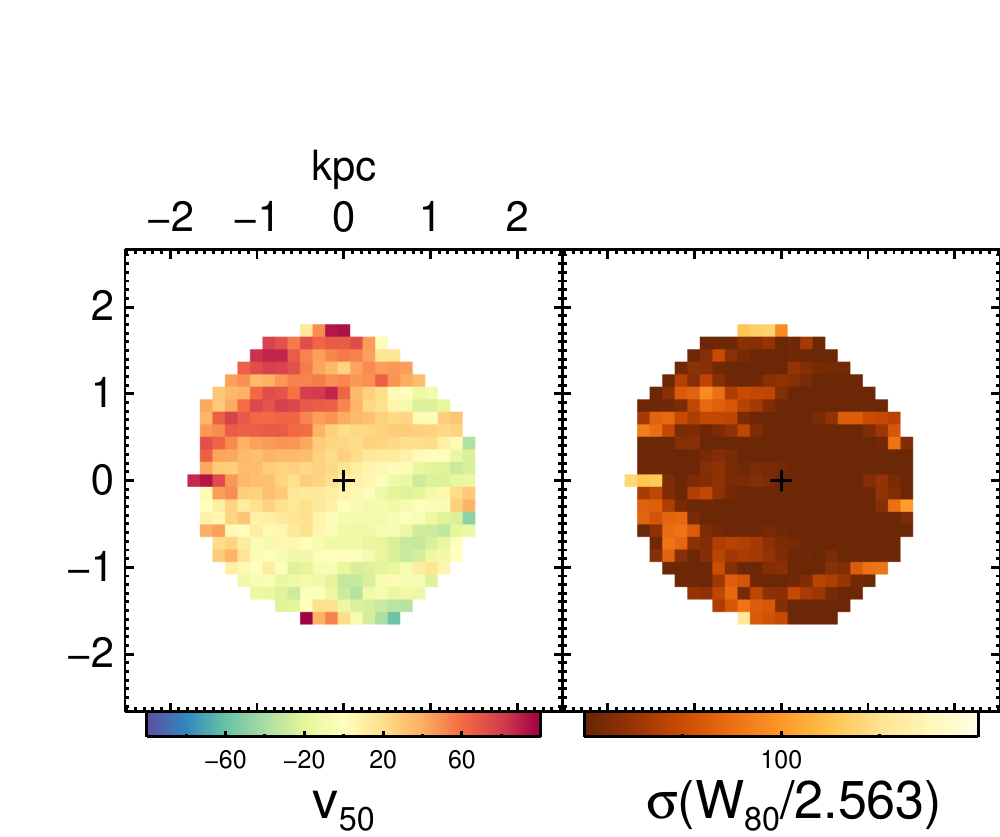}
\caption{Same as Fig. \ref{fig:stelmap1} but for \ta.}
\label{fig:stelmap5}
\end{figure}

\begin{figure*}[!htb]   
\plottwo{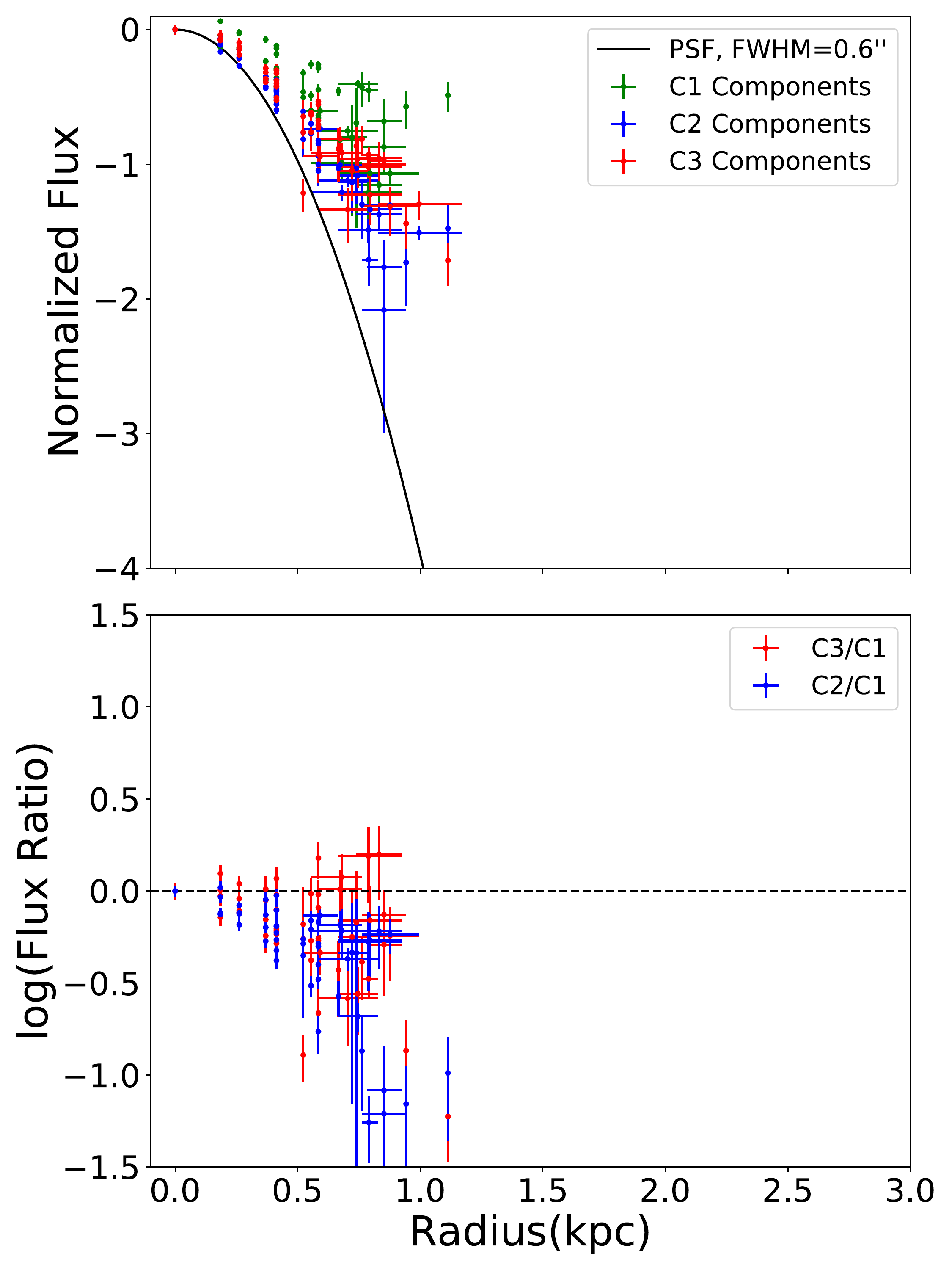}{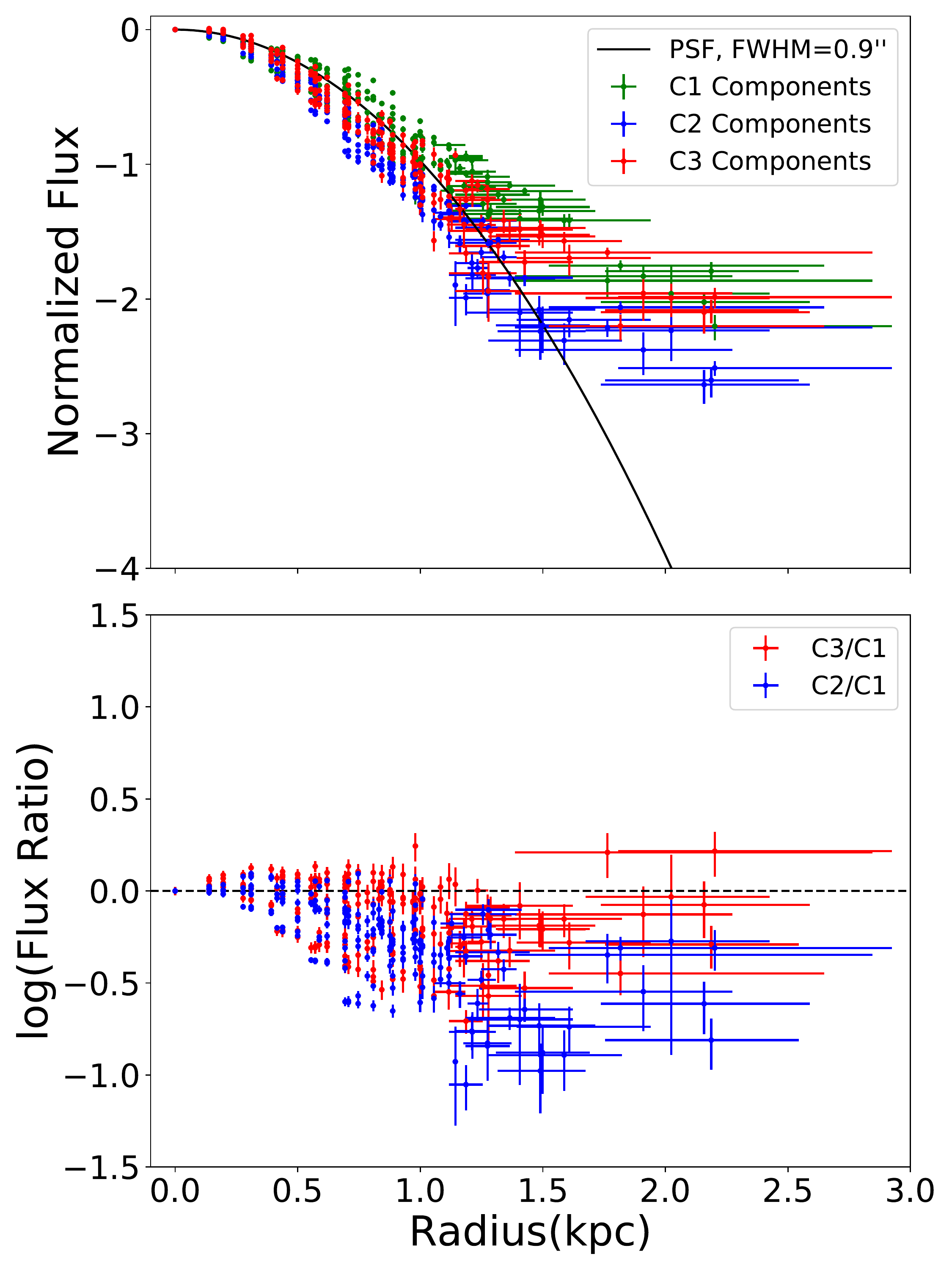}
\caption{Same as Fig. \ref{fig:radial1} but for the GMOS data (left) and KCWI data (right) of \ta.}
\label{fig:radial5}
\end{figure*}

\begin{figure}[!htb]   
\epsscale{1.1}
\plotone{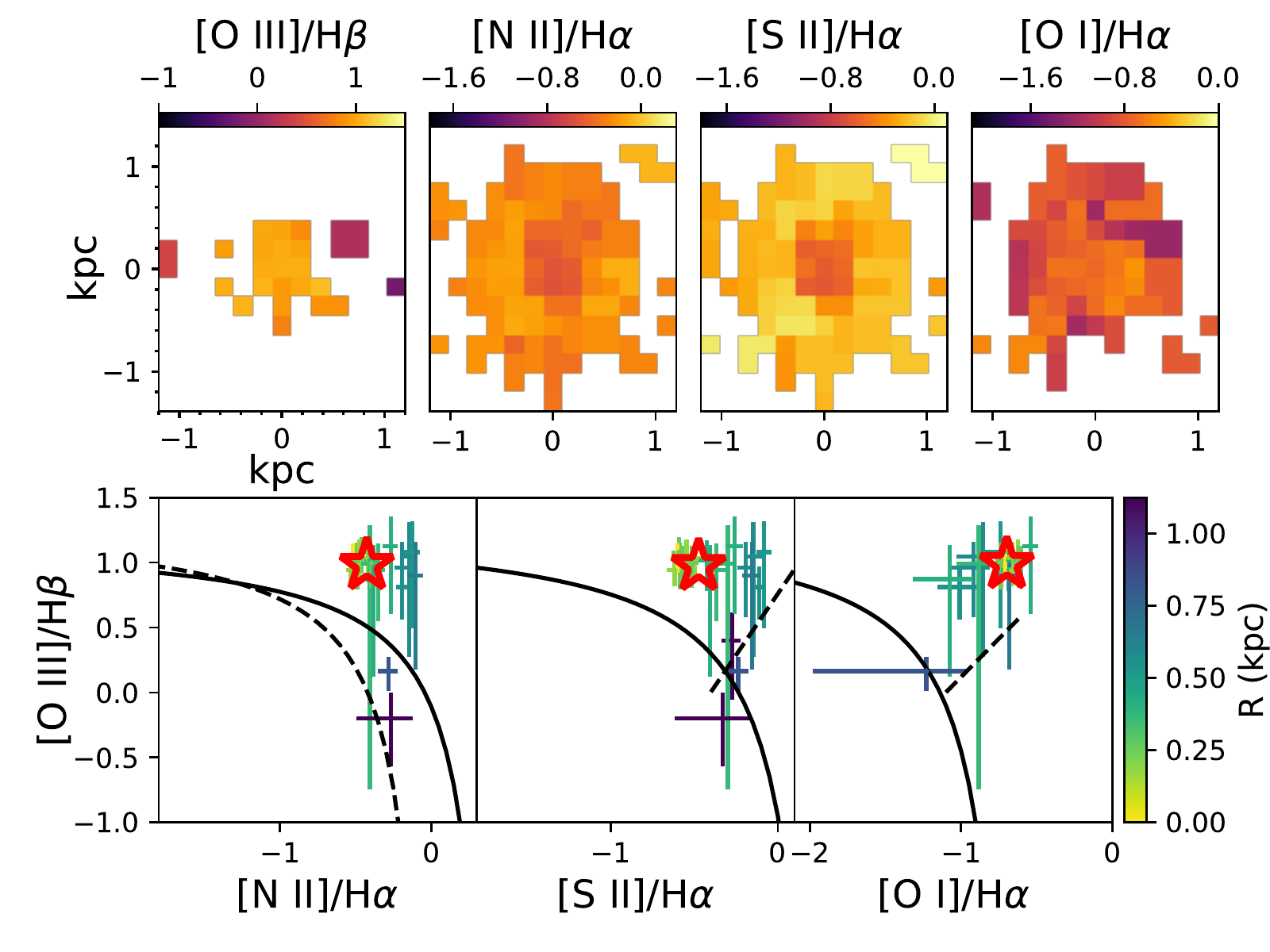}
\caption{Emission line diagnosis for the sum of the host component C1 $+$ outflow component C2 of \ta. Top row: Line ratio maps of \oiiihb, \niiha, \siiha\ and \oiha, from left to right. Bottom row: Standard BPT and VO87 diagrams. The data points are color-coded according to their projected distance (in kpc) to the center of the galaxies. The large red open star in each panel indicates the line ratios derived from the spatially-integrated spectrum. The presentation style and meaning of the symbols are the same as in Figs.\ \ref{fig:J0842bpt1} and \ref{fig:J0842bpt2}.}
\label{fig:J0906bpt2}
\end{figure}

\begin{figure}[!htb]   
\epsscale{1.1}
\plotone{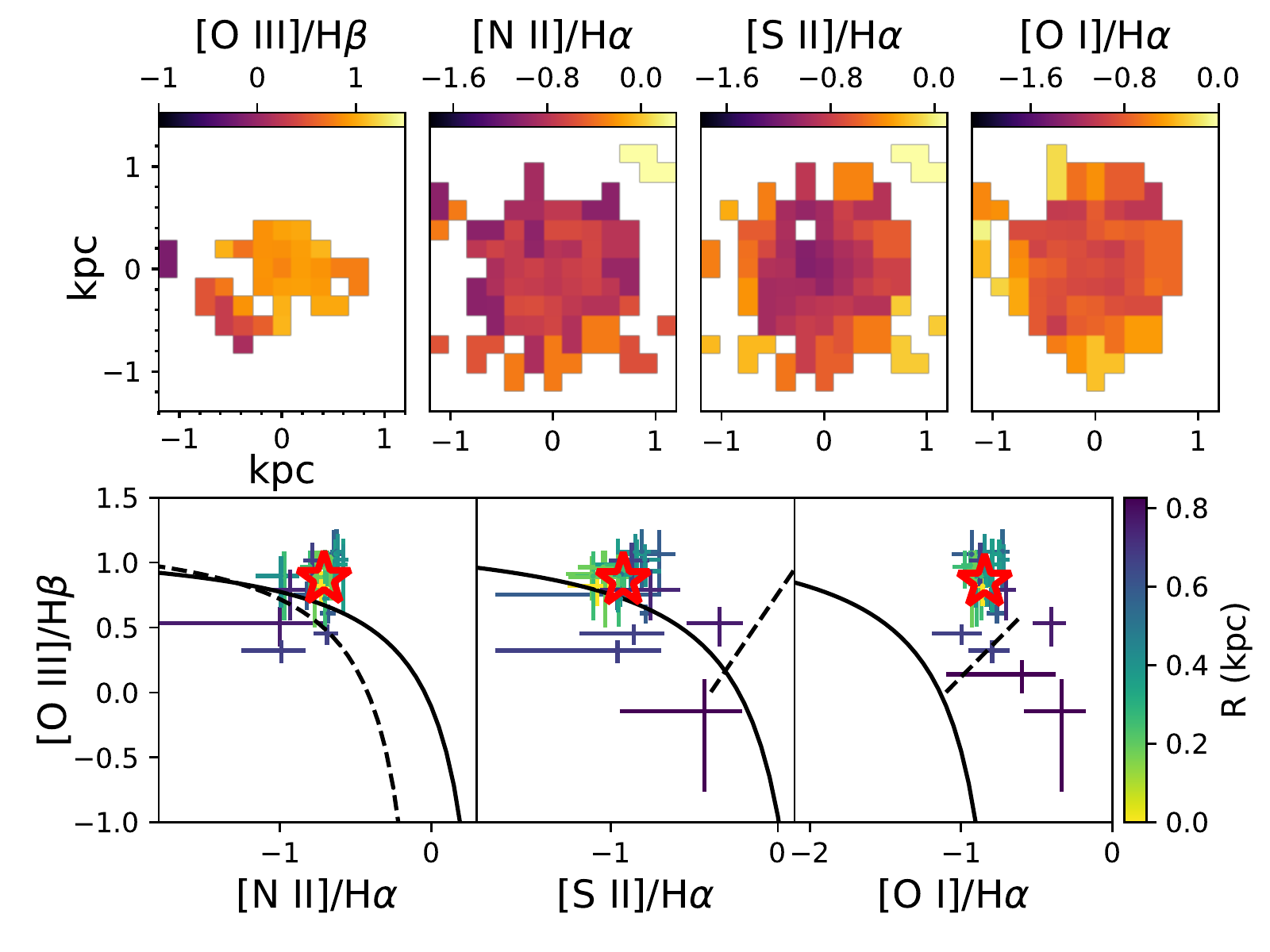}
\caption{Same as Fig.\ \ref{fig:J0906bpt2} but for the outflow component C3.}
\label{fig:J0906bpt3}
\end{figure}

\ta\ was observed with both GMOS and KCWI. While the KCWI data has a larger field of view, and detect emission lines out to a larger physical scale, the results from both data sets agree with each other in general, as in the case of \tb.

The maps of the \oiii\ flux and kinematics, stellar kinematics, as well as \oiii\ flux radial profiles of individual velocity components for this object are shown in Figs. \ref{fig:o3map5}--\ref{fig:radial5}. The line ratio maps and spatially resolved BPT and VO87 diagrams are shown in Figs. \ref{fig:J0906bpt2}--\ref{fig:J0906bpt3}. The map of the stellar kinematics based on the GMOS data is significantly more uncertain than the map based on the KCWI data due to the lower S/N in the stellar continuum in the former. Therefore, only the stellar kinematics maps based on the KCWI data are shown in Figure \ref{fig:stelmap5}.

\subsubsection{Maps of the \oiii\ Flux and Kinematics} \label{411}

The emission line profiles in this object are generally broad, with \wba\ reaching $\sim$650 \kms\ in the GMOS-based maps and $\sim$670 \kms\ in KCWI-based maps. Up to three Gaussian components are needed to properly fit the \oiii\ line emission. The stellar components show a clear velocity gradient with a PA $\simeq$ 30\textdegree\ and a velocity range from $\sim$$-$40 \kms\ to $\sim$$+$90 \kms.

The maps of the velocity-integrated \oiii\ flux and of the individual velocity components all show a roughly circular morphology. There are no clear offsets among the flux peaks of the global and flux individual components except for that of C1 components in the GMOS data, which is slightly offset to the east by $\sim$1 spaxel (0.2\arcsec) with respect to that of the global flux.

The C1 component shows a clear velocity gradient (with \vwu\ varying from $\sim$$-$30 \kms\ to $+$30 \kms\ in the GMOS data, stretching to $\sim$$\pm$50 \kms\ in the KCWI data). The orientation of the velocity gradient (PA $\simeq$ 30\textdegree) is very similar to that of the stellar velocities, while the velocity range of the C1 gas component ($\sim$$\pm$50 \kms) is slightly smaller than that of the stars ($\sim$$-$40 
to $\sim$$+$90 \kms) on the same spatial scale. The line width of the C1 component is generally narrow (median \wba\ $\simeq$ 110 \kms\ in the KCWI data), comparable to the stellar velocity dispersions. The C1 component represents the quiescent rotating material within the host galaxy. 

In contrast, the C3 component is generally blueshifted (with \vwu\ down to $-$50 \kms\ in the GMOS data and $-$70 \kms\ in the KCWI data) with respect to the systemic velocity derived from the stellar velocity field, and the line widths are very large (with \wba\ reaching $\sim$1200 \kms\ in the GMOS data and $\sim$1250 \kms\ in the KCWI data). These characteristics are strong evidence of a fast outflow in this system. 

The C2 component is in general redshifted with respect to the system velocity (with a median \vwu\ of $\sim$$+$30 \kms\ in the GMOs data and $\sim$$+$60 \kms\ in the KCWI data), and the line widths are clearly broader (median \wba\ $\simeq$ 350 \kms\ and $\simeq$ 430\kms in the GMOS and KCWI data, respectively) than those of the C1 components and the stellar velocity dispersions. No clear spatial gradient is seen in either quantities. One possibility is that the C2 component represents the far side of the same outflow traced by the C3 component. The smaller absolute velocities and smaller line widths of the C2 component can be explained in this picture if only a portion of the redshifted gas is visible, where the broader, more redshifted portion of the outflow is blocked by the galaxy itself. 
Alternatively, the C2 component may represent non-outflowing ionized gas in the (extended) narrow-line region (NLR) of the AGN, similar in line widths to NLR gas in other Seyfert 2 Galaxies \citep[e.g.][]{AGN1990}.  

Interestingly, a pc-scale ($\sim$47 pc) radio jet was recently reported in this target \citep{Yang2020}, which might be an important energy source for the outflowing ionized gas on kpc scale \citep[e.g.][]{nadiajenny2014,Morganti2015,Ramos2017}. It is difficult to directlty connect this radio jet to the outflowing ionized gas revealed by our data, due to the large difference in physical scales between the two, and our data do not provide clear information on the orientation of the outflow in the sky plane. Nevertheless, it seems that the radio jet might have enough kinetic energy ($P_{jet} = 10^{42.6\pm{0.7}} erg\ s^{-1}$) to drive the ionized gas outflow, assuming a simple scaling relation between radio luminosity and jet power \citep{Yang2020}.

\subsubsection{Flux Radial Profiles} \label{412}

The radial profiles of the velocity components are all slightly more extended than the PSF based on the GMOS data, where the fluxes are on average larger than the corresponding PSF values at the $\sim$2-$\sigma$ level at $r$ $\simeq$ 0.8 kpc in the GMOS data. For the KCWI data, the more extended PSF of the data makes the flux excesses less significant, although flux excesses are still seen for the C1 and C3 components at $r$ $\simeq$ 1.3 kpc (Fig.\ \ref{fig:radial5}). When comparing the fluxes of individual velocity components, the C3 component has a radial distribution that is very similar to that of the C1 component in both IFU data sets. Since C1 shows a clear, spatially resolved velocity gradient, the C3 component is also likely spatially resolved by our data, However, the C2 component is slightly more compact than the other two, where the C2/C1 flux ratios drops to $\sim$0.1--0.2 at r$\gtrsim$1 kpc in both GMOS and KCWI data.

\subsubsection{Ionization Diagnosis} \label{413}

Here we examine the ionization properties of individual velocity components based on the GMOS data in the same manner as that for \tb.
For the calculations of line ratios below, we combine the fluxes of the C1 and C2 components since i) we are interested in the difference, if any, between the pure, rapidly outflowing gas and the other gas components; ii) the C1 component is significantly fainter and less spectrally resolved, and thus has more uncertain line ratios than the other two components. 
For both the C3 and C1$+$C2 components, the \oiiihb\ ratios are roughly constant across the map. The other three line ratios, \niiha, \siiha, and \oiha, are also roughly constant or show only rather mild radial trends.

Both the C3 and C1$+$C2 components show spatially-integrated line ratios consistent with AGN in all three BPT and VO87 diagrams. For both the C3 and C1$+$C2 components, the line ratios of individual spaxels are also dominated by AGN-like line ratios, at least within r $\simeq$ 0.7 kpc. The \niiha\ and \siiha\ ratios are in general smaller than those measured in the more luminous AGN with more massive host galaxies. This is consistent with the lower gas-phase metallicity expected from this dwarf galaxy (log $M_*/M_\odot$ = 9.36; Table \ref{tab:targets}).

For the C3 component, there is a possible trend that line ratios in spaxels at larger radii are closer to the dividing lines of AGN and star-forming activity in all three BPT and VO87 diagrams. This is perhaps a sign that the ionization parameter decreases with increasing radii, as is generally the case in AGN, and/or that the relative contribution to the ionization/excitation from possible star-forming activity increases at larger radii.

\subsection{\tc} \label{43}

\begin{figure}[!htb]   
\epsscale{0.8}
\plotone{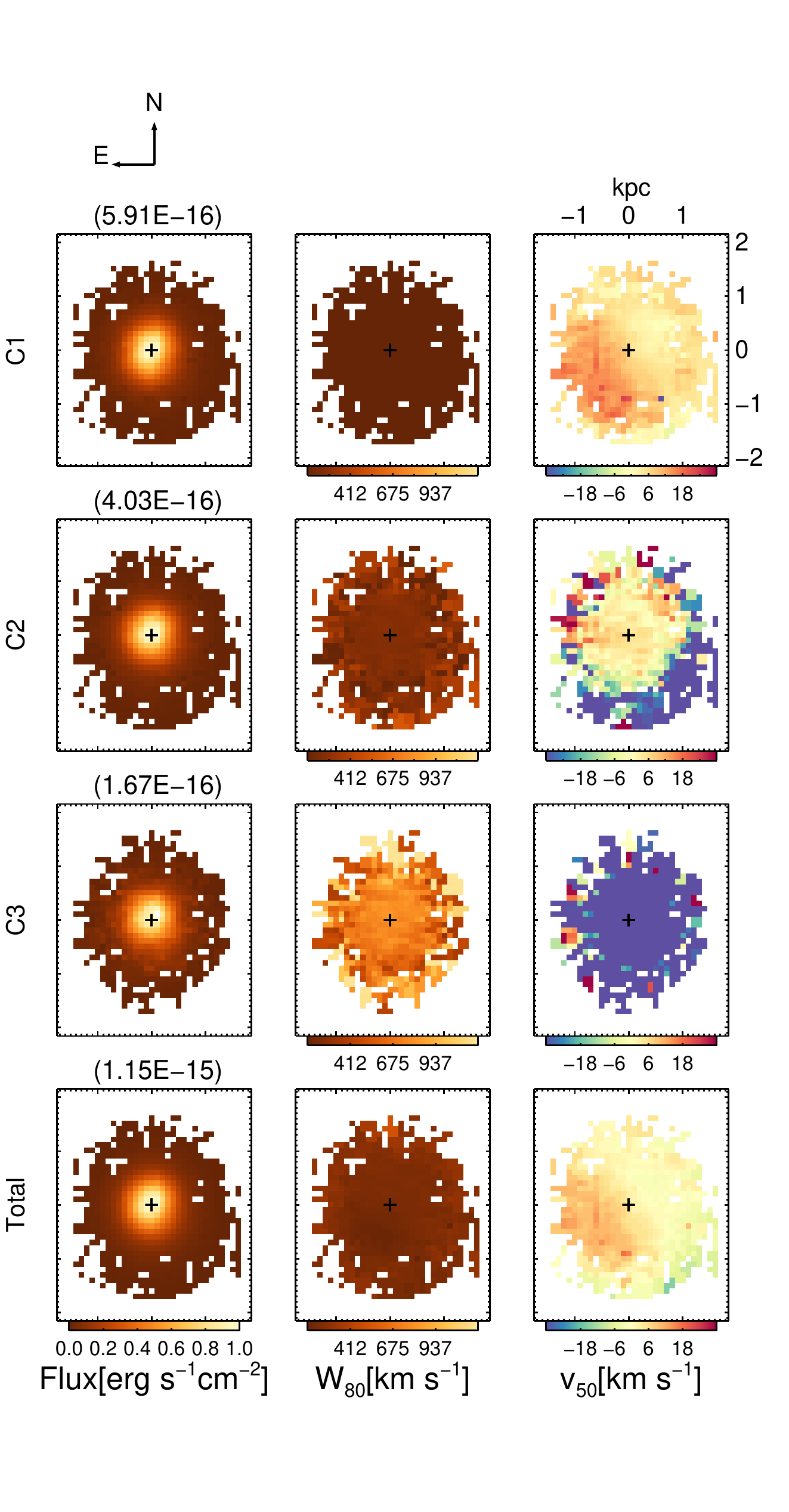}
\caption{Same as Fig.\ \ref{fig:o3map1} but for \tc\ based on the KCWI data.}
\label{fig:o3map3}
\end{figure}

\begin{figure}[!htb]   
\epsscale{0.5}
\plotone{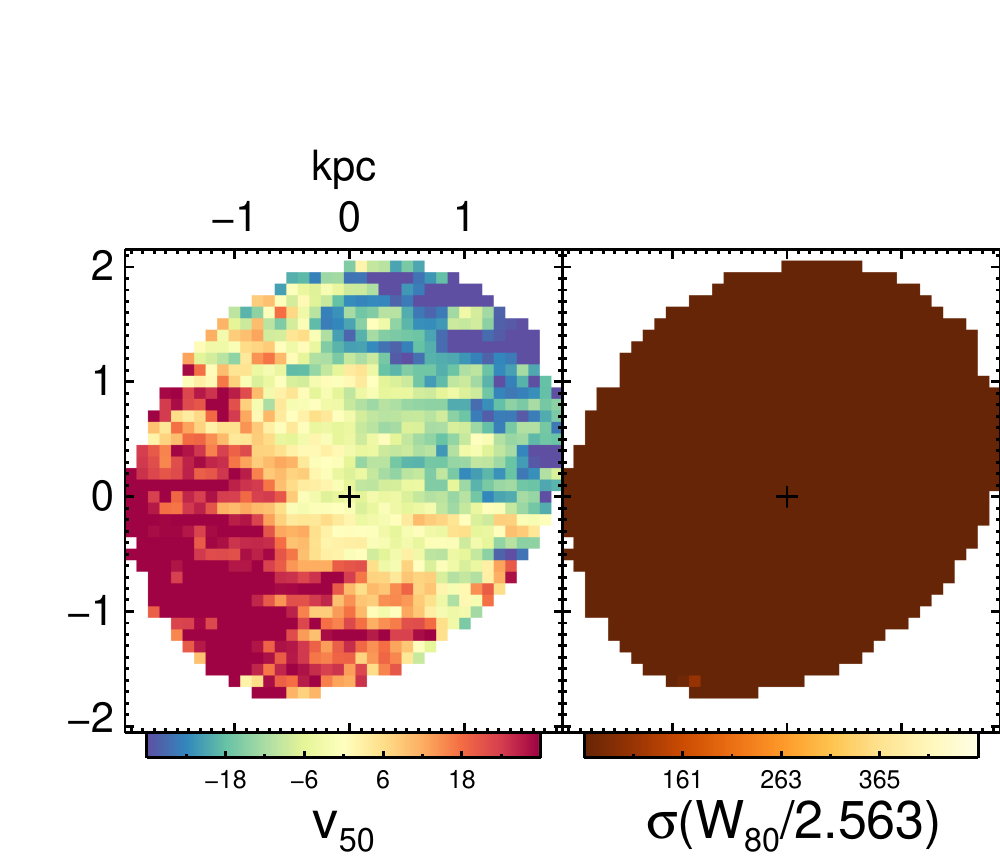}
\caption{Same as Fig.\ \ref{fig:stelmap1} but for \tc.}
\label{fig:stelmap3}
\end{figure}

\begin{figure}[!htb]   
\epsscale{0.5}
\plotone{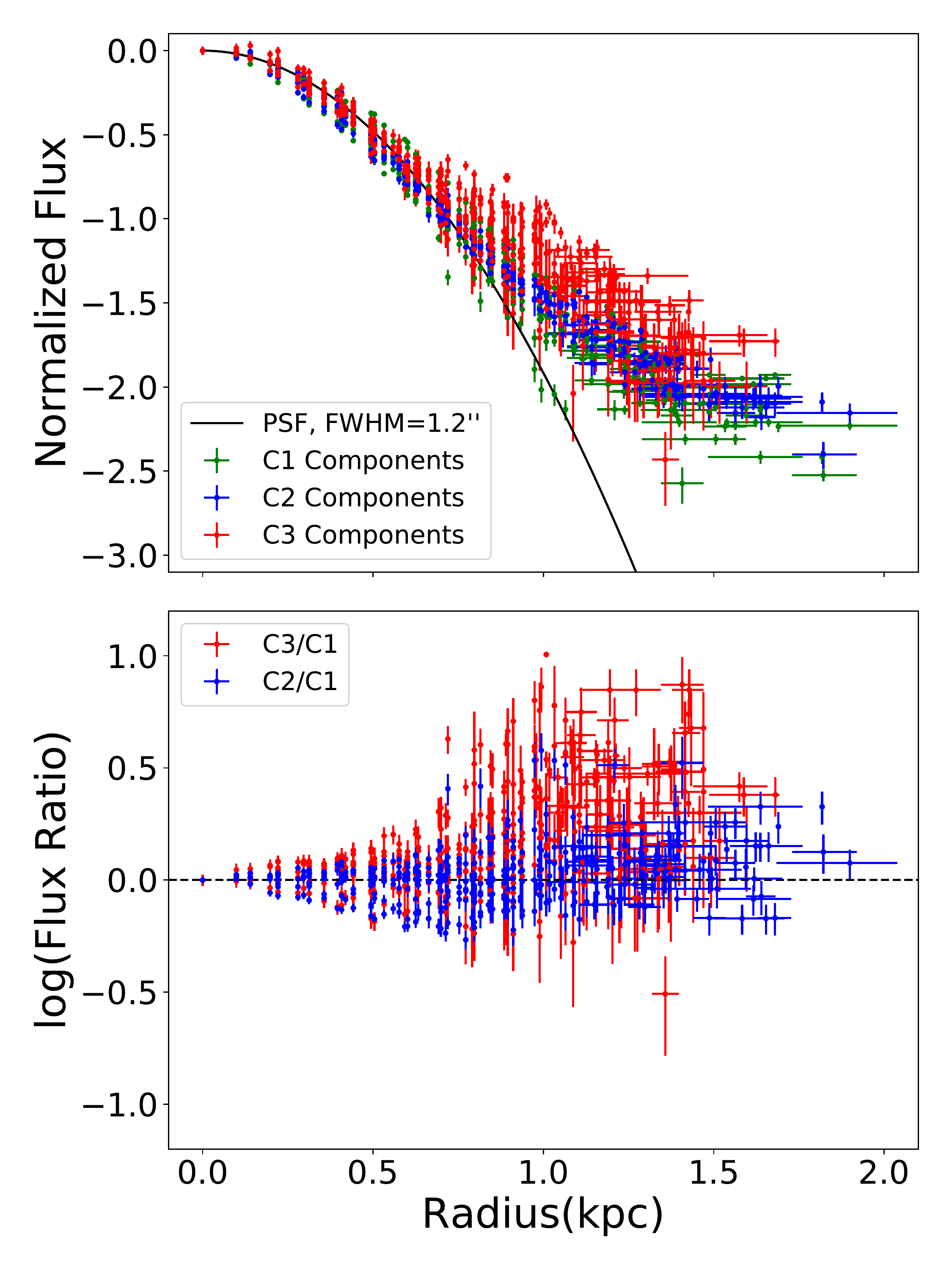} 
\caption{Same as Fig.\ \ref{fig:radial1} but for \tc.
}
\label{fig:radial3}
\end{figure}

The maps of the \oiii\ flux and kinematics, the map of stellar kinematics, and the \oiii\ flux radial profiles of the individual velocity components in this object are presented in Fig. \ref{fig:o3map3}--\ref{fig:radial3}.

\subsubsection{Maps of the \oiii\ Flux and Kinematics} \label{431}

Three velocity components (C1, C2, and C3) are needed to describe the \oiii\ line profiles in this system. The flux maps of the total emission line and individual components show a circular morphology in general. The map of the median velocities \vwu\ of the overall emission line profiles shows a gradient similar to that of the C1 component, although the C1 component is slightly more redshifted on average.

The C1 component shows a clear velocity gradient with \vwu\ varying from $\sim$0 \kms\ to $\sim$$+$20 \kms, with a position angle similar to that of the stellar velocity field (PA $\simeq$ $-$45\textdegree; Fig.\ \ref{fig:stelmap3}). The line widths are small in general (median \wba\ $\simeq$ 70 \kms), similar to the velocity dispersion of the stellar component. These results suggest that the C1 component is largely rotating in the potential well of the galaxy, but with a smaller velocity amplitude (the values of \vwu\ of the \oiii\ C1 component are on average $\sim$10--20 \kms\ smaller in absolute terms than the stellar values), 

The C2 component is generally close to the systemic velocity except in the southwestern region, where they are significantly blueshifted (by as much as $\sim$70 \kms) and slightly broader than in other parts of the galaxy. These blueshifted and broad velocity profiles may indicate the presence of outflowing and/or turbulent gas.    

The C3 component is significantly blueshifted with respect to the systemic velocity (\vwu\ down to $\sim$$-$80 \kms) and show large line width (\wba\ up to $\sim$1100 \kms). Mild radial gradients are seen in these quantities. The large line widths and clear blueshifts of the C3 component strongly suggest that they are associated with a fast outflow.

\subsubsection{Flux Radial Profiles} \label{432}

The \oiii\ flux radial profiles of the individual velocity components are largely consistent with the PSF within 1 kpc, but all show excess emission  (up to $\sim$4-$\sigma$ level) beyond $\sim$1 kpc. The flux ratios of C2/C1 and C3/C1 scatter around unity in general, suggesting that they share a similar radial distribution. The C3 components might thus be spatially resolved by our data, as the C1 and C2 components are very likely so judging from the clear spatial velocity gradients/structures seen in the \vwu\ maps.

\begin{figure}[!htb]   
\epsscale{0.8}
\plotone{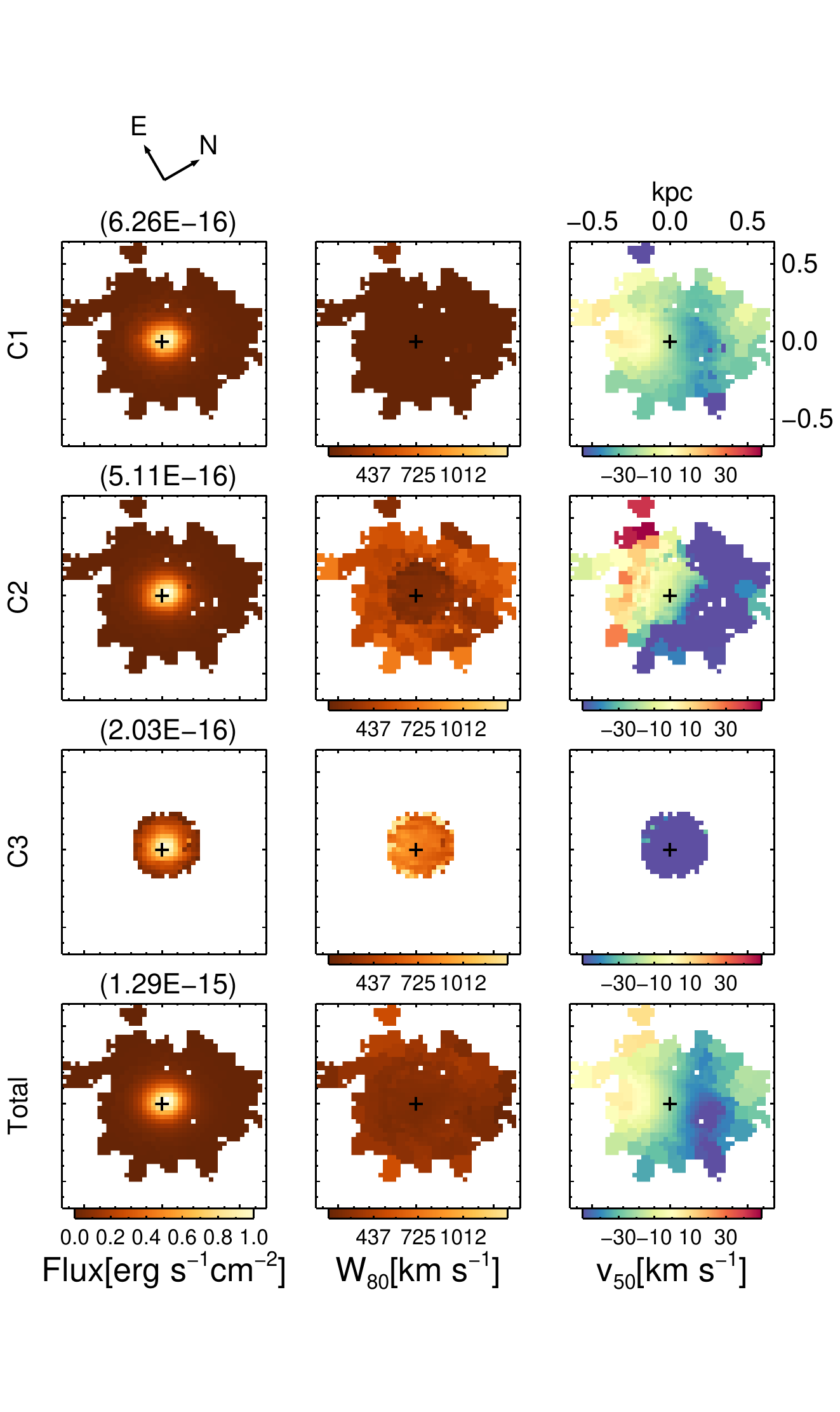}
\caption{Same as Fig.\ \ref{fig:o3map1} but for \td.}
\label{fig:o3map4}
\end{figure}

\begin{figure}[!htb]   
\epsscale{0.5}
\plotone{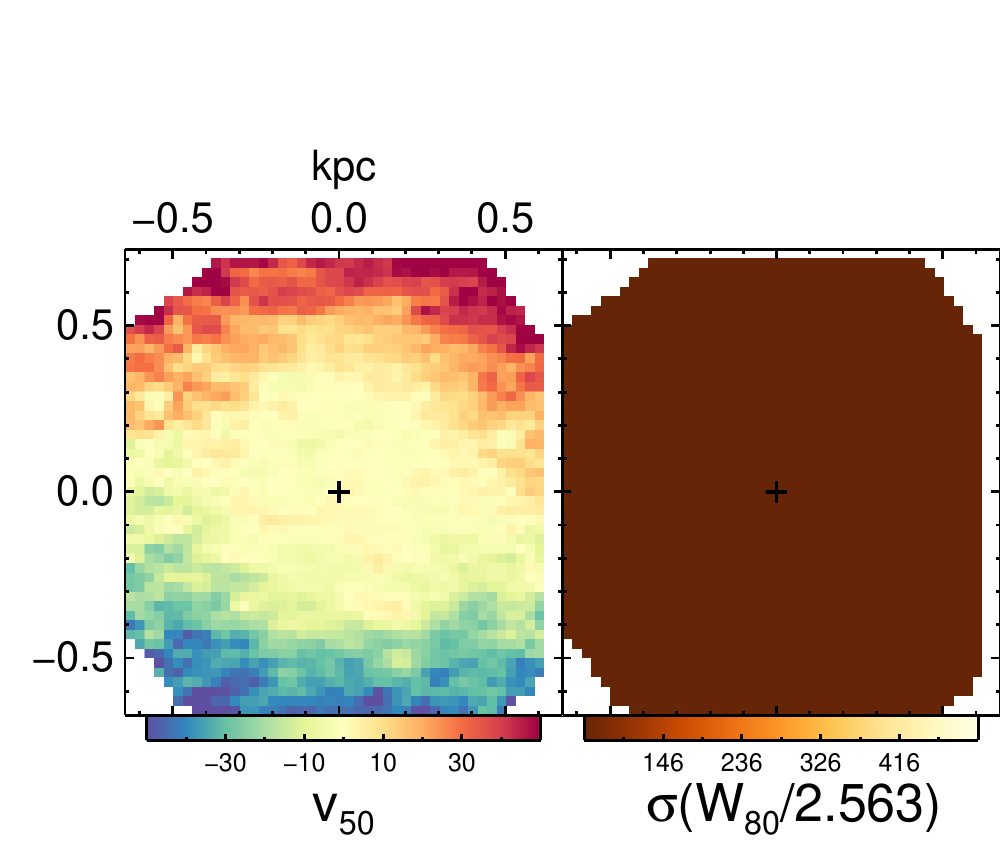}
\caption{Same as Fig.\ \ref{fig:stelmap1} but for \td.}
\label{fig:stelmap4}
\end{figure}

\begin{figure}[!htb]   
\epsscale{0.5}
\plotone{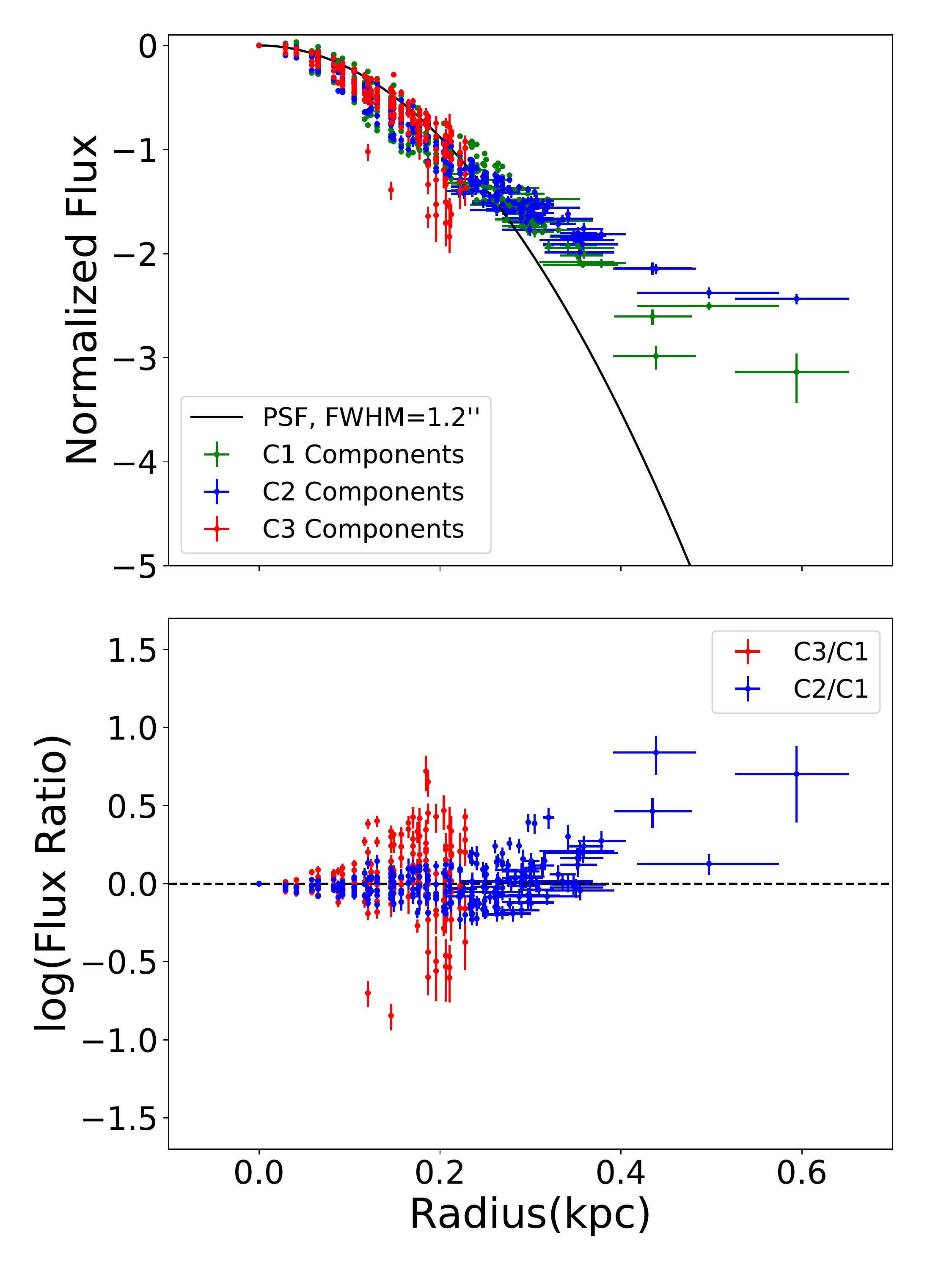}
\caption{Same as Fig.\ \ref{fig:radial1} but for \td. Contrary to Fig.\ \ref{fig:radial1}, the data points of C1 and C2 components beyond the maximal spatial extent of the C3 component are shown instead.}
\label{fig:radial4}
\end{figure}

\subsection{\td} \label{44}

The results on this object are presented in Fig. \ref{fig:o3map4}--\ref{fig:radial4}, in the same format as that for \tc.

\subsubsection{Maps of the \oiii\ Flux and Kinematics} \label{441}

Three velocity components (C1, C2, and C3) are required to fit the \oiii\ line profiles adequately in this object. The values of \vwu\ of the overall emission line profiles are slightly blueshifted ($\sim$$-$30 \kms), and show a gradient along PA $\simeq$ 30\textdegree\ that is very similar to that of the C1 component.

The C1 component shows a clear velocity gradient with \vwu\ ranging from $\sim$$-$40 \kms\ to $\sim$10 \kms, but this gradient is not centered on the systemic velocity and is perpendicular (PA$\simeq$ 150\textdegree) to that seen on a slightly large spatial scale in the stellar velocity field (PA $\simeq$ 60\textdegree). The line widths of the C1 component are in general narrow (median \wba\ $\simeq$ 80 \kms), similar to the stellar velocity dispersions. If we interpret the velocity gradient of the C1 component as a sign of rotation, the different angular momentum of the C1 component relative to that of the stellar component suggests that the C1 component consists of gas acquired externally after the stars in the galaxy were already in place \citep[e.g.][]{Chen2018}. The overall blueshift of the C1 component may also hint at the influence of an outflow (see below). 

The C2 component shows a dramatic velocity gradient where \vwu\ vary from $\sim$$-$100 \kms\ to 50 \kms. The kinematic major axis of the C2 component has a PA $\simeq$ 135\textdegree, offset from those of the C1 component and stellar velocity field. The line widths of the C2 component are also significantly larger (median \wba\ $\simeq$ 440 \kms) than those of the C1 component. The interpretation of the C2 component is unclear; it may represent a mixture of rotating, turbulent and outflowing gas in the galaxy. 

The nature of the C2 component is further explored by fitting the \vwu\ map of this component with \textit{Kinemetry}, following the same procedure as in Appendix \ref{451}. The residual velocities (observed \vwu\ $-$ best-fit circular velocities) show absolute amplitudes similar to those of the best-fit circular velocities. This suggests that a pure, rotating disk cannot explain the kinematics of the C2 component alone, which further supports our statement in the previous paragraph that it is partially affected by outflowing gas.

The C3 component is blueshifted (\vwu\ down to $\sim$$-$200 \kms) and shows large line widths (\wba\ up to $\sim$1200 \kms). Mild radial gradients are seen in these quantities. The large line widths and clear blueshifts of the C3 component suggest that they are most likely due to a fast outflow. 

\subsubsection{Flux Radial Profiles} \label{442}

The \oiii\ flux radial profiles of the individual velocity components are largely consistent with the PSF within $\sim$0.2 kpc, but excess flux is detected (at the $\sim$4-$\sigma$ level on avaerage) beyond 0.2 kpc in C1 and C2 velocity components (Fig.\ \ref{fig:radial4}). The flux ratios among individual velocity components are in general scattered around unity within 0.2 kpc, suggesting no difference of radial flux distributions among the three velocity components. However, the C2 component may have slight flux excess compared to the C1 component beyond $\sim$0.3 kpc.

\subsection{\tf} \label{46}

\begin{figure}[!h]  
\plotone{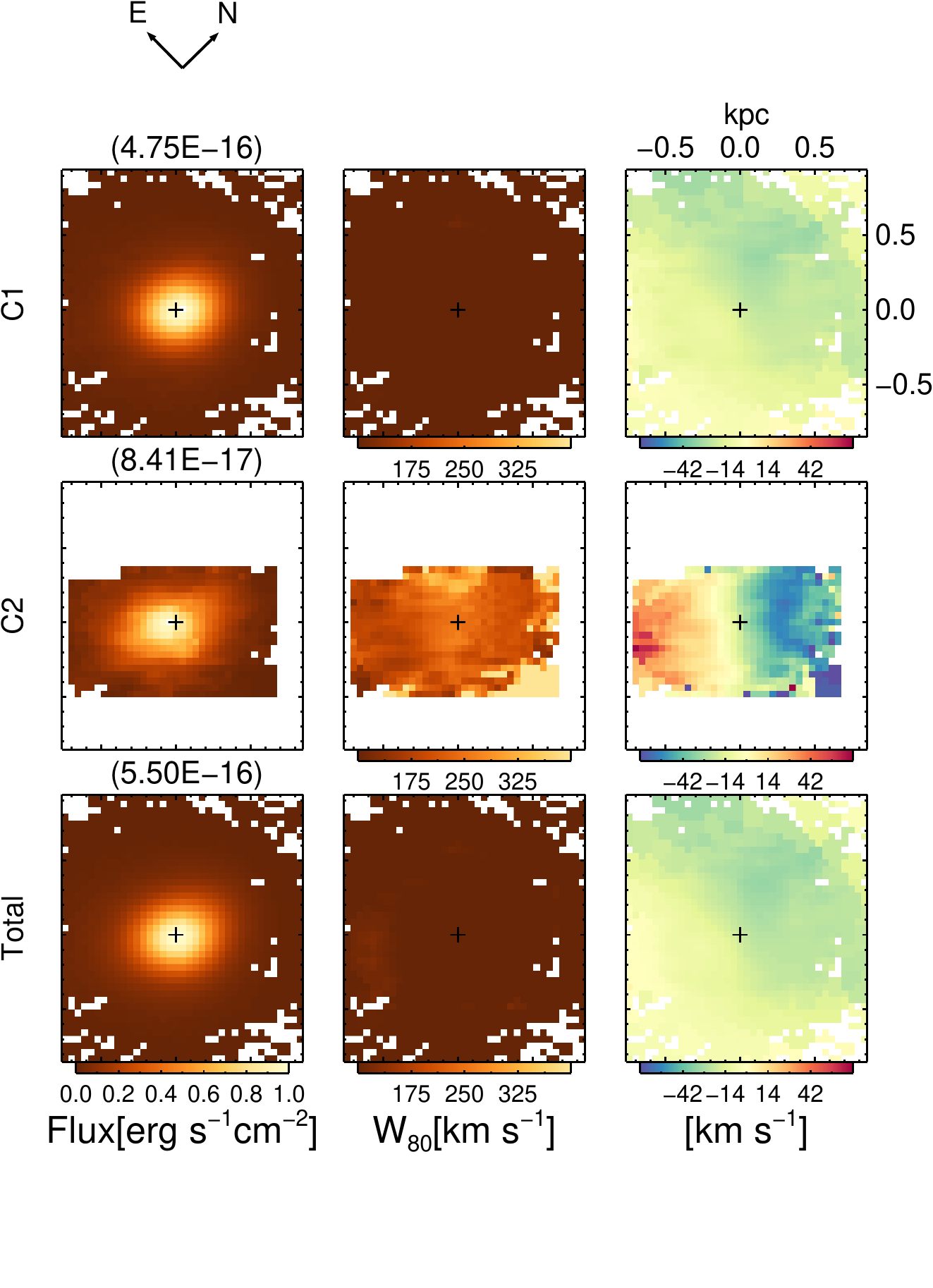}
\caption{Same as Fig.\ \ref{fig:o3map1} but for \tf.}
\label{fig:o3map6}
\end{figure}

\begin{figure}[!htb]   
\epsscale{0.6}
\plotone{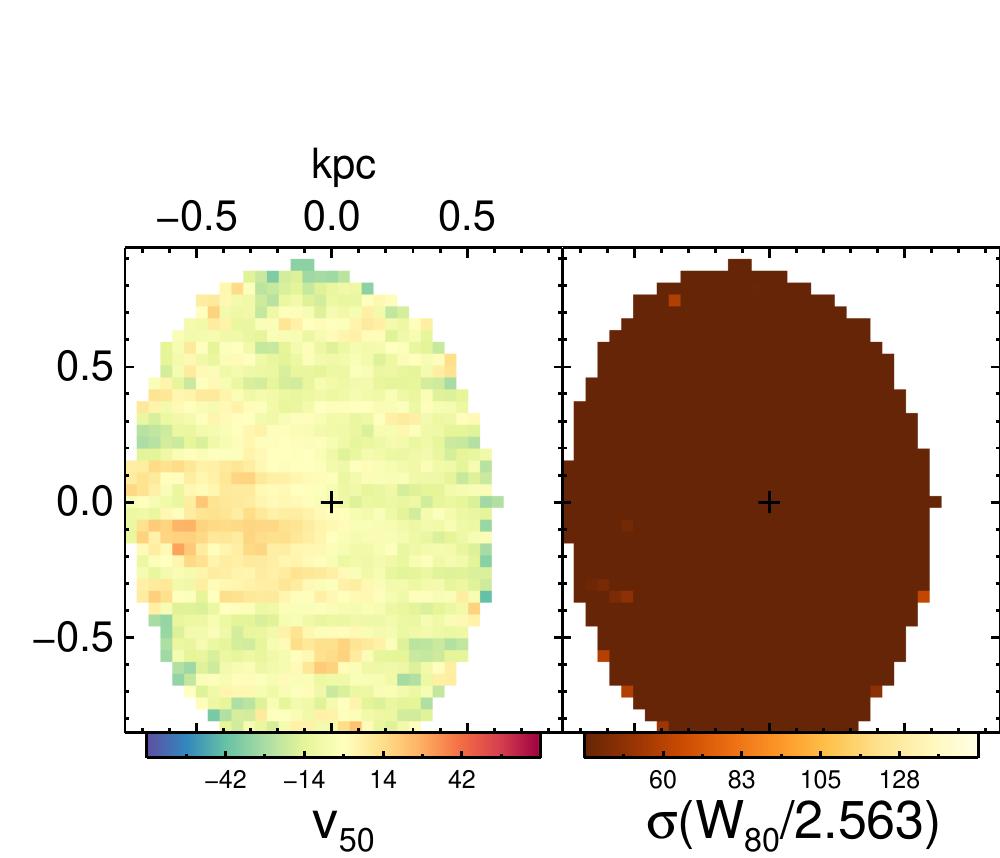}
\caption{Same as Fig.\ \ref{fig:stelmap1} but for \tf.}
\label{fig:stelmap6}
\end{figure}

\begin{figure}[!htb]   
\epsscale{0.5}
\plotone{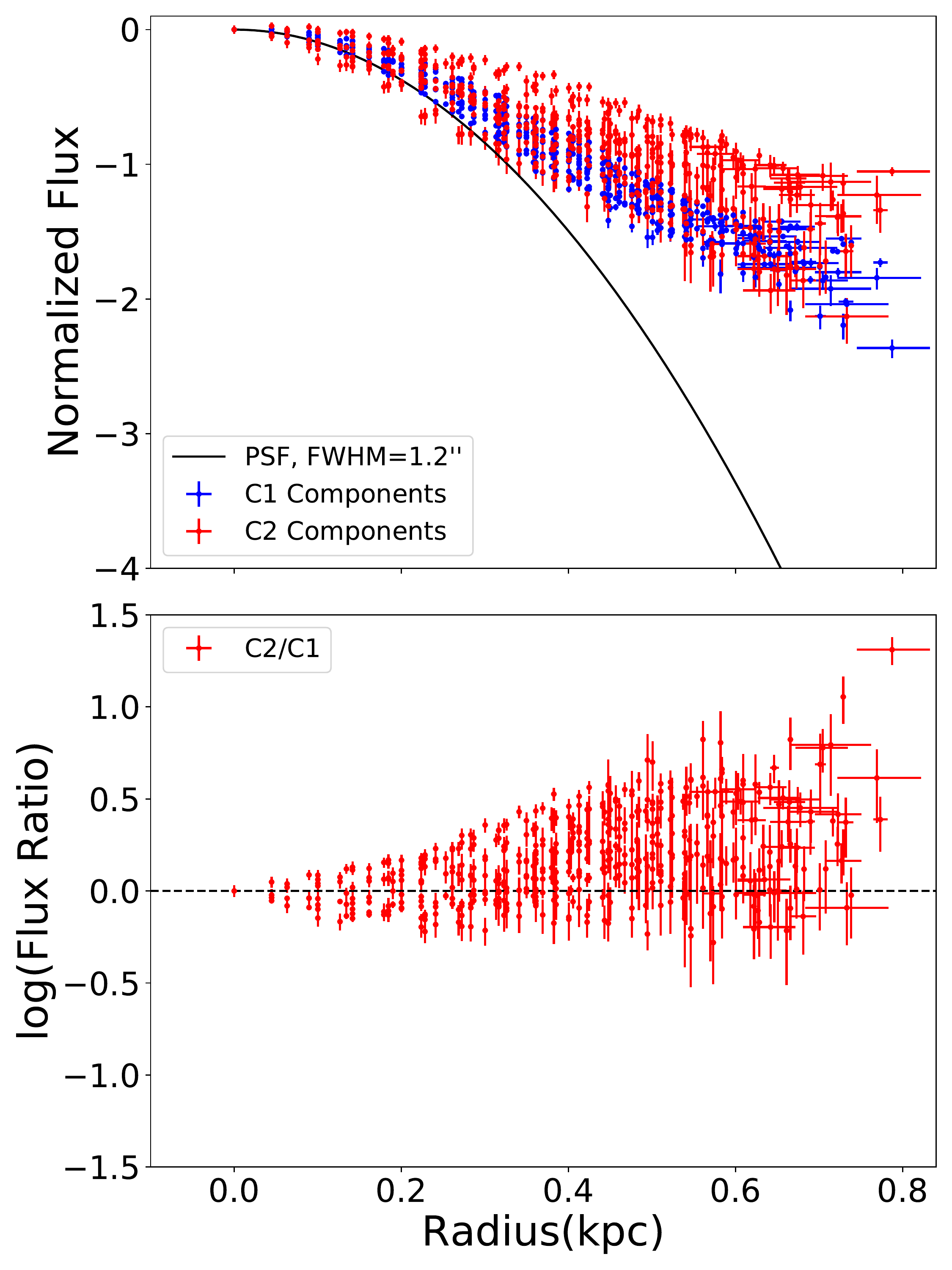}
\caption{Same as Fig.\ \ref{fig:radial1} but for \tf. 
}
\label{fig:radial6}
\end{figure}

The results of our analysis of the KCWI data for \tf\ are presented in Fig. \ref{fig:o3map6}--\ref{fig:radial6}.

\subsubsection{Maps of the \oiii\ Flux and Kinematics} \label{461}

Two Gaussian components are sufficient to describe the \oiii\ emission line profiles in this object. The maps of \vwu\ and \wba\ of the overall line profiles show apparent gradients and structures that are very similar to those of the C1 component (Fig.\ \ref{fig:o3map6}).

The values of \vwu\ in the C1 component are on average slightly blueshifted ($\sim$$-$10 \kms) compared with the stellar velocities (shown in Fig.\ \ref{fig:stelmap6}). The PA of the \vwu\ gradient is $\sim$0\textdegree, similar to that of the gradient of the stellar components. The line widths of the C1 components are once again narrow (median \wba\ $\simeq$ 80 \kms), and similar to the stellar velocity dispersions. The C1 component appears to be associated with gas that is largely rotating within the galaxy, but the slight overall blueshift may be a sign of a small bulk outflow.

The flux peak of the C2 component is slightly offset from that of the C1 component by $\sim$0.15\arcsec\ (1 spaxel) to the southeast. The map of \vwu\ of the C2 component shows a clear gradient along the SE-NW direction (PA $\simeq$ $-$40\textdegree), much steeper (\vwu\ varies from $-$60 \kms\ to 40 \kms) and offset in position angle with respect to that seen in the C1 component.  The line widths of the C2 component are also much broader than those of the C1 component (with \wba\ reaching 480 \kms), and no clear spatial structure is seen. The C2 component may represent a tilted, biconical outflow, like that of \te, where the near (NW) side of the outflow is blueshifted and the far (SE) side is redshifted. Alternatively, the apparent bisymmetry of the velocity field of the C2 component may be interpreted as a rotating structure, but the large line widths indicate that the gas is turbulent. 

This result is confirmed using {\em Kinemetry}: while the residual velocities (observed \vwu\ $-$ best-fit circular velocities) for the C2 component do not show clear patterns indicative of biconical outflowing gas, the best-fit circular velocity field of the C2 component has significantly larger amplitudes and a clearly different position angle when compared to those of the C1 component, as reported above from our visual examination of the observed kinematic maps.

\subsubsection{Flux Radial Profiles} \label{462}

As shown in Fig. \ref{fig:radial6}, the flux radial profiles of both velocity components are clearly more extended than the PSF, which is consistent with the spatial gradient clearly seen in the \vwu\ maps. The C2 component has excess flux relative to the C1 component beyond $\sim$0.3 kpc, where the flux ratios reach a maximum of $\sim$10 at around 0.7 kpc.

 \end{document}